\documentclass{scrartcl}

\usepackage[utf8]{inputenc}
\usepackage[T1]{fontenc}
\usepackage[margin=14mm]{geometry}
\usepackage[svgnames]{xcolor}
\usepackage{multicol,microtype,enumitem,lmodern,dsfont}

\usepackage{mathtools,amssymb,tensor}

\usepackage[colorlinks=true,allcolors=DarkBlue]{hyperref}
\usepackage{cleveref}
\Crefname{equation}{Eq.}{Eqs.}
\crefname{section}{sec.}{secs.}
\crefname{appendix}{app.}{apps.}
\crefname{cond}{condition}{conditions}
\AtBeginDocument{\let\condlabel=\label}
\newcounter{condition}[equation]
\newcommand{\cond}[1]{\refstepcounter{condition}\condlabel[cond]{#1}\thecondition.}

\usepackage[backend=bibtex,sorting=none,citestyle=numeric-comp,giveninits=true,doi=false,date=year,url=false]{biblatex}
\renewbibmacro{in:}{}

\AtEveryBibitem{\clearfield{number}\clearfield{issue}\clearfield{url}}
\DeclareFieldFormat[article]{volume}{\mkbibbold{#1}}
\DeclareFieldFormat{eprint:arxiv}{\href{http://arxiv.org/abs/#1}{#1}}
\DeclareFieldFormat{pages}{\mkfirstpage{#1}}

\usepackage[titletoc,title]{appendix}

\let\Re\relax
\DeclareMathOperator{\Re}{Re}

\newcommand\dif{\mathrm{d}}
\renewcommand\vec[1]{\boldsymbol{#1}}

\DeclareMathOperator\tr{tr}
\DeclareMathOperator\diag{diag}
\DeclareMathOperator\spec{spec}
\DeclareMathOperator\tord{\mathcal{T}}
\newcommand\tran{\mathrm{T}}

\newcommand{\ba}{\begin{eqnarray}}
\newcommand{\ea}{\end{eqnarray}}
\newcommand{\be}{\begin{equation}}
\newcommand{\ee}{\end{equation}}
\newcommand{\nn}{\nonumber}

\newcommand{\te}{(t+\epsilon)}
\newcommand{\q}{\tilde q_}

\newenvironment{alignedeqn}{\begin{equation}\begin{aligned}}{\end{aligned}\end{equation}\ignorespacesafterend}

\begin{document}

\title{Quantum formalism for classical statistics}
\author{C. Wetterich\\
	\small\href{eqn:mailto:c.wetterich@thphys.uni-heidelberg.de}{c.wetterich@thphys.uni-heidelberg.de}}
\date{\small Universität Heidelberg, Institut für Theoretische Physik, Philosophenweg 16, D-69120 Heidelberg}
\maketitle

\begin{abstract}
	\noindent In static classical statistical systems the problem of information transport from a boundary to the bulk finds a simple description in terms of wave functions or density matrices. While the transfer matrix formalism is a type of Heisenberg picture for this problem, we develop here the associated Schrödinger picture that keeps track of the local probabilistic information. The transport of the probabilistic information between neighboring hypersurfaces obeys a linear evolution equation, and therefore the superposition principle for the possible solutions. Operators are associated to local observables, with rules for the computation of expectation values similar to quantum mechanics. We discuss how non-commutativity naturally arises in this setting. Also other features characteristic of quantum mechanics, such as complex structure, change of basis or symmetry transformations, can be found in classical statistics once formulated in terms of wave functions or density matrices. We construct for every quantum system an equivalent classical statistical system, such that time in quantum mechanics corresponds to the location of hypersurfaces in the classical probabilistic ensemble. For suitable choices of local observables in the classical statistical system one can, in principle, compute all expectation values and correlations of observables in the quantum system from the local probabilistic information of the associated classical statistical system. Realizing a static memory material as a quantum simulator for a given quantum system is not a matter of principle, but rather of practical simplicity.
\end{abstract}

\BeforeStartingTOC[toc]{\begin{multicols}{2}}
\AfterStartingTOC[toc]{\end{multicols}}
%{\hypersetup{hidelinks}\tableofcontents}

\newpage
\tableofcontents

\begin{multicols}{2}
\newpage
~~

\newpage
\section{Introduction}
\label{eqn:Introduction}

A key issue for the understanding of memory and computing structures concerns the question to which extent information imprinted on the boundary of a material remains available in the bulk or at some other boundary \cite{SHA,LTP,CWEL,NP,ALL,KWC,LBL,MMW,BRH}. We have proposed \cite{CWIT} a formalism for this problem based on the concept of classical wave functions \cite{CWQP} or the associated density matrix. This formalism resembles the derivation of the wave function from Feynman's path integral for quantum mechanics \cite{FEY}, with time $t$ replaced by the location of hypersurfaces in the bulk of the material.

For static classical statistical systems the problem of information transport can be set up in the transfer matrix formalism \cite{TM,MS,FU}. This formalism corresponds to the Heisenberg picture in quantum mechanics. Our approach based on classical wave functions and density matrix corresponds to the associated Schrödinger picture. Similar to quantum mechanics, wave functions and density matrix keep track of the local probabilistic information.

The problem of information transport is set as the question how a given probabilistic information on a hypersurface at $t$ determines the probabilistic information on the neighboring hypersurface at $t + \epsilon$. Encoding the local information at $t$ in a suitable classical density matrix $\rho^\prime(t)$, one obtains a linear evolution law that expresses $\rho^\prime(t + \epsilon)$ in terms of $\rho^\prime(t)$,
\begin{equation}\label{eqn:evolution}
	\rho^\prime(t + \epsilon)
	= S(t) \, \rho^\prime(t) \, S^{-1}(t),
\end{equation}
with step evolution operator $S$ \cite{CWIT} related to the transfer matrix \cite{TM,MS,FU}. With boundary information set at $t_\text{in}$ by $\rho^\prime(t_\text{in})$, one can follow the evolution to arbitrary $t$ in the bulk, or to $t_\text{f}$ at some other boundary. In the continuum limit \cref{eqn:evolution} entails a von Neumann type equation,
\begin{equation}\label{eqn:von neumann eqn}
	\partial_t \rho^\prime(t)
	= [W,\rho^\prime(t)].
\end{equation}
Operators $A^\prime(t)$ can be associated to local observables $A(t)$ such that the expectation value can be computed according to the ``quantum rule''
\begin{equation}\label{eqn:quantum rule}
	\langle A(t)\rangle
	= \tr\bigl\{A^\prime(t) \, \rho^\prime(t)\bigr\}.
\end{equation}

In the path integral for quantum mechanics, the weight factor is a complex phase $e^{i S_\text{Q}}$. In contrast, for classical statistics the weights are positive real probabilities with weight factor $e^{-S_\text{cl}}$. This leads to three important differences between quantum mechanics and the more general classical statistics. First, instead of a single complex wave function as in quantum mechanics, classical statistical systems are described by a pair of distinct real wave functions. As a consequence, the norm of the wave functions is not conserved by the evolution. This is the basis for the second difference, namely that the operator $W$ in \cref{eqn:von neumann eqn} is in general not antisymmetric. In the presence of a complex structure, where \cref{eqn:von neumann eqn} becomes
\begin{equation}
	i \partial_t \rho
	= [G,\rho],
\end{equation}
the operator $G$ is not Hermitian and the $t$-evolution in classical statistics in general not unitary. Third, not all arbitrary ``initial'' density matrices $\rho^\prime(t_\text{in})$ or $\rho(t_\text{in})$ can necessarily be realized in classical statistics.

Only for particular ``static memory materials'' the operator $G$ becomes Hermitian and can be identified as the Hamiltonian. In these cases also arbitrary $\rho^\prime(t_\text{in})$ can be realized. Even though realized as classical statistical systems, such particular memory materials with unitary $t$-evolution constitute quantum systems satisfying all axioms of quantum mechanics. They may also be dubbed ``quantum simulators'' since the dependence of expectation values $\langle A(t)\rangle$ on the location of the hypersurface at $t$ is the same as the time dependence in a quantum system with identical Hamiltonian. In this sense quantum mechanics emerges as a particular case of classical statistics \cite{CWQM}.

It is the aim of the present paper to explore and exploit the quantum formalism for static classical statistical systems in the general case where the evolution is not necessarily unitary. In particular, we discuss how non-commuting operators arise naturally in classical statistics. We will find many structures familiar from quantum mechanics, such as the possibility to change the basis or symmetry transformations acting on the wave functions or density matrix. For systems admitting a suitable evolution that allows for a complex structure one recovers the complete formalism of quantum mechanics, with the important exception that the evolution is not necessarily unitary and the density matrix is not hermitian.

Furthermore, we will find local similarity transformations that map between equivalent quantum and classical statistical systems. Expectation values and correlations of local observables can be evaluated both in the classical statistical or in the quantum system if the operators representing observables are properly transformed. This weak equivalence demonstrates that there is no difference in principle between the information contained in the corresponding classical statistical and quantum systems. A practical use of this map for the construction of classical statistical static memory materials as quantum simulators depends on the question if the map of density matrices and local observables is sufficiently simple.

This paper is organized as follows: in \cref{sec:classical wave function} we introduce the classical wave functions and establish the evolution law in a discrete formulation. The continuum limit leads to a generalized Schrödinger type evolution equation. Simple generalized Ising models based on particular step evolution operators are introduced in \cref{sec:simple models}. They serve as concrete examples for the general structures discussed in the paper. In \cref{sec:Three-spin chains} we solve the boundary problem for a class of three-spin chain generalized Ising models. They can realize quantum gates for a quantum spin. \Cref{sec:observables} discusses the general structure of local observables whose expectation values can be computed from the wave functions at a given $t$. This reveals the presence of non-commuting operator structures for observables in classical statistics.

\Cref{sec:density matrix} introduces the classical density matrix $\rho^\prime(t)$ and the associated von Neumann type evolution equation. In \cref{sec:solution of bvp} we solve explicitly the boundary problem for the four-state oscillator chain introduced in \cref{sec:simple models} by use of the density matrix. This demonstrates that arbitrary density matrices at the boundary at $t_\text{in}$ cannot always be realized for classical statistical systems. In \cref{sec:evolution for subsystems} we turn to the evolution of subsystems. The situation where a subsystem obeys a unitary evolution describes a quantum system embedded in an environment. It is realized in a wide range of systems. The evolution of the subsystem is given by quantum mechanics, while the evolution of the whole system includes the non-unitary evolution of the environment. Typically, the unitary evolution of suitable subsystems is the only evolution that survives far enough into the bulk. \Cref{sec:complex structure} briefly discusses the realization of a complex structure.

\Cref{sec:change of basis} addresses the change of basis, similarity transformations and symmetries. In \cref{sec:equivalent local probabilistic systems} we show how different classical statistical systems lead to the same continuous evolution equation. In particular, this makes contact to the unitary evolution operator for quantum systems. The general connections between quantum mechanics and classical statistics are discussed in \cref{sec:classical and quantum statistics}. Finally, in \cref{sec:conclusions} we present our conclusions.

\section{Classical wave function}
\label{sec:classical wave function}

In this section we recall the concept of ``classical wave functions'' for classical statistical systems. They encode the local statistical information necessary to describe expectation values of local observables $\langle A(t)\rangle$, together with the evolution law for the $t$-dependence of these expectation values. We consider a $d$-dimensional cubic lattice, with $t$ labeling a sequence of hypersurfaces. The lattice points may be denoted by discrete positions $(t,\vec x)$, where $t$ singles out the coordinate for which we consider the evolution. We discuss generalized Ising models \cite{IM,IS,binder2001ising}, with variables $s_\alpha(t,\vec x) = s_\gamma(t) = \pm 1$.

\subsection{Wave function from ``functional integral''}

The starting point is the probability or weight distribution $w[s]$ which associates to every configuration of Ising spins $\{s\} = \{s_\alpha(t,\vec x)\}$ a positive (or zero) probability $w[s]$. We will use a collective index $\gamma$, i.e. $s_\gamma(t) = s_\alpha(t,\vec x)$. For generalized Ising models one has
\begin{equation}\label{eqn:AF1}
	w[s]
	= Z^{-1} K[s] \, b[s_\text{in},s_\text{f}],
\end{equation}
with $K$ given by the classical action $S_\text{cl}$,
\begin{equation}\label{eqn:AF2}
	K[s]
	= \exp\bigl\{-S_\text{cl}[s]\bigr\}
	= \exp\Bigl\{-\sum_{t^\prime} \mathcal{L}(t^\prime)\Bigr\},
\end{equation}
and $\mathcal{L}(t^\prime)$ involving spins on two neighboring $t$-layers $t^\prime$ and $t^\prime + \epsilon$. The boundary term
\begin{equation}\label{eqn:AF2A}
	b(s_\text{in},s_\text{f})
	= \bar{f}_\text{f}(s_\text{f}) \, f_\text{in}(s_\text{in})
\end{equation}
depends only on the spins at the boundaries at $t_\text{in}$ and $t_\text{f}$, $s_{in,\gamma} = s_\gamma(t_\text{in})$, $s_{f,\gamma} = s_\gamma(t_\text{f})$. It is chosen such that $w[s]$ is positive, $w[s] \geq 0$. The partition function $Z$ is given by
\begin{equation}\label{eqn:partition function def}
	Z
	= \int \mathcal{D} s \, K[s] \, b,
\end{equation}
where $\int \mathcal{D} s$ denotes the sum over all spin configurations $\{s\}$. It normalizes the weight distribution according to
\begin{equation}\label{eqn:AF3}
	\int \mathcal{D} s \, w[s]
	= 1.
\end{equation}
By a suitable additive constant in $\mathcal{L}(t)$, and a suitable normalization of $b$, one can achieve $Z = 1$, what we assume in the following. Classical local observables $A(t) = A \bigl(s(t)\bigr)$ depend only on spins located on the hypersurface $t$. Their expectation values obey the standard rule in classical statistics
\begin{equation}\label{eqn:AF4}
	\langle A(t)\rangle
	= \int \mathcal{D} s \, A(t) \, w[s].
\end{equation}

We split the configuration sum $\int \mathcal{D} s$ into parts involving spins for $t^\prime < t$, $t^\prime > t$, and $t$. One defines the classical wave function $f(t)$ by integrating over the lower part
\begin{equation}\label{eqn:AF5}
	f(t)
	= \int \mathcal{D} s(t^\prime < t) \exp\Bigl\{-\sum_{t^\prime < t}\mathcal{L}(t^\prime)\Bigr\} f_\text{in},
\end{equation}
and the conjugate wave function $\bar{f}(t)$ by the sum over the upper part
\begin{equation}\label{eqn:AF6}
	\bar{f}(t)
	= \int \mathcal{D} s(t^\prime > t) \exp\Bigl\{-\sum_{t^\prime \geq t} \mathcal{L}(t^\prime)\Bigr\} \bar{f}_\text{f}.
\end{equation}
Both $f(t)$ and $\bar{f}(t)$ depend on the ``local spins'' at $t$, $s_\gamma(t)$. For the computation of $\langle A(t)\rangle$ in \cref{eqn:AF4} only the sum over the local spins is left,
\begin{equation}\label{eqn:AF7}
	\langle A(t)\rangle
	= \int \mathcal{D} s(t) \, \bar{f}(t) \, A(t) \, f(t).
\end{equation}
The local probabilities $p(t)$ depend on $s(t)$ and are products of the classical wave function and conjugate wave function
\begin{equation}\label{eqn:AF7A}
	p(t)
	= \bar{f}(t) \, f(t)
	\geq 0,
	\qquad
	\int \mathcal{D} s(t) \, p(t)
	= 1,
\end{equation}
leading to the usual classical statistical expression
\begin{equation}\label{eqn:AF7B}
	\langle A(t)\rangle
	= \int \mathcal{D} s(t) \, p(t) \, A(t).
\end{equation}

The wave functions obey simple linear evolution laws
\begin{alignedeqn}\label{eqn:AF8}
	f(t + \epsilon)
	&= \int \mathcal{D} s(t) \, \mathcal{K}(t) \, f(t),\\
	\bar{f}(t)
	&= \int \mathcal{D} s(t + \epsilon) \, \mathcal{K}(t) \, \bar{f}(t + \epsilon),
\end{alignedeqn}
where
\begin{equation}\label{eqn:AF9}
	\mathcal{K}(t)
	= \exp\bigl\{-\mathcal{L}(t)\bigr\}.
\end{equation}
These evolution laws follow directly from the definitions \labelcref{eqn:AF5,eqn:AF6} by extending the corresponding ranges by one unit. It is the simplicity of the evolution law that singles out the concept of wave functions for the description of evolution in classical statistical systems. Linearity implies that the superposition principle for wave functions holds, in analogy to quantum mechanics. If $f_1(t)$ and $f_2(t)$ are solutions of \cref{eqn:AF8}, then also the linear combinations $\alpha_1 f_1(t) + \alpha_2 f_2(t)$ solve this equation. No such simple linear evolution law involving only the local probabilities $p(t)$ can be formulated. The wave functions contain precisely the additional probabilistic information beyond $p(t)$ that is needed for a simple linear evolution law.

The pair of wave functions $f(t)$ and $\bar{f}(t)$ permits an efficient description of the transport of information. It is sufficient for the computation of expectation values of classical local observables at every $t$ by \cref{eqn:AF7}, as well as of local probabilities \labelcref{eqn:AF7A}. Solving the linear evolution equation \labelcref{eqn:AF8} one can follow how the information at the initial boundary encoded in $f(t_\text{in})$ and $\bar{f}(t_\text{in})$, related to expectation values of observables at $t_\text{in}$, propagates into the bulk at arbitrary $t - t_\text{in}$, or to the other boundary at $t_\text{f}$.

\subsection{Quantum formalism}

With the classical wave functions, the linear evolution law and the bilinear expression \labelcref{eqn:AF7} for expectation values of local observables, many analogies to quantum mechanics arise in a natural way from the problem of information transport in classical statistical systems. The analogy to quantum mechanics becomes even more apparent if we choose a suitable complete set of local basis functions $h_\tau(t) = h_\tau\bigl[s(t)\bigr]$ which depend on the local spins $s_\gamma(t)$. These basis functions obey the orthogonality relation
\begin{equation}\label{eqn:AF10}
	\int \mathcal{D} s(t) \, h_\tau(t) \, h_\rho(t)
	= \delta_{\tau\rho}
\end{equation}
and the completeness relation
\begin{equation}\label{eqn:AF10A}
	h_\tau[s] \, h_\tau[\bar{s}]
	= \delta(s,\bar{s}).
\end{equation}
By virtue of the completeness relation we can expand an arbitrary function of $s(t)$ as
\begin{equation}\label{eqn:AF10B}
	A(t)
	= \alpha_\tau(t) \, h_\tau(t),
	\quad
	\alpha_\tau(t)
	= \int \mathcal{D} s(t) \, A(t) h_\tau(t).
\end{equation}

For the wave functions this yields
\begin{equation}\label{eqn:AF11}
	f(t)
	= \tilde{q}_\tau(t) \, h_\tau(t),
	\qquad
	\bar{f}(t)
	= \bar{q}_\tau(t) \, h_\tau(t),
\end{equation}
while the evolution factor $\mathcal{K}(t)$ depends on $s(t)$ and $s(t + \epsilon)$ and reads
\begin{equation}\label{eqn:AF12}
	\mathcal{K}(t)
	= S_{\tau\rho}(t) \, h_\tau(t + \epsilon) \, h_\rho(t).
\end{equation}
(We assume summations over double indices if not explicitly stated otherwise.) For a given basis the wave functions are given by the vectors $\tilde{q}(t)$ and $\bar{q}(t)$. Their evolution law involves a matrix multiplication
\begin{alignedeqn}\label{eqn:AF13}
	&\tilde{q}_\tau(t + \epsilon)
	= S_{\tau\rho}(t) \, \tilde{q}_\rho(t),\\
	&\bar{q}_\tau(t)
	= \bar{q}_\rho(t + \epsilon) \, S_{\rho\tau}(t).
\end{alignedeqn}
The matrix $S$ can be associated with the transfer matrix \cite{TM,MS,FU}. We choose the additive normalization of $\mathcal{L}(t)$ in \cref{eqn:AF2} such that the eigenvalue of $S$ with largest absolute size is normalized to one, $|\lambda_\text{max}| = 1$. With this normalization $S$ is called the step evolution operator \cite{CWIT}. In the matrix notation \cref{eqn:AF13} reads
\begin{equation}\label{eqn:matrix AF13}
	\tilde{q}(t + \epsilon)
	= S(t) \, \tilde{q}(t),
	\qquad
	\bar{q}(t + \epsilon)
	= (S^{-1})^\tran(t) \, \bar{q}(t).
\end{equation}

We may define
\begin{equation}\label{eqn:AF14}
	\partial_t \tilde{q}(t)
	= \frac{1}{2 \epsilon}\bigl(\tilde{q}(t + \epsilon)-\tilde{q}(t - \epsilon)\bigr)
	= W(t)\tilde{q}(t),
\end{equation}
with
\begin{equation}\label{eqn:FF1}
	W(t)
	= \frac{1}{2 \epsilon} \bigl(S(t) - S^{-1}(t - \epsilon)\bigr).
\end{equation}
If a suitable continuum limit $\epsilon \to 0$ exists, \cref{eqn:AF14} becomes a linear differential equation, generalizing the Schrödinger equation in quantum mechanics. Similarly, the conjugate wave function obeys
\begin{equation}\label{eqn:AF16}
	\partial_t \bar{q}(t)
	= -\tilde{W}^\tran(t) \, \bar{q}(t),
\end{equation}
where
\begin{equation}\label{eqn:AF17}
	\tilde{W}(t)
	= \frac{1}{2 \epsilon}\bigl[S(t - \epsilon)-S^{-1}(t)\bigr]
\end{equation}
equals $W(t)$ if $S$ is independent of $t$.

Without loss of generality, we can formally write
\begin{equation}\label{eqn:FF3}
	W
	= J - i H,
\end{equation}
with $H$ and $J$ Hermitian matrices. The Hamilton operator $H$ is antisymmetric and purely imaginary, while $J$ is real and symmetric. Multiplication with $i$ formally yields the generalized Schrödinger equation in a familiar form
\begin{equation}\label{eqn:46K}
	i \partial_t \tilde{q}
	= H \tilde{q} + i J \tilde{q}.
\end{equation}
For models with $t$-independent orthogonal $S$ the matrix $J$ vanishes. For the conjugate wave function and $t$-independent $S$ one has
\begin{equation}\label{eqn:FF4}
	i\partial_t\bar{q}
	= (H-iJ)\bar{q}.
\end{equation}

For local observables we employ the expansion \labelcref{eqn:AF10B}. We can then associate to each local observable a real symmetric matrix or operator $A^\prime(t)$
\begin{equation}\label{eqn:AF19}
	A_{\tau\rho}^\prime(t)
	= \alpha_\sigma(t)\int \mathcal{D} s(t) \, h_\tau(t) \, h_\rho(t) \, h_\sigma(t)
	= A_{\rho\tau}^\prime(t).
\end{equation}
This expresses the expectation value \labelcref{eqn:AF7} in a form resembling quantum mechanics
\begin{equation}\label{eqn:AF20}
	\langle A(t)\rangle
	= \bar{q}_\tau(t) \, A_{\tau\rho}^\prime(t) \, \tilde{q}_\rho(t)
	= \bar{q}^\tran A^\prime \tilde{q}.
\end{equation}
The operators $A^\prime$ are not necessarily diagonal.

The step evolution operator $S$ or the operators $A^\prime$ associated to local observables depend on the choice of basis functions. A particularly simple basis is the ``occupation number basis'' \cite{CWIT}. It associates to every local spin configuration $[s(t)]$ of $M$ spins $s_\gamma(t)$ ($\gamma \in \{1,\dots,M\}$) a product of $M$ factors $a_\gamma$,
\begin{equation}\label{eqn:AF21}
	h_\tau
	= \prod_{\gamma=1}^M a_\gamma,
\end{equation}
where $a_\gamma$ is either given by $a_\gamma = n_\gamma = (1 + s_\gamma)/2$, or by $a_\gamma = (1-n_\gamma) = (1-s_\gamma)/2$. There are $N = 2^M$ independent basis functions, corresponding to the $N$ different spin configurations. All basis functions equal one precisely for one specific configuration, that we may denote by $\tau$, and vanish for all other configurations. They obey the relations (no index sum here)
\begin{equation}\label{eqn:AF22}
	h_\tau h_\rho
	= h_\tau\delta_{\tau\rho},
	\qquad
	\sum_\tau h_\tau
	= 1,
	\qquad
	\int \mathcal{D} s(t) \, h_\tau
	= 1.
\end{equation}
The relations \labelcref{eqn:AF10,eqn:AF10A} are easily verified. For this basis the operators for classical local observables are diagonal (no index sum)
\begin{equation}\label{eqn:AF23}
	A_{\tau\rho}^\prime(t)
	= A_\tau(t) \delta_{\tau\rho}.
\end{equation}
Here $A_\tau(t)$ is the value that the observable takes for the spin configuration $\tau$. In the occupation number basis the expansion coefficients $\alpha_\tau$ in \cref{eqn:AF10B} coincide with the values that the observable takes for the configuration $\tau$,
\begin{equation}\label{eqn:coincision}
	\alpha_\tau
	= A_\tau.
\end{equation}
The local probabilities are given by (no sum over $\tau$)
\begin{equation}\label{eqn:local probabilities}
	p_\tau
	= \bar{q}_\tau \tilde{q}_\tau.
\end{equation}

In general, the index $\tau$ for quantities such as $A_\tau$ or the local probabilities $p_\tau$ denotes different local configurations of occupation numbers or spins, while for $\bar{q}_\tau$, $\tilde{q}_\tau$, $\alpha_\tau$, $S_{\tau\rho}$ or $\pi_\tau$ in $p(t) = \pi_\tau h_\tau$ it labels the basis functions and associated coefficients. Only in the occupation number basis can we identify these labels, since we can denote by $\tau$ the spin configuration where $h_\tau = 1$, or vice versa. With this association the relation \labelcref{eqn:coincision} or the similar relation $\pi_\tau = p_\tau$ is easily established.

For comparison with some other basis we may discuss the ``spin basis'', for which the factors $a_\gamma$ in \cref{eqn:AF21} are either given by $1$ or $s_\gamma$. For a single spin the operator $n = (1 + s)/2$ corresponds to $\alpha_1 = \alpha_2 = 1/2$, while $1 - n = (1 - s)/2$ is described by $\alpha_1 = 1/2$, $\alpha_2 = -1/2$. The values $A_1 = 1$, $A_2 = 0$ for the observable $n$, and $A_1 = 0$, $A_2 = 1$ for the observable $1 - n$, do not depend on the choice of basis. These values correspond to $\alpha_\tau$ only in the occupation number basis. Positive values of $\alpha_\tau$ in the occupation number basis can result in some $\alpha_\tau$ being negative in the spin basis. The same applies to the wave function or local probabilities, and one infers that some $\tilde{q}_\tau$ or $\pi_\tau$ can be negative in the spin basis, or $S_{\tau\rho}$ can have negative elements. Only in the occupation number basis are $\bar{q}_\tau$, $\tilde{q}_\tau$, $\pi_\tau$, $S_{\tau\rho}$ all positive semidefinite. We will keep the discussion general with an arbitrary choice of basis. The existence of the particular occupation number basis will nevertheless be a very useful tool for establishing general properties.

The formal analogies between the computation of expectation values of local observables $\langle A(t)\rangle$ and their evolution with $t$ in classical statistics on one side, and quantum mechanics on the other side, are the basis for this paper. They will allow us to transfer many methods and concepts of quantum mechanics to the problem of information transport in classical statistics. There remain, nevertheless, also important qualitative differences between classical statistics and quantum mechanics. Quantum mechanics will appear only as a particular special case of classical statistics. A key difference is the presence of two independent wave functions in classical statistics, namely the classical wave function $\tilde{q}(t)$ and the conjugate wave function $\bar{q}(t)$. Only the product of the two wave functions is normalized,
\begin{equation}\label{eqn:AF24}
	\int \mathcal{D} s(t) \, \bar{f}(t) \, f(t)
	= \bar{q}_\tau(t) \, \tilde{q}_\tau(t)
	= Z
	= 1.
\end{equation}
The length of the vector $\tilde{q}$ can therefore change in the course of the evolution if it is accompanied by a compensating change for $\bar{q}$.

\subsection{Quantum mechanics}

For the special case of a ``unitary evolution'' the length of the vectors $\tilde{q}$ and $\bar{q}$ is conserved
\begin{equation}\label{eqn:AF25}
	\tilde{q}_\tau(t) \, \tilde{q}_\tau(t)
	= c^2,
	\qquad
	\bar{q}_\tau(t) \, \bar{q}_\tau(t)
	= d^2.
\end{equation}
This is realized if the step evolution operator $S$ in \cref{eqn:AF13} is an orthogonal matrix, $S^\tran(t) \, S(t) = 1$. In turn, the operator $W$ which characterizes the continuum limit of the evolution in \cref{eqn:AF14} is antisymmetric. The dependence of the classical wave function is then governed by a (real) Schrödinger equation, with Hermitian Hamiltonian $H = i W$.

The weight function \labelcref{eqn:AF1} and the boundary term \labelcref{eqn:AF2A} remain unchanged if we multiply $f_\text{in}$ by $1/c$ and $\bar{f}_\text{f}$ by $c$. We can use this freedom in order to normalize the classical wave function, i.e. setting $c = 1$ in \cref{eqn:AF25}. In the following we consider a normalized initial wave function at the boundary
\begin{equation}\label{eqn:Q2}
	\tilde{q}^\tran(t_\text{in}) \, \tilde{q}(t_\text{in})
	= \sum_\tau \tilde{q}^2_\tau(t_\text{in})
	= 1.
\end{equation}
For orthogonal $S$ this normalization condition is preserved for all $t$,
\begin{equation}\label{eqn:Q3}
	\tilde{q}^\tran(t) \, \tilde{q}(t)
	= 1.
\end{equation}
With the normalization \labelcref{eqn:Q3} the wave function is a normalized vector in a Hilbert space. The fact that it is real is no limitation. Complex wave functions can be obtained from $\tilde{q}$ by introducing a complex structure, cf. \cref{sec:complex structure}, and any complex wave function can be written as a real wave function with twice the number of components. Together with the Schrödinger equation the evolution fulfills the axioms of quantum mechanics.

In order to obtain the normalization of the partition function $Z = 1$ we have to impose for the conjugate wave function at $t_\text{f}$ the condition
\begin{equation}\label{eqn:Q4}
	\bar{q}^\tran(t_\text{f}) \, \tilde{q}(t_\text{f})
	= 1.
\end{equation}
This holds then for all $t$, cf. \cref{eqn:AF24}. For $S^\tran S = 1$ the evolution of $\bar{q}$ is the same as for $\tilde{q}$,
\begin{equation}\label{eqn:Q6}
	i \partial_t \tilde{q}
	= H \tilde{q},
	\qquad
	i \partial_t \bar{q}
	= H \bar{q},
	\qquad
	H
	= H^\dagger.
\end{equation}

Let us now choose
\begin{equation}\label{eqn:Q7}
	\bar{q}(t_\text{f})
	= \tilde{q}(t_\text{f}),
\end{equation}
such that $\bar{q}$ and $\tilde{q}$ can be identified for all $t$
\begin{equation}\label{eqn:176A}
	\bar{q}(t)
	= \tilde{q}(t)
	= q(t).
\end{equation}
The modifications for other choices of $\bar{q}(t_\text{f})$ are discussed in \cref{app:bcs and ivs}. With the choice \labelcref{eqn:Q7} the expectation value of a local observable $A(t)$ obeys by virtue of \cref{eqn:AF20} the quantum rule (for real wave functions),
\begin{equation}\label{eqn:Q8}
	\langle A(t)\rangle
	= q^\tran(t) A^\prime(t) \, q(t).
\end{equation}
Together with the property that the operators associated with product observables $[A(t)]^n$ are the matrix products $[A^\prime(t)]^n$, see \cref{sec:Three-spin chains}, also the axioms of quantum mechanics for observables are obeyed. We can therefore take advantage of the full formalism of quantum mechanics.

In the occupation number basis one has for the local probabilities the standard relation of quantum mechanics
\begin{equation}\label{eqn:Q9}
	p_\tau(t)
	= q_\tau^2(t).
\end{equation}
In any other basis we can still identify the positive semidefinite quantities $q_\tau^2(t)$ with probabilities. They will not coincide with the local probabilities $p_\tau$, however.

\subsection{Weight distribution and step evolution operator}

A \textit{local} probabilistic setting is realized if the product $\bar{q}_\tau(t) \, \tilde{q}_\tau(t)$ is positive semidefinite for all $t$, and for each $\tau$ individually. This condition is sufficient for a well defined local probability distribution. It is necessary but not sufficient for a well defined \textit{overall} probability distribution. The latter requires positivity of $w[s] = p[s]$ for arbitrary spin configurations.

The overall probability distribution \labelcref{eqn:AF1} can be represented as a product of step evolution operators. We use the explicit representation
\begin{equation}\label{eqn:28A}
	w[s]
	= \bar{q}_\tau(t_\text{f}) \, h_\tau(t_\text{f}) \, \mathcal{K}(t_\text{f} - \epsilon) \dots \mathcal{K}(t_\text{in}) \, \tilde{q}_\rho(t_\text{in}) \, h_\rho(t_\text{in}).
\end{equation}
Insertion of \cref{eqn:AF12} expresses $w$ in terms of step evolution operators and products of basis functions. The use of \cref{eqn:AF10} yields for every basis
\begin{equation}\label{eqn:partition function}
	Z
	= \bar{q}_\tau(t_\text{f}) \bigl[S(t_\text{f} - \epsilon) \dots S(t_\text{in} + \epsilon) \, S(t_\text{in})\bigr]_{\tau\rho} \tilde{q}_\rho(t_\text{in}).
\end{equation}
In the occupation number basis $w[s]$ has a particularly simple form, since we can employ the first relation in \cref{eqn:AF22},
\begin{equation}\label{eqn:28B}
	w[s]
	= \sum_{\mathclap{\rho_1,\dots,\rho_{G+1}}}
	w_{\rho_1 \rho_2 \dots \rho_{G+1}} \, h_{\rho_1}(t_1) \, h_{\rho_2}(t_2) \dots h_{\rho_{G+1}}(t_{G+1}),
\end{equation}
with (no summations in the following expression)
\begin{alignedeqn}\label{eqn:28C}
	w_{\rho_1 \rho_2 \dots \rho_{G+1}}
	&= \bar{q}_{\rho_{G+1}}(t_{G+1}) \, S_{\rho_{G+1} \rho_{G}}(t_{G}) \, S_{\rho_{G} \rho_{G-1}}(t_{G-1})\\
	&\hphantom{={}}\times \dots S_{\rho_3 \rho_2}(t_2) \, S_{\rho_2 \rho_1}(t_1) \, \tilde{q}_{\rho_1}(t_1).
\end{alignedeqn}
Here we have numbered the time arguments according to $t_\text{in} = t_1$, $t_\text{in} + \epsilon = t_2$, \dots $t_\text{f} = t_{G+1}$. (With the third relation in \cref{eqn:AF22} one recovers \cref{eqn:partition function}.) Positivity of $w[s]$ requires positivity of all coefficients $w_{\rho_1\dots\rho_{G+1}}$. Conditions for this positivity can be found in ref. \cite{CWIT}.

In our classical statistical setting \labelcref{eqn:AF2,eqn:AF9} the requirement of positivity only restricts the choice of boundary values. We will extend our discussion to arbitrary weight functions $w[s]$ defined by \cref{eqn:AF12,eqn:28A}. Without restrictions on $S$ they are not necessarily real and positive. The classical statistical systems are given by the subclass of positive probability distributions $p[s] = w[s]$.

\section{Simple models}
\label{sec:simple models}

In this section we present three simple models for which the boundary problem is solved exactly in ref.~\cite{CWIT}. They make the general discussion of the preceding section more concrete. We work here in the occupation number basis. For the generalized Ising models \eqref{eqn:AF2} we expand ${\cal L}(t')$ in the basis functions
\be\label{52AA}
{\cal L}(t')=M_{\tau\rho}(t')h_\tau(t'+\epsilon)h_\rho(t').
\ee
This yields for the step evolution operator the matrix elements
\be\label{52AB}
S_{\tau\rho}(t')=\exp \big(-M_{\tau\rho}(t')\big).
\ee
All elements are positive, $S_{\tau\rho}\geq 0$, such that the step evolution operator is a nonnegative matrix. Inversely, every $S$ with positive elements can be associated with suitable elements $M_{\tau\rho}(t')=-\ln \big(S_{\tau\rho}(t')\big)$, and therefore be expressed by interactions between Ising spins on neighboring $t$-layers by virtue of \cref{52AA}. Vanishing elements $S_{\tau\rho}=0$ correspond to $M_{\tau\rho}\to\infty$. For the corresponding neighboring spin configurations ${\cal L}(t')$ diverges, such that ${\cal K}(t')=\exp\big(-{\cal L}(t')\big)$ vanishes. Such configurations contribute zero weight and are ``forbidden'' in this sense. 

Our first simple generalized Ising model is the four-state oscillator chain. It describes two Ising spins, $M=2$, such that wave functions have four real components and $S$ is a $4\times 4$-matrix. The evolution of the wave function obeys an approach to an equilibrium state by damped oscillations. We will discuss the evolution of the density matrix for this model in detail in \cref{sec:solution of bvp}. It will serve as an illustration that classical statistical systems do not always allow for arbitrary ``initial values'' of the density matrix at $t_\text{in}$. A limiting case of this model is a unique jump chain that can be associated to a cellular automaton \cite{CA,ICJ,TH,TH2,EL}. Our second example are unique jump chains that account for a discrete quantum evolution. The third model is the two-dimensional asymmetric diagonal Ising model. It describes a quantum field theory of free relativistic fermions \cite{CWFGI}. Complementing these simple examples, a detailed discussion of the classical wave function for the one-dimensional Ising model \cite{IM,IS} can be found in ref. \cite{CWIT}. In \cref{sec:solution of bvp} we add the evolution of the classical density matrix for the Ising model, extending the complete solution of the boundary problem to mixed classical states. The following section \cref{sec:Three-spin chains} provides for a simple example for non-commuting operators in classical statistics. There we will present ``three-spin chains'' that show many analogies to the behavior of a single quantum spin.

\subsection{Four-state oscillator chain}

It is possible that the step evolution operator $S$ has complex eigenvalues. This can occur if $S$ is not symmetric. In this case one expects oscillatory behavior. The boundary information may still be lost far inside the bulk if $S$ has a unique largest eigenvalue, corresponding to a unique equilibrium state. This ``extinction of information'' is not monotonic, however, as in many classical statistical systems. The local probabilities, and correspondingly the expectation values of local observables, reach their equilibrium values in an oscillating fashion. 

An example is the four-state oscillator chain, as realized by the four by four matrix
\begin{equation}\label{eqn:GD}
	S
	= \begin{pmatrix}
		1-\eta & 0 & \eta & 0\\
		\eta & 1-\eta & 0 & 0\\
		0 & 0 & 1-\eta & \eta\\
		0 & \eta & 0 & 1-\eta
	\end{pmatrix}.
\end{equation}
For $0 \leq \eta \leq 1$ all entries of $S$ are positive and the model can be realized with Ising spins, cf. ref. \cite{CWIT} for details. Since some elements of $S$ vanish it is a type of ``constrained Ising model'' for which certain combinations of neighboring spin configurations are prohibited due to a divergent action.

The four eigenvalues of $S$ are
\begin{equation}\label{eqn:GE}
	\lambda_0
	= 1,
	\qquad
	\lambda_{1 \pm}
	= 1 - \eta \pm i \eta,
	\qquad
	\lambda_2
	= 1 - 2 \eta,
\end{equation}
and the (unnormalized) eigenvectors to the eigenvalues $\lambda_0$, $\lambda_{1+}$, $\lambda_{1-}$, $\lambda_2$ read
\begin{alignedeqn}\label{eqn:GF}
	v_0
	&= \begin{pmatrix}
		1\\1\\1\\1
	\end{pmatrix}\!,
	\qquad
	&&v_{1+}
	= \begin{pmatrix}
		1\\-i\\i\\-1
	\end{pmatrix}\!,\\
	v_{1-}
	&= \begin{pmatrix}
		1\\i\\-i\\-1
	\end{pmatrix}\!,
	&&\hphantom{_+}v_{2}
	= \begin{pmatrix}
		1\\-1\\-1\\1
	\end{pmatrix}\!.
\end{alignedeqn}
For $0 < \eta < 1$ the absolute value of $\lambda_{1\pm}$ is smaller than one. One therefore encounters damped oscillations of the wave function if the initial wave function contains components $\sim v_{1\pm}$.

Whenever a unique largest eigenvalue one of $S$ is separated from the other eigenvalues $\lambda_i$, with $|\lambda_i| < 1$, one may suspect a loss of memory of boundary conditions with a finite correlation length. Far enough inside the bulk the classical wave function will attain an equilibrium value $\tilde q_*$ which is independent of $t$. We will see in \cref{sec:density matrix} that the situation is subtle, however. The reason is that an exponentially decaying $\tilde{q}-\tilde{q}_\ast$ may be compensated by an exponentially increasing $\bar{q}-\bar{q}_\ast$. This could lead to undamped oscillations of the density matrix. The general solution of the evolution equation for the density matrix admits indeed undamped oscillatory behavior. One then has to investigate if such density matrices can be realized by suitable boundary conditions. We will present an explicit exact solution of the boundary problem for the four state oscillator chain in \cref{sec:solution of bvp}. For this and similar systems interesting cases arise for finite systems for which the information inside the bulk is not erased in practice. 

For small $\eta$, we define $\omega = \eta/\epsilon$ and extract the continuum limit \labelcref{eqn:46K} with
\begin{alignedeqn}\label{eqn:HE}
	&J
	= \frac{1}{2}
	\begin{pmatrix}
		-2 \omega & \omega & \omega & 0\\
		\omega & -2 \omega & 0 & \omega\\
		\omega & 0 & -2 \omega & \omega\\
		0 & \omega & \omega & -2 \omega
	\end{pmatrix},\\
	&H
	= \frac{i}{2}
	\begin{pmatrix}
		0 & -\omega & \omega & 0\\
		\omega & 0 & 0 & -\omega\\
		-\omega & 0 & 0 & \omega\\
		0 & \omega & -\omega & 0
	\end{pmatrix}.
\end{alignedeqn}
This constitutes a concrete example for the generalized Schr\"odinger equation \eqref{eqn:46K}. The general solution of the evolution equation \labelcref{eqn:46K} is straightforward \cite{CWIT}. It shows damped oscillations, as expected.

\subsection{Unique jump chains}

An example of complete information transport is given by the limiting case of \cref{eqn:GD} for $\eta = 1$. It corresponds to $\tilde{q}_2(t + \epsilon) = \tilde{q}_1(t)$, $\tilde{q}_4(t + \epsilon) = \tilde{q}_2(t)$, $\tilde{q}_3(t + \epsilon) = \tilde{q}_4(t)$, $\tilde{q}_1(t + \epsilon) = \tilde{q}_3(t)$. This evolution is periodic with period $4 \epsilon$. With $S^\tran S = 1$ the step evolution operator
\begin{equation}\label{eqn:GN}
	S
	= \begin{pmatrix}
		0 & 0 & 1 & 0\\
		1 & 0 & 0 & 0\\
		0 & 0 & 0 & 1\\
		0 & 1 & 0 & 0
	\end{pmatrix}
\end{equation}
describes a rotation. The eigenvalues of $S$ are $\pm 1$, $\pm i$. The conjugate wave function $\bar{q}(t)$ follows the same evolution as $\tilde{q}(t)$ according to \cref{eqn:matrix AF13}. Therefore the local probabilities $p(t)$ oscillate with period $4\epsilon$, according to $p_2(t + \epsilon) = p_1(t)$, $p_4(t + \epsilon) = p_2(t)$, $p_3(t + \epsilon) = p_4(t)$, $p_1(t + \epsilon) = p_3(t)$.

The step evolution operator \labelcref{eqn:GN} is an example of a ``unique jump chain''. Each spin configuration at $t$ is mapped to precisely one other spin configuration at $t + \epsilon$. The limit $\eta \to 1$ of \cref{eqn:GD} can therefore be considered a deterministic system, associated to a cellular automaton \cite{CA,ICJ,TH,TH2,EL}. It is an example of discrete quantum mechanics, where the continuous evolution according to the Schrödinger equation is replaced by a discrete sequence of evolution operators between $t$ and $t + \epsilon$. We can also consider the step evolution operator \labelcref{eqn:GN} as the realization of a static memory material \cite{CWIT}. Information about boundary conditions is transported into the bulk because $S$ has more than one eigenvalue with $|\lambda| = 1$. The values of observables far inside the bulk depend on the boundary conditions. 

These simple examples demonstrate already that the issue of information transport is expected to differ substantially from a simple approach to equilibrium in the bulk whenever more than one eigenfunction to eigenvalues of $S$ with $|\lambda| = 1$ exists. Memory of boundary conditions is expected to occur due to a ``degeneracy'' of generalized equilibrium wave functions.

\subsection{Ising spins for free massless fermions}

Our third examples are two-dimensional memory materials where the coordinate $t$ is supplemented by a second coordinate $x$. Here $t$ and $x$ are treated equally as locations on a two dimensional lattice. The coordinate $t$ indicates the direction in which the evolution between different hypersurfaces is studied. We discuss Ising spins $s(t,x)$, with $x$ corresponding to $\gamma$. For periodic boundary conditions in $x$ we consider $M$ different values for $x$, $x = x_\text{in} + (m-1) \epsilon$, $1 \leq m \leq M$, $x_\text{f} = x_\text{in} + (M-1) \epsilon$, $x_\text{in} + M \epsilon = x_\text{in}$.

A static memory material can be realized as an Ising type model with interactions only among diagonal neighbors in one direction
\begin{equation}\label{eqn:M13}
	\mathcal{L}(t)
	= -\frac{\beta}{2} \sum_x \bigl\{s(t + \epsilon,x + \epsilon) \, s(t,x) - 1\bigr\}.
\end{equation}
For $\beta \to \infty$ this ``asymmetric diagonal Ising model'' can be solved exactly in terms of the free propagation of an arbitrary number of Weyl fermions \cite{CWIT,CWFGI}.

This solution is easily understood by considering two neighboring sequences of Ising spins, e.g. $\{s(t + \epsilon,x)\}$ and $\{s(t,x)\}$. Whenever the spins $s(t + \epsilon,x + \epsilon)$ and $s(t,x)$ are the same, the factor $\mathcal{K}(t)$ in \cref{eqn:AF9} equals one. If they are different, $s(t + \epsilon,x + \epsilon) \neq s(t,x)$, \cref{eqn:AF9} gives a factor $e^{-\beta}$. The leading term in $\mathcal{K}(t)$ thus copies the sequence $s(t,x)$ from $t$ to $t + \epsilon$, now translated by one unit in the positive $x$-direction. For this configuration of two neighboring displaced sequences one has $\mathcal{K}(t) = 1$. Each copy error in one spin reduces $\mathcal{K}(t)$ by a factor $e^{-\beta}$. In the limit $\beta \to \infty$ the probability for configurations with copy errors goes to zero. This realizes a unique jump chain, where every sequence $\{s(x)\}$ at $t$ is mapped uniquely to a sequence $\{s^\prime(x)\}$ at $t + \epsilon$.

In the limit $\beta \to \infty$ the step evolution operator $S$ becomes simple. In the occupation number basis a given sequence $[s(x)]$ corresponds to a given basis function $h_\tau$. The sequence displaced by one unit, e.g. $s^\prime(x) = s(x + \epsilon)$, corresponds to a different basis function $h_{\alpha(\tau)}$. The transfer matrix $\bar{S}_{\rho\tau}$ equals one whenever $\rho = \alpha(\tau)$, while all other elements are suppressed by factors $e^{-k\beta}$, with $k$ the ``number of errors''. For $\beta \to \infty$ one has $\bar{S} = S$, where $S_{\rho\tau} = 1$ for $\rho = \alpha(\tau)$, and $S_{\rho\tau} = 0$ for $\rho \neq \alpha(\tau)$. We recognize a unique jump chain similar to \cref{eqn:GN}. The evolution of the wave function is given by
\begin{equation}\label{eqn:M14}
	\tilde{q}_{\alpha(\tau)}(t + \epsilon)
	= \tilde{q}_\tau(t),
	\qquad
	\tilde{q}_\tau(t + \epsilon)
	= \tilde{q}_{\alpha^{-1}(\tau)}(t).
\end{equation}
The evolution of the conjugate wave function $\bar{q}_\tau(t)$ is the same as for $\tilde{q}_\tau(t)$. We consider boundary conditions such that $\bar{q}$ and $\tilde{q}$ can be identified, $\tilde{q}(t) = \bar{q}(t) = q(t)$ and refer to \cref{app:bcs and ivs} for more general boundary conditions.

We may describe our model in terms of fermionic particles, with occupation numbers $n(x) = [s(x) + 1]/2 \in \{0,1\}$. The displacement by one $x$-unit of the copied $\bigl[n(x)\bigr]$-sequence does not change the number of ones in this sequence. If we associate each $n(x) = 1$ with a particle present at $x$ (and $n(x) = 0$ with no particle present), the total number of particles $F = \sum_x n(x)$ is the same at $t$ and $t + \epsilon$. The step evolution operator conserves the particle number. We can therefore decompose the wave function $q_\tau(t)$ into sectors with different particle numbers $F$. They do not mix by the evolution and can be treated separately.

For the sector with $F = 0$ the wave function is independent of $t$. The sector with $F = 1$ describes the propagation of a single particle. It is characterized by a single particle wave function $q(t,x)$, where $x$ denotes the location of the single particle or the position of the unique one in the sequence $[s(x)]$. The evolution of the one-particle wave function obeys
\begin{equation}\label{eqn:one-particle evolution}
	q(t + \epsilon,x)
	= q(t,x - \epsilon).
\end{equation}

The sector with $F = 2$ is characterized by the positions $x$ and $y$ of the two particles. There is no distinction which particle sits at $x$ and which one at $y$, and we may use an antisymmetric wave function
\begin{equation}\label{eqn:M15}
	q(t;x,y)
	= -q(t;y,x).
\end{equation}
The origin of the antisymmetric wave functions for fermions becomes apparent if we employ the general map between the functional integral for Ising spins and an equivalent Grassman functional integral \cite{CWFGI}. Using variables
\begin{equation}\label{eqn:M16}
	s
	= \frac{x + y}{2},
	\qquad
	r
	= x-y
\end{equation}
the evolution obeys
\begin{equation}\label{eqn:M17}
	q(t + \epsilon;s,r)
	= q(t;s-\epsilon,r).
\end{equation}
The distance $r$ between the occupied sites remains the same for all $t$. This evolution equation has a similar structure as \cref{eqn:one-particle evolution}, with $s$ replacing $x$ and $r$ considered an additional property that is not affected by the evolution.

The evolution equations \labelcref{eqn:one-particle evolution} or \labelcref{eqn:M17} admit a continuum limit $\epsilon \to 0$ if the wave functions are sufficiently smooth. The one particle wave function obeys
\begin{equation}
	\partial_t q(x)
	= -\partial_x q(x),
\end{equation}
while for the two-particle wave function one has
\begin{equation}\label{eqn:M18}
	\partial_t q
	= -\partial_s q
	= -(\partial_x + \partial_y) q.
\end{equation}
This generalizes to an arbitrary number of particles $F$,
\begin{equation}\label{eqn:M19}
	\partial_t q
	= W \, q,
	\qquad
	W
	= -i P
\end{equation}
with total momentum operator
\begin{equation}
	P
	= -i \sum_{j=1}^F \frac{\partial}{\partial x_j}.
\end{equation}

Associating formally $i W$ with the energy $E$ of the $F$-particle system, we recover the energy-momentum relation for relativistic fermions $E = P$. It is possible to define for this model a complex structure \cite{CWIT}, such that \cref{eqn:M19} becomes the usual complex Schrödinger equation with hermitian Hamiltonian $H = P$. The particles are ``right-movers'' and can be identified with Weyl fermions in two dimensions. A more detailed discussion including the complex structure and the realization of Lorentz symmetry can be found in ref.~\cite{CWFGI}. The diagonal asymmetric Ising model \labelcref{eqn:M13} realizes for $\beta \to \infty$ a two-dimensional quantum field theory. It constitutes a simple example that the realization of quantum mechanics according to the discussion following \cref{eqn:Q2} is possible within a classical statistical system.

\section{Three-spin chains}
\label{sec:Three-spin chains}

In this section we present the solution of the boundary value problem for a chain of three Ising spins at each site. We follow how the information at the two ends of the chains at $t_{in}$ and $t_f$ propagates to arbitrary $t$ in the bulk. We will admit interactions between the spins that can vary for different $t$. This can represent the presence of external fields, which may be used to influence the outcome of a sequence of computational steps if the spins are taken as classical bits. One could also associate the spins with a subsystems of neurons whose interactions are influenced by potentials reflecting the state the environment. Interestingly, we find partial quantum features already for this simple setting. Our main emphasis concerns, however, the information transport in the classical statistical systems. 

We label the three classical Ising spins by $s_k(t), k=1\dots 3$, replacing here the index $\gamma$ by $k$ since the three spins may be associated with spins having each a ``direction'' in one of the three space dimensions. We will find many analogies with the three components of a single quantum spin. We work in the occupation number basis and number the eight states $\tau$ by the values of $s_1,s_2,s_3$, e.g. $\tau=1\dots 8$ corresponding to $(-1,-1,-1)$, $(-1,-1,1)$, $(-1,1-1)$, $(-1,1,1)$, $(1,-1,-1)$, $(1,-1,1)$, $(1,1,-1)$, $(1,1,1)$. The step evolution operator $S$ is an $8\times 8$-matrix. 

\subsection{Unique jump operations}

Consider first a simple constrained Ising model for $\beta\to\infty$, 
\ba\label{J1}
{\cal L}_H(t)&=&-\beta
\big(s_1\te s_3(t)+s_3\te s_1(t)\nn\\
&&-s_2\te s_2(t)-3\big).
\ea
One finds ${\cal K}(t)=1$ only if $s_1\te=s_3(t)$, $s_3\te=s_1(t)$ and $s_2\te=-s_2(t)$. For all other configurations one has ${\cal K}(t)=0$. Correspondingly, the step evolution is given by 
\be\label{XAX}
S_{13}=S_{31}=S_{27}=S_{72}=S_{45}=S_{54}=S_{68}=S_{86}=1,
\ee
with all other elements zero. The evolution of the wave function obeys
\ba\label{J2}
\q1\te&=&\q3(t),~\q2\te=\q7(t),\nn\\
\q3\te&=&\q1(t),~\q4\te=\q5(t),\nn\\
\q5\te&=&\q4(t),~\q6\te=\q8(t),\nn\\
\q7\te&=&\q2(t),~\q8\te=\q6(t),
\ea
and similar for the components of the conjugate wave function $\bar q_\tau(t)$, and therefore also for the local probabilities $p_\tau(t)$. 

The expectation value of $s_1$ is interpreted as the expectation value of a spin in the $x$-direction, 
\be\label{J3}
\rho_1=\langle s_1\rangle =-(p_1+p_2+p_3+p_4)+p_5+p_6+p_7+p_8.
\ee
The corresponding classical spin operator $A_1$ is a diagonal matrix, $A'_1=$diag$(-1,-1,-1,-1,1,1,1,1)$. For the other two spin directions one obtains similarly $A'_2=$diag$(-1,-1,1,1,-1,-1,1,1)$ and $A'_3=$diag$(-1,1,-1,1,-1,1,-1,1)$, or 
\ba\label{J4}
\rho_2&=&\langle s_2\rangle\nn\\
&=&-p_1-p_2+p_3+p_4-p_5-p_6+p_7+p_8,\nn\\
\rho_3&=&\langle s_3\rangle\\
&=&-p_1+p_2-p_3+p_4-p_5+p_6-p_7+p_8.\nn
\ea
With the unique jump evolution \eqref{J2} one has
\be\label{J5}
\rho_1\te=\rho_3(t),~\rho_3\te=\rho_1(t),~\rho_2\te=-\rho_2(t).
\ee

This simple transformation can be mapped to a unitary transformation in (discrete) quantum mechanics for a single spin. We define \cite{CWQM} a complex $2\times 2$ ``quantum density matrix'' $\rho$,
\be\label{J6}
\rho=\frac12(1+\rho_k\tau_k),~\rho^\dagger=\rho,~\tr \rho=1.
\ee
The three Pauli matrices $\tau_k$ can be associated with the spin operators $\hat S_k$ in the three space directions $(\hbar/2=1)$, such that the expectation values of the quantum spin agrees with the classical Ising spins
\be\label{J7}
\rho_k=\langle s_k\rangle =\text{Tr}(\rho\hat S_k)=\text{Tr}(\rho\tau_k).
\ee
The positivity of the quantum density matrix $\rho$ is realized if we restrict the classical probability distributions (by suitable initial conditions) to ones for which $\sum_k\rho^2_k\leq 1$. A pure state quantum density matrix obtains for the limiting case $\sum_k\rho^2_k=1$. 

The unique jump evolution \eqref{J5} constitutes a discrete unitary evolution of the quantum density matrix
\be\label{J8}
\rho\te =U(t)\rho(t)U^\dagger(t),~U^\dagger (t)U(t)=1,
\ee
with ``Hadamard gate''
\be\label{J9}
U_H=\frac{1}{\sqrt{2}}
\left(\begin{array}{rr}
1&1\\1&-1
\end{array}
\right).
\ee
Other quantum operators can be realised in a similar way. For example,
\ba\label{J10}
{\cal L}_{31}(t)&=&\beta
\big(s_1\te s_3(t)-s_3\te s_1(t)\nn\\
&&-s_2\te s_2(t)+3\big)
\ea
induces for $\beta\to\infty$
\be\label{J11}
\rho_1\te=-\rho_3(t),~\rho_3\te=\rho_1(t),~\rho_2\te =\rho_2(t),
\ee
or the unitary transformation \eqref{J8} with 
\be\label{J12}
U_{31}=\frac{1}{\sqrt{2}}\left(\begin{array}{rr}
1&1\\-1&1
\end{array}
\right).
\ee
Further quantum operations as $U_X,U_Y,U_Z=\tau_1,i\tau_2,\tau_3$ are realized by simultaneous sign flips of $s_2$ and $s_3$, $s_1$ and $s_3$, or $s_1$ and $s_2$, respectively. A jump $\rho_1\te=\rho_2(t)$, $\rho_2\te =-\rho_1(t)$, $\rho_3\te=\rho_3(t)$ results in 
\be\label{J13}
U_{12}=
\left(\begin{array}{rr}
1&0\\0&-i
\end{array}
\right).
\ee

A sequence of two jumps $S_AS_B$ is reflected by the multiplication of the corresponding unitary matrices. The non-commutative matrix product of step evolution operators maps to the non-commutative matrix product for unitary quantum evolution operators. A sequence of discrete quantum operations, as needed for quantum computing, can be realized as a static memory material with $t$-dependent $S$, e.g. $S_A(t+2\epsilon)S_B\te S_C(t)$ results in $U_A(t+2\epsilon)U_B\te U_C(t)$. While all matrix elements of $S$ are positive, the matrix elements of $U$ can be negative or imaginary. In case of a quantum density matrix for a pure quantum state, e.g. $\rho^2=\rho$, we can employ the usual complex wave functions $\psi_\alpha$, $\rho_{\alpha\beta}=\psi_\alpha\psi^*_\beta$. Their evolution obeys a discrete Schr\"odinger equation. 

Arbitrary products of unique jump operators can also be realized as a unique jump operator for a single time step. Those that can be represented as unitary transformations on the quantum density matrix $\rho$ form a discrete subgroup of $SU(2)\times U(1)$. Here the overall phase of the unitary transformation, which corresponds to the $U(1)$-factor, plays no role, since it leaves $\rho$ invariant. Examples for products are $(U_{31})^2=i\tau_2$, which flips the sign of $\rho_1$ and $\rho_3$, or
\be\label{J13A}
U_{31}U_{12}=\frac{1}{\sqrt{2}}
\left(\begin{array}{rr}
1&-i\\-1&-i
\end{array}
\right),
\ee
which transforms
\be\label{J13B}
\rho_1\te =-\rho_3(t),~\rho_2\te =-\rho_1(t),~\rho_3\te =\rho_2(t).
\ee
A rotation in the 2-3-plane obtains for 
\be\label{J13C}
U_{23}=U_YU_{12}U_{31}U_{12}=\frac{i}{\sqrt{2}}
\left(\begin{array}{rr}
1&i\\i&1
\end{array}
\right),
\ee
which realizes
\be\label{J13D}
\rho_1\te =\rho_1(t),~\rho_2\te=\rho_3(t),~\rho_3\te=-\rho_2(t). 
\ee

The probabilistic information contained in the three numbers $\rho_3$ can be viewed as characterizing a subsystem. With respect to the subsystem the information about the expectation values of the Ising spins $s_k$ can be recovered by use of the non-commuting operators $\hat S_k=\tau_k$ in \cref{J7}. In contrast, information about classical correlations, as $\langle s_k,s_l\rangle$, is not available in the subsystem.

\subsection{Partial loss of information}

Unique jump operations are a particular limiting case. For more general step evolution operators one typically encounters a partial or complete loss of boundary information far inside the bulk. As a simple example we may consider constrained Ising models of the type
\be\label{J13E}
{\cal L}(t)={\cal L}_0(t)-\kappa
\big(s_1\te s_1(t)+s_3\te s_3(t)-2\big).
\ee
In the limit $\kappa\to\infty$ the values of the spins $s_1$ and $s_{3}$ is conserved. Non-vanishing elements of the step evolution operator can occur only on the diagonal and for the elements $S_{13},S_{31},S_{24},S_{42},S_{57},S_{75},S_{68},S_{86}$. For the moment we only require for ${\cal L}_0(t)$ that the equipartition state $\tilde q_* = (1/2\sqrt{2})(1,1,1,1,1,1,1,1)$ is left invariant by the evolution. (This happens indeed for a large class of ${\cal L}_0$.) Invariance of equipartition is realized if the sum of all elements of $S$ in a given column equals one, 
\be\label{J13F}
\sum_\tau S_{\tau\rho}=1.
\ee

For this general class of models the step evolution operator takes the form 
\ba\label{J14}
S_{pl}&=&
\left(\begin{array}{rr}
S_+&0\\0&S_-
\end{array}
\right),\\
S_\pm&=&
\left(\begin{array}{cccc}
1-a_\pm&0&c_\pm&0\\
0&1-b_\pm&0&d_\pm\\
a_\pm&0&1-c_\pm&0\\
0&b_\pm&0&1-d_\pm
\end{array}
\right),\nn
\ea
with $0\leq a_\pm,b_\pm,c_\pm,d_\pm\leq 1$. Four eigenvalues of this step evolution operator equal one, while the other four are given by
\be\label{J15}
\lambda_\pm=(1-a_\pm-c_\pm),~\lambda'_\pm=(1-b_\pm-d_\pm).
\ee
The partial conservation of boundary information is related to the four-dimensional space of eigenfunctions to the eigenvalues $\lambda=1$. Information imprinted in this sector by the boundary conditions will not be lost in the bulk. 

Additional information is preserved if $a_+=c_+=0$ or similarly, since additional eigenvalues equal one. Additional information is also preserved if some eigenvalues equal minus one, e.g. for $a_+=c_+=1$. In particular, $a_\pm=b_\pm=c_\pm=d_\pm=1$ realizes a unique jump operator with symmetric $S$, $S_{13}=S_{24}=S_{67}=S_{68}=1$. This transformation corresponds to a flip of $s_2$ and maps 
\be\label{J16}
\rho_2\leftrightarrow - \rho_2,~\rho_1,\rho_3 \textup{ invariant}.
\ee
On the level of the quantum density matrix it amounts to complex conjugation, $\rho\to\rho^*$, and is not represented by a unitary transformation \eqref{J8}.

We are interested here in values of $a_\pm,b_\pm,c_\pm,d_\pm$ for which the eigenvalues obey $|\lambda_\pm|<1, ~|\lambda'_\pm|<1$. For simplicity, we also exclude zero eigenvalues, e.g. $a_\pm+c_\pm\neq 1$, such that $S_{pl}$ is invertible. A basis of linearly independent eigenfunctions for the eigenvalues one of $S_{pl}$ is given by 
\ba\label{J17}
v^{(1+)}&=&
\begin{pmatrix}
v^{(1+)} \\0
\end{pmatrix},~
v^{(1-)}=
\begin{pmatrix}
0\\ w^{(1-)}
\end{pmatrix},\nn\\
v^{(2+)}&=&
\begin{pmatrix}
w^{(2+)} \\ 0
\end{pmatrix},~
v^{(2-)}=
\begin{pmatrix}
0 \\ w^{(2-)}
\end{pmatrix},
\ea
with 
\be\label{J18}
w^{(1\pm)}=\frac{1}{\sqrt{a^2_\pm+c^2_\pm}}
\begin{pmatrix} c_\pm\\0\\a_\pm\\0 \end{pmatrix},~
w^{(2\pm)}=\frac{1}{\sqrt{b^2_\pm+d^2_\pm}}
\begin{pmatrix} 0\\d_\pm\\0\\b_\pm \end{pmatrix}.
\ee
The information in the space spanned by these four eigenfunctions is preserved. The eigenfunctions for the eigenvalues $\lambda_\pm$ and $\lambda'_\pm$, eq. \eqref{J15}, are given by
\ba\label{J19}
u^{(1+)}&=&
\begin{pmatrix}
\tilde u^1\\0
\end{pmatrix},~
u^{(1-)}=
\begin{pmatrix}
0\\\tilde u^1
\end{pmatrix},\nn\\
u^{(2+)}&=&
\begin{pmatrix}
\tilde u^{2}\\0
\end{pmatrix},~
u^{(2-)}=
\begin{pmatrix}
0\\ \tilde u^2
\end{pmatrix},
\ea
with 
\be\label{J20}
\tilde u^1=\frac{1}{\sqrt{2}}
\begin{pmatrix}
1\\0\\-1\\0 
\end{pmatrix},~
\tilde u^2=\frac{1}{\sqrt{2}}
\begin{pmatrix}
0\\1\\0\\-1 
\end{pmatrix}.
\ee

For $S(t)$ independent of $t$ one has far inside the bulk for large $P$
\be\label{J21}
\tilde q(t_{in}+P\epsilon)=S^P\tilde q(t_{in}),
\ee
The part of the wave function that can be written as a linear combination of the basis functions \eqref{J19} will be suppressed by factors $|\lambda_\pm|^P,~|\tilde \lambda_\pm|^P$. If these eigenvalues are not close to $1$ or $-1$, this part vanishes far inside the bulk. Memory of the initial information in $\tilde q(t_{in})$ contained in this subspace is lost. 

In order to understand the character of the information kept far inside the bulk we observe that the classical spin operators $A'_1$ and $A'_3$ commute with $S_{pl}$
\be\label{J22}
[A'_1,S_{pl}]=0,~[A'_3,S_{pl}]=0.
\ee
The expectation values $\langle s_1\rangle$ and $\langle s_3\rangle$, as well as the classical correlation $\langle s_1s_3\rangle$, are preserved by the evolution. This is easily seen by relations of the type
\ba\label{J23}
\langle s_3\te \rangle&=&\bar q^T\te A'_3\tilde q\te\nn\\
&=&\bar q^T\te A'_3 S(t)\tilde q(t)\nn\\
&=&\bar q^T\te S(t) A'_3\tilde q(t)=\bar q^T(t)A'_3\tilde q(t)\nn\\
&=&\langle s_3(t)\rangle.
\ea
The step evolution operator \eqref{J14} is characterized by conserved quantities. With fixed values of $\langle s_1\rangle$, $\langle s_3\rangle$, $\langle s_1s_3\rangle$ the non-trivial part of the evolution concerns the Ising spin $s_2$. 

More interesting evolutions can be described if we combine the step evolution operator \eqref{J14} with suitable unit jump operators. Let us restrict for a moment the discussion to the family of ${\cal L}_0(t)$ in \cref{J13E} which is symmetric in the exchange of $s_k\te$ and $s_k(t)$. This results in a symmetric step evolution operator, $S^T=S$, or $c_\pm=a_\pm,~d_\pm=b_\pm$. The eigenvectors \eqref{J17} are then given by 
\be\label{J24}
w^{(1\pm)}=\frac{1}{\sqrt{2}}
\begin{pmatrix}
1\\0\\1\\0 
\end{pmatrix},~
w^{(2\pm)}=\frac{1}{\sqrt{2}}
\begin{pmatrix}
0\\1\\0\\1 
\end{pmatrix}.
\ee 
The eight eigenvectors $(v^{(1+)}$, $v^{(2+)}$, $v^{(1-)}$, $v^{(2-)}$, $u^{(1+)}$, $u^{(2+)}$, $u^{(1-)}$, $u^{(2-)}$) form an ON-basis. In this basis $S_{pl}$ is diagonal, $S_{pl}=$diag$(1,1,1,1,1-2a_+$, $1-2b_+$, $1-2a_-$, $1-2b_-)$. Unit jump operators that act separately on the two subsystems $(v^{(1+)}$, $v^{(2+)}$, $v^{(1-)}$, $v^{(2-)})$ and $(u^{(1+)}$, $u^{(2+)}$, $u^{(1-)}$, $u^{(2-)})$ can be considered as separate orthogonal transformations within each of the two systems. For example, the unit jump transformation $S_H$, which represents the Hadamard gate \eqref{J2}, \eqref{J9}, maps $v^{(2+)}\leftrightarrow v^{(1-)}$, while leaving $v^{(1+)}$ and $v^{(2-)}$ invariant. In the other sector it maps $u^{(2+)}\leftrightarrow -u^{(1-)}$ and changes the sign of $u^{(1+)}$ and $u^{(2-)}$. The combined step evolution operator $S_HS_{pl}$ acts in the memory-subsector $(v^{(1+)}$, $v^{(2+)}$, $v^{(1-)}$, $v^{(2-)})$ as a unit-jump ``Hadamard transformation'', while the action in the other subsector is more complicated. 

We may define a reduced quantum system for a two-component quantum spin with components $\hat S_1$ and $\hat S_3$. This leaves out in eq. \eqref{J6} the part $\sim \tau_2$ such that $\rho$ is now a real symmetric $2\times 2$ matrix. Since a general $S_{pl}$ leaves $\rho_1$ and $\rho_3$ invariant, and $S_H$ changes $\rho_1\leftrightarrow\rho_3$, the combined transformation $S_HS_{pl}$ also exchanges $\rho_1\leftrightarrow\rho_3$. For the reduced quantum system $S_HS_{pl}$ results in the unitary transformation \eqref{J9} for the Hadamard gate, independently of the details of $S_{pl}$. As it should be, the corresponding unitary matrix $U_H$ is a real orthogonal matrix. A similar discussion holds for the transformation $S_{31}S_{pl}$, realized in the $(\rho_1,\rho_3)$ sector by the transformations \eqref{J11}, \eqref{J12}, or for the transformations $U_X,U_Y,U_Z$. The gate $U_Y$ switches the sign of both $\rho_1$ and $\rho_3$. For all unique jump operations $S$ for which $\rho_1\te$ and $\rho_3\te$ can be expressed uniquely in terms of $\rho_1(t)$ and $\rho_3(t)$, the combination $SS_{pl}$ is represented as a transformation of the real density matrix $\rho$. This property is not realized by the transformation $S_{12}$, cf. \cref{J13}, which therefore does not act as a transformation within the reduced quantum system. 

\subsection{Boundary problem} 
For a practical demonstration of the solution of the boundary problem we consider a generalized Ising model \eqref{J13E} with time independent ${\cal L}$, such that the step evolution operator \eqref{J14} is the same for all $t$. For $S_{pl}$ the expectation values $\langle s_1\rangle$, $\langle s_3\rangle$ and $\langle s_1s_3\rangle$ do not change with $t$. The boundary problem has two parts. The first is the computation of $\langle s_1\rangle$, $\langle s_3\rangle$, $\langle s_1s_3\rangle$ for given initial and final boundary conditions. The second concerns the information about $\langle s_2(t)\rangle$, as well as the correlation functions $\langle s_1(t)s_2(t)\rangle$, $\langle s_2(t)s_3(t)\rangle$, $\langle s_1(t)s_2(t)s_3(t)\rangle$. The $t$-dependence of these quantities depends on the values of $\langle s_1\rangle$, $\langle s_3\rangle$ and $\langle s_1s_3\rangle$. Our three-spin chain can therefore realize a conditional evolution of the expectation value and correlations of $s_2$. We will concentrate on interesting limits, and indicate the procedure and qualitative outcome for the general case. 

We first consider the case where the absolute value of all four eigenvalues $\lambda_\pm$ and $\lambda'_\pm$ in \cref{J15} is smaller than one. The infinite volume limit takes an infinite number of time steps both between $t$ and $t_{in}$, and between $t_f$ and $t$. As a result, for both $\tilde q(t)$ and $\bar q(t)$ the part involving the basis vectors $u^{(\alpha)}$ in \cref{J19} vanishes. This identifies $\tilde q_3(t)=\tilde q_1(t)$, $\tilde q_4(t)=\tilde q_2(t)$, $\tilde q_7(t)=\tilde q_5(t)$, $\tilde q_8(t)=\tilde q_6(t)$, and similar for $\bar q_\tau(t)$ and the local probabilities $p_\tau(t)$. One infers from \eqref{J4} that $\langle s_2(t)\rangle$ vanishes. For the correlations one finds
\ba\label{J25}
\langle s_1s_2\rangle&=&p_1+p_2-p_3-p_4-p_5-p_6+p_7+p_8\nn\\
&=&0,\nn\\
\langle s_2s_3\rangle&=&p_1-p_2-p_3+p_4+p_5-p_6-p_7+p_8\nn\\
&=&0,\nn\\
\langle s_1s_2s_3\rangle&=&-p_1+p_2+p_3-p_4+p_5-p_6-p_7+p_8\nn\\
&=&0.
\ea
Far enough inside the bulk the expectation value of $s_2$ and its correlations all vanish. The meaning of ``far enough'' depends, however, on the size of the eigenvalues $\lambda_\pm,\lambda'_\pm$. 

In the infinite volume limit we can determine explicitly $\langle s_1\rangle$, $\langle s_3\rangle$ and $\langle s_1s_3\rangle$. They depend both on the initial and final boundary conditions. Since they are constant we determine them at $t_{in}$. For $\bar q(t_{in})$ the coefficients of the basis functions $u^{(\alpha)}$ vanish. With $\bar q(t_f)=f^{(\alpha)}w^{(\alpha)}+g^{(\alpha)}u^{(\alpha)}$ one finds that $p_\tau(t_{in})$ keeps only the information stored in $f^{(\alpha)}$, 
\ba\label{J26}
p_1&=&\frac{f^{(1+)}c_+\tilde q_1}{\sqrt{a^2_++c^2_+}}~,~
p_2=\frac{f^{(2+)}d_+\tilde q_2}{\sqrt{b^2_++d^2_+}},\nn\\
p_3&=&\frac{f^{(1+)}a_+\tilde q_3}{\sqrt{a^2_++c^2_+}}~,~
p_4=\frac{f^{(2+)}b_+\tilde q_4}{\sqrt{b^2_++d^2_+}},\nn\\
p_5&=&\frac{f^{(1-)}c_-\tilde q_5}{\sqrt{a^2_-+c^2_-}}~,~
p_6=\frac{f^{(2-)}d_-\tilde q_6}{\sqrt{b^2_-+d^2_-}},\nn\\
p_7&=&\frac{f^{(1-)}a_-\tilde q_7}{\sqrt{a^2_-+c^2_-}}~,~
p_8=\frac{f^{(2-)}b_-\tilde q_8}{\sqrt{b^2_-+d^2_-}},
\ea
where $\tilde q_\tau$ and $p_\tau$ stands for $\tilde q_\tau(t_{in})$ and $p_\tau(t_{in})$. The initial wave function $\tilde q_\tau(t_{in})$ and final wave function $\bar q_\tau(t_f)$ have to be normalized such that $Z=1$, implying $\sum_\tau p_\tau(t_{in})=1$. Despite the simple structure of $S_{pl}$ the interplay between initial and final boundary information is already rather complex. It simplifies for symmetric $S_{pl}$ for which one infers
\ba\label{J27}
\langle s_1(t_{in})\rangle&=&-p_{13}-p_{24}+p_{57}+p_{68},\nn\\
\langle s_3(t_{in})\rangle&=&-p_{13}+p_{24}-p_{57}+p_{68},\nn\\
\langle s_1(t_{in})s_3(t_{in})\rangle&=&p_{13}-p_{24}-p_{57}+p_{68},
\ea
with 
\ba\label{J28}
p_{13}&=&p_1+p_3=\frac{f^{(1+)}(\tilde q_1+\tilde q_3)}{\sqrt{2}},\nn\\
p_{24}&=&p_2+p_4=\frac{f^{(2+)}(\tilde q_2+\tilde q_4)}{\sqrt{2}},\nn\\
p_{57}&=&p_5+p_7=\frac{f^{(1-)}(\tilde q_5+\tilde q_7)}{\sqrt{2}},\nn\\
p_{68}&=&p_6+p_8=\frac{f^{(2-)}(\tilde q_6+\tilde q_8)}{\sqrt{2}}.
\ea
The expectation values \eqref{J27} depend on the initial and final wave functions, but no longer on the precise form of ${\cal L}_0$. 

We next consider a different limit for which the evolution of the expectation value and correlations of $s_2$ strongly depends on $\langle s_1\rangle$, $\langle s_3\rangle$, $\langle s_1s_3\rangle$. Let us take $S_{pl}$ with $a_+=c_+=0$, $b_-=d_-=1$, $c_-=a_-$, $d_+=b_+$, such that it depends on two parameters $a_-$ and $b_+$. The eigenvalues $\lambda_+=1$ and $\lambda'_-=-1$ imply that additional boundary information is kept, while for $|\lambda_-|<1$, $|\lambda'_+|<1$ part of the boundary information is still lost far inside the bulk. The additional transported information results in 
\ba\label{J29}
\tilde q_1\te &=&\tilde q_1(t)~,~\tilde q_3\te =\tilde q_3(t),\nn\\
\tilde q_6\te &=&\tilde q_8(t)~,~\tilde q_8\te =\tilde q_6(t),
\ea
and similar for $\bar q_\tau$ and $p_\tau$. Thus $p_1$ and $p_3$ do not depend on $t$, while for every $t$-step one observes a switch $p_6\leftrightarrow p_8$. 

This situation describes a conditional evolution of $s_2(t)$, depending on the values of $s_1(t)$ and $s_3(t)$. If both $s_1$ and $s_3$ are positive, $s_2(t)$ switches sign at every $t$-step,
\be\label{J30}
s_1=s_3=1~:~\langle s_2 (t)\rangle=-p_6(t)+p_8(t).
\ee
On the other hand, if $s_1$ and $s_3$ are both negative, $s_2(t)$ remains constant, according to
\be\label{J31}
s_1=s_3=-1~:~\langle s_2(t)\rangle=-p_1(t)+p_3(t).
\ee
For an opposite sign of $s_1$ and $s_3$, $s_1s_3=-1$, the expectation value $|\langle s_2\rangle |$ decreases as one moves from the boundary further inside the bulk,
\be\label{J32}
s_1s_3=-1~:~\langle s_2(t)\rangle=-p_2(t)+p_4(t)-p_5(t)+p_7(t).
\ee
In the limiting case $a_-=b_+=1/2$, where $\lambda_-=\lambda'_+=0$, the information about $s_2$ is immediately erased for $s_1s_3=-1$, since
\ba\label{J33}
\tilde q_2\te&=&\q4\te=\frac12\big(\tilde q_2(t)+\tilde q_4(t)\big),\nn\\
\q5\te&=&\q7\te=\frac12\big(\q5(t)+\q7(t)\big),
\ea
implies that the information about the differences $\q2(t_{in})-\q4(t_{in})$ and $\q5(t_{in})-\q7(t_{in})$ does not influence $\tilde q(t_{in}+\epsilon)$. With $p_2\te=p_4\te$, $p_5\te=p_7\te$ one finds for $s_1s_3=1$ a vanishing expectation value $\langle s_2\te\rangle =0$. Similar considerations hold for the correlations involving $s_2$. 

For the initial conditions $s_1(t_{in})=s_3(t_{in})=1$, $s_1(t_{in})=s_3(t_{in})=-1$, and $s_1(t_{in})s_3(t_{in})=-1$ we find rather different evolutions of $s_2(t)$. This demonstrates how a conditional evolution can be implemented for suitable generalized Ising models. General initial conditions $\tilde q(t_{in})$ can be decomposed into the eigenfunctions $v^{(\alpha)}$ and $u^{(\alpha)}$ which evolve according to the respective eigenvalues of $S_{pl}$. 

There are other similar limiting cases. The amount of local information at $t$, encoded in the classical wave functions $\tilde q(t)$ and $\bar q(t)$, depends on the size of $\lambda^P_j$ and $\lambda^{P'}_j$, with $\lambda_j=\lambda_\pm,~\lambda'_\pm$ and $t-t_{in}=P\epsilon$, $t_f-t=P'\epsilon$. If the products of eigenvalues are close to zero the corresponding boundary information is lost, if they differ substantially from zero it is preserved. For given parameters in $S_{pl}$ the transport of information can conveniently be followed by solving the evolution equation for the classical wave functions \eqref{eqn:AF13}. This extends to situations where $S_{pl}$ is combined with a unique jump operator, and for $t$-dependent $S(t)$. 

In principle, the results of this section could also be obtained from the transfer matrix formalism. For $S(t)$ independent of $t$ they correspond to a diagonalization of the transfer matrix. For many aspects the concepts of classical wave functions allow for a simpler understanding, however. For example, the initial probabilities $p_\tau(t_{in})$, given by eq. \eqref{J26}, are rather easily understood in terms of information from the boundary at $t_f$ transported by the conjugate wave function $\bar q(t_{in})$. Similar to quantum mechanics, an important strength of the Schr\"odinger picture with wave functions, as compared to the Heisenberg picture, concerns the evolution for $S(t)$ depending on $t$. (This corresponds to a time dependent Hamiltonian in quantum mechanics.) The advantage of wave functions is the availability of the local probabilistic information at every $t$, such that the effect of the evolution to neighboring $t$ can easily be followed in a local manner.

\section{Observables and non-commuting operators}
\label{sec:observables}

In this section we discuss non-commuting product structures between observables in classical statistical systems. Following the main emphasis of this paper we are concerned here with the quantum formalism for classical statistics. While the non-commuting operators for the quantum subsystems in the preceding section give a first flavor for the occurrence of non-commutativity, they are not the subject of the present section. We show here how non-commutativity arises naturally in classical statistical systems if the information is reduced to a ``local subsystem'' at a given $t$. This is independent of the presence or absence of quantum subsystems.

As a key point, the local information encoded in the pair of wave functions $\tilde{q}(t)$ and $\bar{q}(t)$ or, more generally in the density matrix, permits the computation of expectation values of observables $A(t)$ beyond the classical local observables. For these observables the probabilistic information in the wave functions or density matrix is incomplete. It is sufficient to compute for a local observable $A(t)$ the expectation value $\langle A(t)\rangle$, as well as arbitrary powers $\langle A^n(t)\rangle$. For two local observables $A_1(t)$ and $A_2(t)$ it is not always sufficient, however, for a computation of the classical correlation function $\langle A_1(t) \, A_2(t)\rangle_\text{cl}$. The reason is that information about the joint probabilities of finding a given possible value for $A_1$, simultaneously with a given possible value of $A_2$, is missing. In this case $A_1$ and $A_2$ will be represented by operators that do not commute. As a typical example for a pair of ``non-commuting'' observables we will discuss a classical local observable $A(t)$ together with the observable for its $t$-derivative $\dot{A}(t)$. In this section we turn back to an arbitrary basis obeying \cref{eqn:AF10,eqn:AF10A}.

\subsection{Local observables}
\label{sec:local observables}

We start our discussion of the relation between observables and operators, and of the non-commuting structure of operators, by extending the class of local observables beyond the classical local observables. The classical local observables take a fixed value for every \textit{local} configuration of spins $s(t)$. This is no longer the case for the extended class of local observables. While they may still have fixed values for every overall spin configuration $\{s\}$, e.g. with spins for all $t$ specified, a value for a local spin configuration is not defined. Nevertheless, expectation values can be computed for the extended class of local observables by using only the local probabilistic information contained in the wave functions or density matrix.

As a first necessary condition for a local observable we require that the expectation value can be computed in terms of the classical wave functions $\tilde{q}(t)$ and $\bar{q}(t)$ by virtue of the relation \labelcref{eqn:AF20}. To every local observable $A(t)$ one can associate an operator $A^\prime(t)$ such that the local information contained in the wave functions is sufficient for the computation of the expectation value $\langle A(t)\rangle$. We have already discussed the classical local observables for which $A(t)$ is a function of the local spins $s_\gamma(t)$. The local information contained in the wave functions exceeds the information needed for the expectation values of classical local observables. For those only the local probabilities $p(t) = \bar{f}(t) f(t)$ are needed according to \cref{eqn:AF7B}. The additional local information contained in the wave functions permits us to compute the expectation values of further observables involving spins at different $t$ or derivative observables. We will see that the associated operators do not commute, in general. Our description of classical statistical systems will encounter the non-commuting operator structures which are characteristic for quantum mechanics. Non-commuting operators arise naturally from the use of wave functions to encode the local probabilistic information. They are not specific for quantum mechanics.

For a local observable we further require that it can be ``measured locally'' at a given $t$. This is understood in the sense that the possible outcomes of measurements of the observable $A$ are related to expectation values by probabilities $p_i^{(A)}(t)$ in a standard way
\begin{equation}\label{eqn:70A}
	\langle A\rangle
	= \sum_i A_i \, p_i^{(A)},
	\qquad
	\langle F(A)\rangle
	= \sum_i F(A_i) \, p_i^{(A)}.
\end{equation}
Here $p_i^{(A)}(t) \geq 0$ denote the probabilities to find a given possible measurement value $A_i$ of the observable $A$. The quantities $p_i^{(A)}(t)$ should be computable in terms of the wave functions $f(t)$ and $\bar{f}(t)$. They can be specific for a given observable $A$, such that they are not necessarily given by the local probabilities $p(t)$ in \cref{eqn:AF7A}. The requirement \labelcref{eqn:70A} is considered necessary but perhaps not sufficient for the notion of a local observable. In a more vague sense a measurement of a local observable at $t$ should involve spins in some neighborhood of $t$. It should not involve spins $s(t^\prime)$ for $t^\prime$ ``far away'' from $t$.

A set of classical correlation functions for classical observables obeys Bell's inequalities \cite{BE},while ideal measurements of correlations in quantum systems violate these inequalities. There are two ways to account for the violation of Bell's inequalities in our classical statistical setting. For the first, the observables $A$ and $B$ for which the correlation is computed are both classical observables. The correlation describing ideal measurements may, however, not be given by the classical correlation function, $A \cdot B$, but rather based on a different product of observables. For the second way $A$ or $B$ may be an observable with well defined expectation value and correlations, but not a classical observable in the sense that it takes a fixed value for every spin configuration. For inequalities involving several correlations it is sufficient that one of the correlations is not a classical correlation of classical observables in order to invalidate the corresponding bound. We will indeed encounter in classical statistics correlations relevant for measurements that violate Bell's inequalities.

\subsection{Local operators}
\label{sec:local operators}

We associate to each local observable $A$ a local operator $A^\prime(t)$. The operators $A^\prime$ are chosen such that the eigenvalues $\lambda_i^{(A)}$ of the matrix $A^\prime$ coincide with the possible measurement values $A_i$ of the observable $A$. Thus \cref{eqn:70A} becomes
\begin{equation}\label{eqn:68A}
	\langle A\rangle
	= \sum_i \lambda_i^{(A)} \, p_i^{(A)},
	\quad
	\langle F(A)\rangle
	= \sum_i F\bigl(\lambda_i^{(A)}\bigr) \, p_i^{(A)}.
\end{equation}
Furthermore, we require that the expectation value $\langle A\rangle$ can be computed in terms of $A^\prime$ by the ``quantum rule'' \labelcref{eqn:AF20}.

The matrix $A^\prime$ can be diagonalized by a similarity transformation,
\begin{equation}\label{eqn:71A}
	A_D^\prime
	= D A^\prime D^{-1}
	= \diag(a_\tau)
	= \diag\bigl(\lambda_i^{(A)}\bigr).
\end{equation}
We can therefore write \cref{eqn:AF20} as
\begin{alignedeqn}\label{eqn:71B}
	\langle A\rangle
	&= \bar{q}^\tran A^\prime \tilde{q}
	= \bar{q}^\tran D^{-1} A_D^\prime D\tilde{q}\\
	&= \sum_\tau \bigl[D^{-1})^\tran \bar{q}\bigr]_\tau \bigl[D\tilde{q}\bigr]_\tau a_\tau.
\end{alignedeqn}
We conclude that the matrix $A^\prime(t)$ represents a local observable obeying the relations \labelcref{eqn:70A} if the combinations
\begin{equation}\label{eqn:71C}
	w_\tau^{(A)}
	= \bigl[(D^{-1})^\tran \bar{q}\bigr]_\tau \, \bigl[D\tilde{q}\bigr]_\tau
\end{equation}
are positive for all $\tau$. For a non-degenerate spectrum we can then identify $p_i^{(A)} = w_\tau^{(A)}$. In case of degenerate eigenvalues the probabilities $p_i^{(A)}$ obtain as
\begin{equation}\label{eqn:71D}
	p_i^{(A)}
	= \sum_{\tau(\lambda_i) }w_\tau^{(A)}
\end{equation}
where the sum extends over all $\tau$ for which $a_\tau = \lambda_i^{(A)}$. The second part of \cref{eqn:70A} is obeyed provided one associates to the product observables $A^n$ the matrix products $\bigl[A^\prime(t)\bigr]^n$.

Inversely, if the spectrum of $A^\prime$ consists of the possible measurement values of $A$, and the combinations $w_\tau^{(A)}$ in \cref{eqn:71C} can be associated by \cref{eqn:71D} with the positive probabilities $p_i^{(A)}$ for all possible states (pairs $\bar{q}$, $\tilde{q}$ that can be realized by boundary conditions), we conclude that for all $n$ the operator associated to $A^n$ is given by the matrix product $[A^\prime(t)]^n$. We can infer the second equation \labelcref{eqn:68A}, such that for all $n$ one has
\begin{equation}\label{eqn:73A}
	\langle A^n\rangle
	= \bigl(\lambda_i^{(A)}\bigr)^n \, p_i^{(A)}.
\end{equation}
Indeed, the operator $B_n^\prime(t)$ associated to $A^n$ has to obey $\langle A^n\rangle = \langle B_n^\prime(t)\rangle$. Since $\bigl\langle[A^\prime(t)]^n\bigr\rangle$ equals the r.h.s. of \cref{eqn:73A} one infers
\begin{equation}
	\bigl\langle B_n^\prime(t) - [A^\prime(t)]^n\bigr\rangle
	= 0.
\end{equation}
If this holds for arbitrary states one can identify $B_n^\prime(t)$ with $[A^\prime(t)]^n$.

We summarize that for an operator $A^\prime(t)$ a central condition for being associated with a local observable is
\begin{equation}\label{eqn:71E}
	w_\tau^{(A)}
	\geq 0.
\end{equation}
This allows the standard probabilistic setting \labelcref{eqn:68A} with positive probabilities \labelcref{eqn:71D}. The normalization of the probabilities is guaranteed by the definition \labelcref{eqn:71C},
\begin{equation}\label{eqn:71F}
	\sum_\tau w_\tau^{(A)}
	= 1.
\end{equation}
The condition \labelcref{eqn:71E} depends on both the operator $A^\prime$ (which determines $D$) and the possible values of wave functions $\tilde{q},\bar{q}$ that characterize the allowed states. In short, any operator with real eigenvalues that can be diagonalized by a matrix $D$ for which the expression \labelcref{eqn:71C} is always positive, can possibly be associated with a local observable. Since the formal criteria are met, it is sufficient that a prescription of ``local measurements'' for this observable exists. The family of such local observables extends largely beyond the classical local observables. We will provide examples in the later parts of this section.

If the condition \labelcref{eqn:71E} is realized, and $A^n$ is associated to $(A^\prime)^n$, the identification of the spectrum of $A^\prime$ with the possible measurement values of $A$ is no separate condition. We exploit that the relation
\begin{equation}\label{eqn:OP8}
	\langle A^n(t) \rangle
	= \bar{q}^\tran D^{-1} A^{\prime n}_D D\tilde{q}
	= \sum_\tau w_\tau^{(A)} a_\tau^n
\end{equation}
holds for arbitrary $n$. For positive $w_\tau^{(A)} = p_\tau^{(A)}$ and a non-degenerate spectrum this implies indeed that the possible values of $A(t)$ that can be found in measurements are the eigenvalues $a_\tau$ of the matrix $A^\prime$. The extension of these statements to a degenerate spectrum of $A^\prime$ proceeds by considering degenerate spectra as limiting cases of non-degenerate spectra.

In summary, local observables can be characterized by four conditions for the associated operators:
\begin{equation}\label{eqn:76A}
	\begin{array}{ll@{\qquad}ll}
		\cond{cnd:expec} & \langle A\rangle = \langle A^\prime\rangle, & \cond{cnd:spec} & \spec(A^\prime) = \{\lambda_i^{(A)}\},\\
		\cond{cnd:non-neg} & w_i^{(A)} \geq 0, & \cond{cnd:corresp} & A^n \to (A^\prime)^n.
	\end{array}
\end{equation}
Here $\{\lambda_i^{(A)}\}$ are the possible real values that can be found by measurements of $A$. The \cref{cnd:non-neg} involves $w_i^{(A)}$ as defined by \cref{eqn:71C}. It should hold for all states (pairs of wave functions or density matrix that can be realized for appropriate boundary conditions). The \cref{cnd:corresp} has to hold for all $n$. The four conditions are not independent. One has
\begin{alignedeqn}\label{eqn:76B}
	&\labelcref{cnd:expec} \land \labelcref{cnd:spec} \land \labelcref{cnd:non-neg}
	\enskip\Rightarrow\enskip \labelcref{cnd:corresp}\\
	&\labelcref{cnd:expec} \land \labelcref{cnd:non-neg} \land \labelcref{cnd:corresp}
	\enskip\Rightarrow\enskip \labelcref{cnd:spec}
\end{alignedeqn}
In other words, in addition to the \cref{cnd:expec,cnd:non-neg}, either \labelcref{cnd:spec} or \labelcref{cnd:corresp} are sufficient. Beyond the formal aspects a local observable needs a practical prescription how it is measured ``locally''.

\subsection{Classical local observables}

We first discuss the classical local observables and show that they obey indeed the criterion \labelcref{eqn:71E}. Classical local observables $A(t)$ depend only on local spins $s(t)$. Their representation as operators is given by \cref{eqn:AF19}. We will next establish that the classical product of two classical local observables is represented by the operator product or matrix multiplication of the associated operators. For this purpose we first write the expectation value of the classical product of two classical local observables $A(t)$ and $B(t)$ in the form
\begin{alignedeqn}\label{eqn:OP1}
	\langle A(t) B(t)\rangle
	&= \int \mathcal{D} s(t)\bar{f}(t) A(t) B(t) f(t)\\
	&= \bar{q}_\tau(t) \, (A\cdot B)_{\tau\rho}^\prime(t) \, \tilde{q}_\rho(t),
\end{alignedeqn}
with
\begin{equation}\label{eqn:OP2}
	(A\cdot B)_{\tau\rho}^\prime
	= \int \mathcal{D} s(t) \, h_\tau(t) \, h_\rho(t) \, h_\sigma(t) \, h_\lambda(t) \, \alpha_\sigma(t) \, \beta_\lambda(t)
\end{equation}
the operator associated to the product observable $A(t) B(t)$. On the other hand the matrix product of the associated operators $A^\prime$ and $B^\prime$ reads according to \cref{eqn:AF19}
\begin{align}\label{eqn:OP3}
	A_{\tau\alpha}^\prime B_{\alpha\rho}^\prime
	= \int \mathcal{D} s \int \mathcal{D} \bar{s} \, &h_\tau(s) \, h_\sigma(s) \, h_\alpha(s) \, h_\alpha(\bar{s})\\
	&\times h_\lambda(\bar{s}) \, h_\rho(\bar{s}) \, \alpha_\sigma(s) \, \beta_\lambda(\bar{s}).\notag
\end{align}
Employing the completeness relation \labelcref{eqn:AF10A} one indeed finds that the expression \labelcref{eqn:OP2} is given by the matrix product \labelcref{eqn:OP3},
\begin{equation}\label{eqn:OP4}
	(A \cdot B)^\prime
	= A^\prime B^\prime.
\end{equation}
The explicit representation \labelcref{eqn:OP2} shows that the classical local observables are represented by commuting operators,
\begin{equation}\label{eqn:OP5}
	[A^\prime,B^\prime]
	= 0.
\end{equation}

The representation of products of classical local observables by the matrix products of the associated operators can be extended to an arbitrary number of factors. In particular, $A^n(t)$ is represented by $[A^\prime(t)]^n$, such that
\begin{equation}\label{eqn:OP6}
	\langle A^n(t) \rangle
	= \bar{q}^\tran A^{\prime n}\tilde{q}.
\end{equation}
The symmetric matrix $A^\prime$ can be diagonalized by an orthogonal transformation, $D^\tran D = 1$,
\begin{equation}\label{eqn:OP7}
	A_D^\prime
	= D A^\prime D^{T}
	= \diag(a_\tau),
\end{equation}
with $a_\tau$ the eigenvalues of $A^\prime$. If the criterion \labelcref{eqn:71E} holds the eigenvalues $a_\tau$ determine the possible measurement values of $A(t)$. In turn, these possible measurement values coincide with the possible values that the classical local observable $A[s(t)]$ can take for the various configurations of local spins $[s(t)]$. The quantum rule that possible measurement values of an observable are given by the spectrum of the associated operator is therefore directly rooted in the standard setting of classical statistics.

We want to show the relation (no sum over $\tau$)
\begin{equation}\label{eqn:OP9}
	p_\tau^{(A)}(t)
	= (D\bar{q})_\tau(t) (D\tilde{q})_\tau(t) \geq 0.
\end{equation}
For a non-degenerate spectrum the quantities $p_\tau^{(A)}$ can be interpreted as the probabilities to find the value $a_\tau$ for the local observable $A(t)$. We first establish the important property \labelcref{eqn:OP9} for classical local observables in the occupation number basis. In this basis the local probabilities for a given spin configuration $\tau$ simply obey \cite{CWIT} (no sum over $\tau$)
\begin{equation}\label{eqn:OP10}
	p_\tau(t)
	= \bar{q}_\tau(t) \tilde{q}_\tau(t).
\end{equation}
with $p_\tau(t) \geq 0$. Classical local observables are diagonal in this basis, such that $A^\prime$ coincides with $A_D^\prime$, $D = 1$. In this case $A_\tau(t)$ is the value that the observable $A(t)$ takes for the spin configuration $\tau$ for which $h_\tau[s] = 1$. With \cref{eqn:AF23} it indeed coincides with an eigenvalue of $A^\prime, A_\tau = a_\tau$, such that our statements hold for classical local observables in the occupation number basis.

For a proof of the property \labelcref{eqn:71E} in an arbitrary basis we proceed by a basis transformation, starting from the occupation number basis. A general complete orthogonal basis obeying \cref{eqn:AF10,eqn:AF10A} can be obtained from the occupation number basis by an orthogonal transformation,
\begin{equation}\label{eqn:OP17}
	h_\alpha^\prime
	= V_{\alpha\tau} \, h_\tau,
	\qquad
	V^\tran \, V
	= 1,
\end{equation}
where $h^\prime$ refers to a general basis function and $h$ to the one in the occupation number basis. With
\begin{equation}\label{eqn:PP1}
	\tilde{q}_\tau^\prime \, h_\tau^\prime
	= \tilde{q}_\tau \, h_\tau,
	\qquad
	\bar{q}_\tau^\prime \, h_\tau^\prime
	= \bar{q}_\tau \, h_\tau
\end{equation}
the transformation of the wave vectors reads
\begin{equation}\label{eqn:PP2}
	\tilde{q}_\alpha^\prime
	= V_{\alpha\tau}\tilde{q}_\tau,
	\qquad
	\bar{q}_\alpha^\prime
	= V_{\alpha\tau}\bar{q}_\tau,
\end{equation}
while operators transform as
\begin{equation}\label{eqn:PP3}
	A_V^\prime
	= VA^\prime V^\tran,
\end{equation}
such that the expectation value \labelcref{eqn:AF20} does not depend on the choice of basis. The spectrum of the operator remains invariant under the orthogonal transformation \labelcref{eqn:PP3}. For classical local observables the possible outcomes of measurements are therefore given in any basis by the eigenvalues of the associated operator.

For an arbitrary basis, classical local observables are, in general, no longer represented by diagonal operators. They correspond to operators $A_V^\prime$ that are related to the diagonal operators $A^\prime$ in the occupation number basis by the relation \labelcref{eqn:PP3}. Transforming the wave functions $\tilde{q}^\prime$ and $\bar{q}^\prime$ in some arbitrary basis to a form where $A_D^\prime$ is diagonal according to \cref{eqn:OP7} amounts precisely to a change of basis to the occupation number basis, $D = V^\tran$. Thus $p_\tau^{(A)}$ in \cref{eqn:OP9} corresponds to \cref{eqn:OP10} in the occupation number basis and is therefore positive semi definite. We conclude that for classical local observables the possible measurement values are given by the eigenvalues of $A^\prime$, and the associated probabilities $p_\tau^{(A)}$ are positive. This holds for an arbitrary basis.

\subsection{Observables involving spins at different \texorpdfstring{$t$}{t}}

A first class of local observables $A(t)$ beyond the classical local observables at $t$ can be related to classical local observables at some $t^\prime \neq t$. Indeed, the local information contained in the wave functions $\tilde{q}(t)$ and $\bar{q}(t)$ is sufficient for the computation of expectation values of classical local observables $A(t^\prime)$ for $t^\prime \neq t$. We start with the classical local observable $A(t + \epsilon)$ and show that its expectation value can be computed from the wave functions at $t$. It therefore belongs to the wider class of local observables at $t$, while not being a classical local observable at $t$. In the occupation number basis the operator $A^\prime(t)$ associated to $A(t + \epsilon)$ will, in general, no longer be diagonal. The operators for the classical observables $A(t + \epsilon)$ and $B(t)$ typically do not commute and we observe the natural appearance of non-commuting operators familiar from quantum mechanics.

From \cref{eqn:AF20},
\begin{equation}\label{eqn:PP4}
	\langle A(t + \epsilon)\rangle
	= \bar{q}^\tran(t + \epsilon) \, A^\prime(t + \epsilon) \, \tilde{q}(t + \epsilon),
\end{equation}
and \cref{eqn:AF13} one infers
\begin{equation}\label{eqn:PP5}
	\langle A(t + \epsilon)\rangle
	= \bar{q}^\tran(t) S^{-1}(t) A^\prime(t + \epsilon) S(t) \tilde{q}(t).
\end{equation}
The operator $A^\prime(t)$ associated to $A(t + \epsilon)$ therefore reads
\begin{equation}\label{eqn:PP6}
	A^\prime(t)
	= S^{-1}(t) \, A^\prime(t + \epsilon) \, S(t).
\end{equation}
In the occupation number basis a classical observable $A(t + \epsilon)$ is represented by diagonal $A^\prime(t + \epsilon)$. Therefore the operator representation for this observable at $t$, $A^\prime(t)$, typically is not diagonal. For a classical local observable at $t$, $B(t)$, the corresponding operator $B^\prime(t)$ is diagonal. One typically finds for the commutator of operators at $t$ which represent classical observables at different $t^\prime \neq t$,
\begin{equation}\label{eqn:PP7}
	\bigl[A^\prime(t),B^\prime(t)\bigr]
	\neq 0.
\end{equation}

We emphasize that these simple findings imply three important consequences:
\begin{enumerate}[label=\roman*)]
	\item The expectation values of classical local observables at $t^\prime$ different from $t$ can be computed from the information contained in the local wave functions at $t$. This is not possible with the local information contained in the local probabilities at $t$.
	
	\item Local observables that are represented by off-diagonal operators in the occupation number basis can have physical meaning. Their expectation value is given by \cref{eqn:AF20}. In our case the possible measurement values are again given by the spectrum of the associated operator. Indeed, the spectrum of $A^\prime(t + \epsilon)$ and $A^\prime(t)$ is the same, since the relation \labelcref{eqn:PP6} is a similarity transformation. If $S$ is not orthogonal, $A^\prime(t)$ needs not be symmetric.
	
	\item Operators representing physical observables do not need to commute. This may generate typical uncertainty relations as familiar from quantum mechanics.
\end{enumerate}

The possibility to compute expectation values of classical local observables at $t^\prime$ from the wave functions at $t$ extends to arbitrary $t^\prime$. For example, $A(t + 2 \epsilon)$ is represented at $t$ by the operator
\begin{equation}\label{eqn:PP8}
	A(t^\prime)
	= S^{-1}(t) S^{-1}(t + \epsilon) A^\prime(t + 2 \epsilon)S(t + \epsilon) S(t).
\end{equation}
This construction is analogous to the Heisenberg picture for operators in quantum mechanics. It can be used for the whole range of $t^\prime$ for which the step evolution operator is known between $t$ and $t^\prime$, and becomes particularly simple for $S$ independent of $t$. For all non-diagonal operators in the occupation number basis that can be obtained as representations of classical local observables at some $t^\prime \neq t$ we are guaranteed that the spectrum corresponds to the possible measurement values.

The association of classical local observables at $t^\prime$ with operators at $t$ is not invertible. Many different $A(t^\prime)$ may be mapped to the same $A^\prime(t)$. For given local wave functions $\tilde{q}(t)$ and $\bar{q}(t)$ they all have the same expectation value. This extends to functions $F\bigl(A(t^\prime)\bigr)$. Also the possible measurement values, as given by the spectrum of $A^\prime(t)$, are the same. We conclude that $A^\prime(t)$ defines an equivalence class of observables that all have the same possible measurement values and expectation values.

We may associate to the operator $A^\prime(t)$ a local observable without any reference to classical local observables at $t^\prime$. Indeed, the condition \labelcref{eqn:71E} is obeyed without any reference to a particular $A(t^\prime)$. In the occupation number basis we may consider the transformation \labelcref{eqn:PP6} as the similarity transformation that diagonalizes $A^\prime(t)$ in \cref{eqn:71A}, e.g. $D = S$, $A_D^\prime = A^\prime(t + \epsilon)$. The combinations \labelcref{eqn:71C} are given by the local probabilities at $t + \epsilon$,
\begin{alignedeqn}\label{eqn:90A}
	w_\tau^{(A)}
	&= \bigl((S^{-1})^\tran \bar{q}(t)\bigr)_\tau \bigl(S\tilde{q}(t)\bigr)_\tau\\
	&= \bar{q}_\tau(t + \epsilon)\tilde{q}_\tau(t + \epsilon)\\
	&= p_\tau(t + \epsilon)
	\geq 0.
\end{alignedeqn}
This generalizes to \cref{eqn:PP8} and further to local observables that can be obtained from arbitrary $A(t^\prime)$.

We can use the local operators $A^\prime(t)$ in an arbitrary basis. The relations of the type \labelcref{eqn:PP6,eqn:PP8} for the diagonalization of $A^\prime(t)$ are operator relations that remain preserved by similarity transformations. The same holds for the commutation relations as \cref{eqn:PP7}.

If one can find a prescription for a local measurement of the expectation value $\langle A(t^\prime)\rangle$, and more generally $\bigl\langle \bigl[A(t^\prime)\bigr]^n\bigr\rangle$, we can consider the observable defined by this prescription as a genuine local observable. The fact that the same operator $A^\prime(t)$ can also be associated to other observables as, for example, classical observables at $t^\prime \neq t$, even for $t^\prime$ very different from $t$, is no argument against the local character of the observable defined by the local measurement prescription. It simply states that other observables in the equivalence class have the same expectation values. The existence of classical observables at $t^\prime \neq t$ that belong to the same equivalence class can be considered merely as a technical tool in order to establish that the condition \labelcref{eqn:71E} holds.

\subsection{Derivative observables}

In quantum mechanics the perhaps most famous pair of observables represented by non-commuting operators is position and momentum. For non-relativistic particles momentum corresponds to velocity up to the proportionality factor mass, such that the pair may also be seen as operators for position $x$ and time derivative of position $\dot{x}$. In classical statistics the $t$-derivative $\dot{A}$ of a local observable $A$ is typically also a local observable. The operators for $A$ and $\dot{A}$ do not commute, similar to quantum mechanics \cite{CWPT}.

Let us consider a classical local observable $A(t)$. We may define the classical derivative observable
\begin{equation}\label{eqn:DO1}
	\dot{A}_m(t)
	= \frac{1}{2 \epsilon} \bigl[A(t + \epsilon) - A(t - \epsilon)\bigr].
\end{equation}
We use the subscript $m$ for the particular definition of the discrete time derivative. This particular definition will play no role for the continuum limit where the subscript can be dropped. The expectation value of $\dot{A}_m(t)$ can be computed from the operator $\dot{A}_m^\prime(t)$,
\begin{equation}\label{eqn:DO2}
	\langle\dot{A}_m(t)\rangle
	= \bar{q}^\tran(t) \, \dot{A}_m^\prime(t) \, \tilde{q}(t),
\end{equation}
with $\dot{A}_m^\prime$ formed similar to \cref{eqn:PP6},
\begin{alignedeqn}\label{eqn:DO3}
	\dot{A}_m^\prime(t)
	= \frac{1}{2 \epsilon} \bigl\{&S^{-1}(t) \, A^\prime(t + \epsilon) \, S(t)\\
	&- S(t - \epsilon) \, A^\prime(t - \epsilon) \, S^{-1}(t - \epsilon)\bigr\}.
\end{alignedeqn}

We concentrate on $S$ independent of $t$, and also on an observable $A(t)$ that has no explicit $t$-dependence, such that $A^\prime(t + \epsilon)$ and $A^\prime(t -\epsilon)$ correspond to the same operator $A^\prime$. The operator $\dot{A}_m^\prime$ is then independent of $t$ as well,
\begin{alignedeqn}\label{eqn:DO4}
	\dot{A}_m^\prime(t)
	&= \frac{1}{2 \epsilon} \bigl(S^{-1} \, A^\prime \, S - S \, A^\prime \, S^{-1}\bigr)\\
	&= \frac{1}{2 \epsilon} \bigl(S^{-1} \, [A^\prime,S] + [A^\prime,S] \, S^{-1}\bigr)\\
	&= \frac{1}{2 \epsilon} \bigl\{S^{-1},[A^\prime,S]\bigr\}.
\end{alignedeqn}
One infers for the commutator of $A^\prime$ and $\dot{A}_m^\prime$
\begin{equation}\label{eqn:DO5}
	[A^\prime,\dot{A}_m^\prime]
	= \frac{1}{2 \epsilon} \Bigl(\{A^\prime,S^{-1}\} \, [A^\prime,S] - \{A^\prime,S\} \, [A^\prime,S^{-1}]\Bigr).
\end{equation}

We may also express $\dot{A}_m^\prime$ in terms of $W$ as given by \cref{eqn:FF1},
\begin{equation}\label{eqn:DO6}
	\dot{A}_m^\prime
	= S \, A \, W - W A \, S.
\end{equation}
Acting on sufficiently smooth wave functions such that $S \, \tilde{q} = \tilde{q} + \mathcal{O}(\epsilon)$, one obtains for the continuum limit
\begin{equation}\label{eqn:DO7}
	\dot{A}^\prime
	= -[W,A^\prime].
\end{equation}
This relation can also be obtained directly from the continuum limit, using
\begin{alignedeqn}\label{eqn:DO8}
	\langle\dot{A}(t)\rangle
	= \frac{1}{2 \epsilon} \bigl\{&\bar{q}^\tran(t + \epsilon) \, A^\prime(t + \epsilon) \, \tilde{q}(t + \epsilon)\\
	&- \bar{q}^\tran(t - \epsilon) \, A^\prime(t - \epsilon) \, \tilde{q}(t - \epsilon)\bigr\},
\end{alignedeqn}
and \cref{eqn:AF14,eqn:AF16},
\begin{alignedeqn}\label{eqn:DO9}
	&\tilde{q}(t + \epsilon)
	= \tilde{q}(t - \epsilon) + 2 \epsilon \, W \, \tilde{q}(t),\\
	&\bar{q}^\tran(t + \epsilon)
	= \bar{q}^\tran(t - \epsilon) - 2 \epsilon \, \bar{q}^\tran(t) \, W.
\end{alignedeqn}
Taking the limit $\epsilon \to 0$ with $\tilde{q}(t + \epsilon) = \tilde{q}(t) + \mathcal{O}(\epsilon)$, $\bar{q}(t + \epsilon) = \bar{q}(t) + \mathcal{O}(\epsilon)$ yields \cref{eqn:DO7}.

With the formal relation \labelcref{eqn:FF3} we can write \cref{eqn:DO7} as
\begin{equation}\label{eqn:DO10}
	\dot{A}^\prime
	= i [H,A^\prime] - [J,A^\prime].
\end{equation}
The first piece yields the standard expression of quantum mechanics. The second piece arises from the symmetric part of $W$. Operators for conserved quantities obey $[W,A^\prime] = 0$, such that $\dot{A}^\prime = 0$. For general classical observables $W$ will not commute with $A^\prime$. Typically $[W,A^\prime]$ will not commute with $A^\prime$ either, such that $[\dot{A}^\prime,A^\prime] \neq 0$.

One may investigate a different choice of the $t$-derivative of a classical local observable, as
\begin{equation}\label{eqn:DO11}
	\dot{A}_+(t)
	= \frac{1}{\epsilon} \bigl(A(t + \epsilon) - A(t)\bigr).
\end{equation}
For $t$-independent $S$, and $A$ not depending explicitly on $t$ such that $A(t + \epsilon)$ is represented by $S^{-1} \, A^\prime \, S$, this yields
\begin{equation}\label{eqn:DO12}
	\dot{A}_+^\prime(t)
	= \frac{1}{\epsilon} S^{-1} [A^\prime,S],
\end{equation}
or
\begin{equation}\label{eqn:DO13}
	\frac{1}{2} \bigl(\dot{A}_+^\prime(t) + S \, \dot{A}_+^\prime \, S\bigr)
	= -[W,A^\prime] S.
\end{equation}
In the continuum limit one finds again the relation \labelcref{eqn:DO7}. The distinction between $\dot{A}_m^\prime$ and $\dot{A}_+^\prime$ is no longer needed.

In the continuum limit the derivative observables $\dot{A}_m$ and $\dot{A}_+$ cannot be distinguished by measuring their expectation values. Both classical observables are represented by the same operator $\dot{A}^\prime$. They may therefore be considered as equivalent. Nevertheless, $\dot{A}_m$ and $\dot{A}_+$ are different classical observables. Consider a classical local observable $A(t)$ with possible measurement values $\pm 1$, corresponding to the spectrum of $A^\prime$. The possible measurement values of $\dot{A}_m(t)$ are then $\{\pm 1/2 \epsilon,0\}$, while the possible measurement values of $\dot{A}_+(t)$ differ, being given by $\{\pm 1/\epsilon,0\}$. Neither for $\dot{A}_m(t)$ nor for $\dot{A}_+(t)$ do the possible measurement values coincide with the spectrum of the associated operator $\dot{A}^\prime$ as given by \cref{eqn:DO7}. The eigenvalues of $\dot{A}^\prime$ do not diverge for $\epsilon \to 0$ since both $A^\prime$ and $W$ remain finite in this limit. Thus $\dot{A}_m(t)$ and $\dot{A}_+(t)$ do not obey the requirements of a local observable.

Measurements of the $t$-derivative of $A$ in the continuum limit will not be able to resolve the difference between $\dot{A}_m(t)$ and $\dot{A}_+(t)$. Continuum measurements typically involve a finite range of $t$ even in the limit $\epsilon \to 0$. Formally, the measured observables are linear combinations of classical observables belonging to a large class of possible $t$-derivative observables. The precise linear combination typically even differs from one measurement to another, such that only a probability distribution of individual linear combinations is specified. For ``ideal continuum measurements'' of $\dot{A}$ we require, however, that $\langle\dot{A}(t)\rangle$ is reproduced faithfully, such that one can employ the operator $\dot{A}^\prime$ for the computation of the outcome.

This type of ``coarse-grained'' measurements will yield a finite value $\dot{A}(t)$ for each individual measurement even in the limit $\epsilon \to 0$. We associate to an ideal continuum measurement a local observable $\dot{A}(t)$ whose possible measurement values are given by the spectrum of the operator $\dot{A}^\prime$. This can be done provided that the probabilities $p_i$ defined by \cref{eqn:71C,eqn:71D} are positive for the operator $\dot{A}^\prime$. This new local observable is typically not a classical observable whose value is fixed for every overall spin configuration. The coarse graining rather produces a type of probabilistic observable \cite{CWPO,CWQM,CWEQM}, for which only probability distributions of values are defined on the level of the overall statistical ensemble. For continuum measurements at $t$ the coarse-grained observable $\dot{A}$ is, however, simply a local observable represented by $\dot{A}^\prime$ according to \cref{eqn:DO7}. This operator is typically not diagonal in the occupation number basis. Correlations for this type of observables need not obey Bell's inequalities.

\subsection{Classical correlations}

We can also compute the classical correlation function $\langle A(t + \epsilon) \, B(t)\rangle_\text{cl}$ from the local probabilistic information at $t$. This is done by representing $A(t + \epsilon) \, B(t)$ by an operator $C^\prime(t)$ at $t$,
\begin{equation}\label{eqn:PP9}
	A(t + \epsilon) \, B(t)
	\to C^\prime(t)
	= A^\prime(t) \, B^\prime(t).
\end{equation}
Here the r.h.s. of \cref{eqn:PP9} is given by the matrix product of the operators $A^\prime(t)$ and $B^\prime(t)$ representing at $t$ the observables $A(t + \epsilon)$ and $B(t)$. Indeed, from the general definition of the expectation value in the overall classical statistical ensemble \labelcref{eqn:AF4} we infer
\begin{alignedeqn}\label{eqn:PP10}
	\langle A(t + \epsilon) \, B(t)\rangle
	= \int \mathcal{D} s(t + \epsilon) \int \mathcal{D} s(t) \, \bar{f}(t + \epsilon)\\
	\times A(t + \epsilon) \, \mathcal{K}(t) \, B(t) \, f(t).
\end{alignedeqn}
Using \cref{eqn:AF19}, the expansion in basis functions yields
\begin{alignedeqn}\label{eqn:PP11}
	&\langle A(t + \epsilon)B(t)\rangle\\
	&= \smash{\int} \mathcal{D} s(t + \epsilon) \, \mathcal{D} s(t) \, h_\tau(t + \epsilon) \, h_\gamma(t + \epsilon)\\
	&\hphantom{={} \int \mathcal{D}}\times h_\sigma(t + \epsilon) \, h_\mu(t) \, h_\delta(t) \, h_\rho(t) \, \bar{q}_\tau(t + \epsilon)\\
	&\hphantom{={} \int \mathcal{D}}\times \alpha_\gamma(t + \epsilon) \, S_{\sigma\mu}(t) \, \beta_\delta(t) \, \tilde{q}_\rho(t)\\
	&= \bar{q}_\tau(t + \epsilon) A_{\tau\sigma}^\prime(t + \epsilon) S_{\sigma\mu}(t) B_{\mu\rho}^\prime(t) \tilde{q}_\rho(t)\\
	&= \bar{q}_\tau(t) S_{\tau\nu}^{-1}(t) A_{\nu\sigma}^\prime(t + \epsilon) S_{\sigma\mu}(t) B_{\mu\rho}^\prime(t) \tilde{q}_\rho(t)\\
	&= \bar{q}_\tau(t) \bigl(A^\prime(t) B^\prime(t)\bigr)_{\tau\rho} \tilde{q}_\rho(t).
\end{alignedeqn}
The classical correlation function is therefore represented by the $t$-ordered operator product, where the operator representing the observable at the larger $t$ stands on the left.

These arguments extend to classical correlations $A(t^\prime)\cdot B(t) $ for arbitrary $t^\prime,t$. At $t$ they are represented by
\begin{equation}\label{eqn:PP12}
	A(t^\prime) \cdot B(t)
	\to \tord\bigl\{A^\prime(t) \, B^\prime(t)\bigr\}.
\end{equation}
Here $\tord$ denotes the $t$-ordering, in the sense that the operator for the observable with the larger $t$-argument on the l.h.s. stands to the left. (The $t$ argument on the r.h.s. denotes the hypersurface on which the observables are expressed by operators. More precisely, we could use a notation $A^\prime(t_1,t_2)$ where the first argument labels the $t$-argument of the classical local observable $A(t_1)$, and the second the reference point on which we choose to evaluate the expectation value by \cref{eqn:AF20}. The r.h.s. of \cref{eqn:PP12} becomes then $A^\prime(t^\prime,t) \, B^\prime(t,t)$, and the $t$-ordering refers to the first argument.) We recall that the $t$-ordered operator product is commutative, reflecting the commutative classical product of observables. For a proof of \cref{eqn:PP12} the expression \labelcref{eqn:PP10} is generalized to an integration over all spin configurations between $t$ and $t^\prime$, with an appropriate chain of $\mathcal{K} $ factors between $A(t^\prime)$ and $B(t)$. The integration of the chain of $\mathcal{K} $-factors over intermediate spins produces the ordered product of step evolution operators, with larger $t$ to the left \cite{CWIT}.

By a similar line of arguments we can represent arbitrary classical $n$-point functions of classical local observables by suitable operator products at $t$, and evaluate expectation values according to \cref{eqn:AF20}. For example, for $t_1 < t_2 < t_3$ the classical three point function is represented by the product of three matrices
\begin{equation}\label{eqn:PP13}
	A(t_3) \cdot B(t_2) \cdot C(t_1)
	\to A^\prime(t) \, B^\prime(t) \, C^\prime(t).
\end{equation}
For a given $t$-independent step evolution operator the local information contained in the wave functions $\tilde{q}(t) ,\bar{q}(t)$ is sufficient for the computation of expectation values and classical $n$-point functions of classical local observables at arbitrary $t^\prime$. At $t$, the operator for an observable at $t^\prime = t + G\epsilon$ simply reads
\begin{equation}\label{eqn:PP14}
	A(t^\prime)
	\to A^\prime(t)
	= S^{-G} A^\prime(t^\prime) \, S^G.
\end{equation}
Products of observables are mapped to $t$-ordered products of operators, with ordering given by the $t$-labels of the observables.

This extends to linear combinations of classical observables that are represented by the corresponding linear combinations of operators. In this way we can compute the expectation values of all classical observables by using \cref{eqn:AF20} for the appropriate operators. This high predictive power of the local information is familiar from quantum mechanics. It requires the wave functions and is no longer given if one restricts the local information to local probabilities.

\subsection{Incomplete statistics}

The occurrence on non-commuting operator products is tightly connected to incomplete statistics. Incomplete statistics does not permit the computation of simultaneous probabilities to find a measurement value $\lambda_A$ for an observable $A$ and a measurement value $\lambda_B$ for a different observable $B$. This situation is characteristic for subsystems if the information in the subsystem is insufficient for the computation of simultaneous probabilities. A simple example is given by the quantum subsystem discussed in \cref{sec:Three-spin chains}. A local description of classical statistics at a given $t$ constitutes a subsystem of the overall statistical ensemble. One should therefore not be surprised that non-commuting operator products appear quite genuinely. 

We first discuss the non-commutativity of operators associated to products $A(t_2)B(t_1)$. As we have seen, one can associate to the classical product observable $A(t_2) \, B(t_1)$ an operator $C^\prime(t)$ that allows the computation of its expectation value -- the classical correlation -- from the probabilistic information contained in the wave functions,
\begin{equation}\label{eqn:101A}
	\langle A(t_2) \, B(t_1)\rangle
	= \langle C^\prime(t)\rangle,
	\quad
	C^\prime
	= \tord\bigl\{A^\prime(t) \, B^\prime(t)\bigr\}.
\end{equation}
If $A(t_2)B(t_1)$ would be a local observable we could compute simultaneous probabilities from the information at $t$ and statistics would be complete in this respect. The classical product is, in general, not a local observable, however. Typically the spectrum of $C^\prime(t)$ does not coincide with the possible measurement values of the classical product observable $A(t_2) \, B(t_1)$. As one of the indications, the \cref{cnd:corresp} in \cref{eqn:76A} is not obeyed. The expectation value of the product observable $\bigl(A(t_2) \, B(t_1)\bigr)^2$ can be computed from an associated operator $D^\prime(t)$, which is the time-ordered operator product of the operators for $A^{\prime 2}(t)$ and $B^{\prime 2}(t)$. Assuming $t_2 > t_1$ one has
\begin{alignedeqn}\label{eqn:101B}
	&\bigl\langle\bigl(A(t_2) \, B(t_1)\bigr)^2\bigr\rangle
	= \langle D^\prime(t)\rangle,\\
	&D^\prime(t)
	= \bigl(A^\prime(t)\bigr)^2 \, \bigl(B^\prime(t)\bigr)^2.
\end{alignedeqn}
If $A^\prime$ and $B^\prime$ do not commute, the operator $D^\prime$ differs from $\bigl(C^\prime(t)\bigr)^2$,
\begin{equation}\label{eqn:101C}
	D^\prime - C^{\prime 2}
	= A^{\prime 2} \, B^{\prime 2} - A^\prime \, B^\prime \, A^\prime \, B^\prime
	= A^\prime \, [A^\prime,B^\prime] \, B^\prime.
\end{equation}
If the expectation value of this expression does not vanish for all states the \cref{cnd:corresp} in \cref{eqn:76A} is indeed violated. According to our discussion in \cref{sec:local operators} we conclude that $A(t_2) \, B(t_1)$ cannot be a local observable. Either the spectrum of $C^\prime(t)$ does not coincide with the possible measurement values of the classical correlation, or the weights $w_i^{(A)}$ are not all positive, or both.

The observation that the classical product $A(t_2) \, B(t_1)$ is not a local observable is directly rooted in incomplete statistics. This can be seen by showing that the assumption of complete statistics contradicts the difference between $D^\prime$ and $C^{\prime 2}$ according to \cref{eqn:101C}. If the probabilistic information contained in the wave functions would be sufficient for the computation of the joint probabilities $p_{ij}^{(AB)}$ to find for $A$ the value $\lambda_i^{(A)}$, and for $B$ the value $\lambda_j^{(B)}$, we could compute
\begin{equation}\label{eqn:101D}
	\langle A(t_2) \, B(t_1)\rangle
	= \sum_{i,j} \lambda_i^{(A)} \, \lambda_j^{(B)} \, p_{ij}^{(AB)}.
\end{equation}
We could then interpret $A(t_2) \, B(t_1)$ as a new observable $C$ for which the possible measurement values are products of possible measurement values for $A$ and $B$,
\begin{equation}\label{eqn:101E}
	\lambda_k^{(C)}
	= \lambda_i^{(A)} \, \lambda_j^{(B)}.
\end{equation}
The probabilities to find $\lambda_k^{(C)}$ are given by
\begin{equation}\label{eqn:101F}
	p_k^{(C)}
	= p_{ij}^{(AB)},
\end{equation}
assuming for simplicity the non-degenerate case. Thus $C$ would be a local observable, represented by the operator $C^\prime(t)$ in \cref{eqn:101A}. In turn, this implies $D^\prime = C^{\prime 2}$, establishing the contradiction. We conclude that the joint probabilities $p_{ij}^{(AB)}$ are not computable from the local probabilistic information contained in the wave functions. They may be computable from the overall probability distribution $p[s]$, but the necessary information is lost by the process of partial integration leading to the wave functions.

The non-commutativity of operators and incomplete statistics are intrinsically related. Consider two arbitrary classical observables $A$, $B$ that take fixed values for any overall spin configuration $\{s\}$. Let $A^\prime$ and $B^\prime$ be the non-commuting operators corresponding to these classical observables $A$ and $B$, with possible measurement values $\lambda_i^{(A)}$ and $\lambda_j^{(B)}$. On the level of the classical observables we can implement the notions of linear combinations and products of observables in a standard way. These structures cannot be transferred, however, to corresponding structures among local observables. The possible measurement values of $\alpha A + \beta B$ are given by $\alpha\lambda_i^{(A)} + \beta\lambda_j^{(B)}$. They do not correspond to the spectrum of the operator $\alpha A^\prime + \beta B^\prime$. Also the possible measurement values of $AB$ are $\lambda_i^{(A)} \lambda_j^{(B)}$. They differ, in general, from the spectrum of the operator product $A^\prime B'$.

We conclude that for a wide class of classical observables $C$, as $C = \alpha A(t^\prime) + \beta B(t)$ or $C = A(t_1)B(t_2)E(t_3)F(t_4)$, one can find an operator $C^\prime(t)$ such that $\langle C\rangle = \bar{q}^\tran C^\prime \tilde{q}$. The operator $C^\prime(t)$ does not correspond necessarily to a local observable, however. The spectrum of $C^\prime$ may not coincide with the possible measurement values of $C$. The association $C^n \to (C^\prime)^n$ typically does not hold because of the $t$-ordering. (For example, one has for $t_2 > t_1$ the association $A(t_2) \to A^\prime$, $B(t_1) \to B^\prime$, $\bigl(A(t_2) + B(t_1)\bigr)^2 \to A^{\prime 2} + B^{\prime 2} + 2 A^\prime B^\prime \neq (A^\prime + B^\prime)^2$.) Furthermore, it is not guaranteed that $w_\tau^{(C)}(t)$, as defined by \cref{eqn:71C} for the matrix $D$ diagonalizing $C^\prime(t)$, is positive. In summary, we can compute $\langle C\rangle$ from the local probabilistic information contained in the wave functions, but $C$ is typically not a local observable.

The map between observables and operators defines equivalence classes of observables. All observables that are mapped to the same operator $A^\prime(t)$ belong to the same equivalence class. They all have the same expectation value. The classical correlation function is not a structure that respects the equivalence classes. For $A_1$ and $A_2$ belonging to the same equivalence class their classical correlations with a third observable $B$ may differ,
\begin{equation}
	\langle A_1 \cdot B\rangle
	\neq \langle A_2 \cdot B\rangle.
\end{equation}
The computation of classical correlations requires therefore information beyond the local system, e.g. beyond the definition of the associated operators $A^\prime$ and $B^\prime$ and the local probabilistic information in the wave functions. Even for the simple case where $A$ and $B$ are classical local observables at different times $t_1$ and $t_2$ one needs the information about these times in order to fix the $t$-ordering. This is information beyond the local system. In \cref{sec:equivalent local probabilistic systems} we define different ``quantum correlations'' that are computable from the information in the local system and respect the equivalence classes between observables.

\subsection{General local observables}

Within the local equivalence class of $C$ one may define a new local observable $C_l(t)$ which is represented by the same operator $C^\prime(t)$ as $C$. Its expectation value equals the one for $C$, i.e. $\langle C_l(t)\rangle = \langle C\rangle$. For $C_l(t)$ being a local observable we require that $C_l^n(t)$ is represented by $[C^\prime(t)]^n$. Thus products and correlations of $C_l(t)$ will be different from the ones of $C$. If $w^{(C)}(t)$ is positive, $C_l(t)$ indeed obeys all characteristics for a local observable. Its possible measurement values are determined by the spectrum of the operator $C^\prime(t)$.

We may therefore consider the association of local operators for observables at $t^\prime \neq t$ or of classical correlations as a mere technical tool to obtain genuine new local observables $C_l(t)$. These local observables often are no longer classical observables that take definite values for each spin configuration $\{s\}$. Among the extended class of local observables we may find in a complex formulation the Hamiltonian $\hat{H}$ or the momentum operator $\hat{P} = -i \partial_x$ for the Ising models describing fermions. In more complex systems one can realize in a similar way angular momentum as the generator of rotations.

The information contained in the wave functions is sufficient for the computation of expectation values $\langle C_l^n(t)\rangle$ for arbitrary $n$. Still, for $[A^\prime,C^\prime] \neq 0$, it does not contain sufficient information to find simultaneous probabilities for $C_l(t)$ taking a value $\smash{\lambda_i^{(C)}}$ and some other observable $A(t)$ taking a value $\smash{\lambda_j^{(A)}}$. The classical correlation function between $A(t)$ and $C_l(t)$, if it exists at all, cannot be computed from the local information in the wave functions. One may define more suitable measurement correlations, cf. \cref{sec:equivalent local probabilistic systems}. They can violate Bell's inequalities.

\section{Classical density matrix}
\label{sec:density matrix}

Instead of computing the time evolution of the wave function $\tilde{q}_\tau(t)$ and the conjugate wave function $\bar{q}_\tau(t)$ separately, it is often more convenient to follow directly the time evolution of an associated density matrix $\rho_{\tau\rho}^\prime(t)$. As is well known from quantum mechanics the density matrix is particularly useful for the discussion of subsystems. In the occupation number basis the local probabilities can be extracted as the diagonal elements of $\rho^\prime$. The density matrix allows the description of mixed states, as characteristic for subsystems. It is convenient to switch from now on the notation from Ising spins to occupation numbers, $n_\gamma = (s_\gamma + 1)/2$. They take values one and zero. In the occupation number basis the basis functions $h_\tau$ in \cref{eqn:AF21} involve factors $n_\gamma$ or $1-n_\gamma$.

\subsection{Properties of the density matrix}

For ``pure states'' the real classical density matrix $\rho_{\tau\rho}^\prime(t)$ obtains by multiplication of the wave function with its conjugate
\begin{equation}\label{eqn:D1a}
	\rho_{\tau\rho}^\prime(t)
	= \tilde{q}_\tau(t) \, \bar{q}_\rho(t).
\end{equation}
(Primes are used here in order to make the distinction to an equivalent complex formulation or to density matrices for quantum subsystems more easy to follow.) For pure states the quantum expression \labelcref{eqn:quantum rule} for the computation of expectation values of local observables from the density matrix follows directly from \cref{eqn:AF20}. From \cref{eqn:AF24} we infer
\begin{equation}\label{eqn:102A}
	\tr \rho^\prime
	= 1.
\end{equation}
In general, the density matrix needs not be symmetric since $\tilde{q}$ and $\bar{q}$ are different. This constitutes an important difference of the general classical density matrix to the special case of quantum mechanics for which $\rho^\prime$ is symmetric. In the occupation number basis the diagonal elements of the classical density matrix are the local probabilities (no sum here)
\begin{equation}\label{eqn:D2a}
	p_\tau(t)
	= \rho_{\tau\tau}^\prime(t).
\end{equation}
This holds by virtue of \cref{eqn:local probabilities}, and no particular conditions on $\tilde{q}$ or $\bar{q}$ need to be imposed except for the normalization $Z = 1$.

The pure state density matrix has $N - 1$ eigenvalues equal to zero, while one eigenvalue equals one. Indeed, the construction of $\rho^\prime$ as the product of two vectors implies that at most one eigenvalue differs from zero. The relation
\begin{equation}\label{eqn:OP12}
	\rho_{\tau\rho}^\prime\tilde{q}_\rho
	= \tilde{q}_\tau\bar{q}_\rho\tilde{q}_\rho
	= \tilde{q}_\tau
\end{equation}
directly establishes the eigenvalue one (with eigenvector $\tilde{q}$). We conclude that all eigenvalues of $\rho^\prime$ are positive semidefinite. A pure state classical density matrix obeys
\begin{equation}\label{eqn:104A}
	\rho^{\prime 2}
	= \rho^\prime,
\end{equation}
similar to quantum mechanics. This follows from \cref{eqn:AF24} by
\begin{equation}\label{eqn:104B}
	(\rho^{\prime 2})_{\tau\rho}
	= \tilde{q}_\tau \, \bar{q}_\sigma \, \tilde{q}_\sigma \, \bar{q}_\rho
	= \tilde{q}_\tau \, \bar{q}_\rho
	= \rho_{\tau\rho}^\prime.
\end{equation}

Consider next a change of basis by an orthogonal matrix $D$,
\begin{equation}\label{eqn:OP13}
	\rho_D^\prime
	= D \rho^\prime D^{-1},
	\qquad
	D^\tran D
	= 1.
\end{equation}
Orthogonal transformations do not change the eigenvalues, such that the eigenvalues of $\rho_D^\prime$ are again one and zero and therefore positive semidefinite. We next discuss the sign of the diagonal elements in an arbitrary basis (no sum over $\tau$)
\begin{equation}\label{eqn:OP15}
	(\rho_D^\prime)_{\tau\tau}
	= (D\bar{q})_\tau(D\tilde{q})_\tau
	= \sum_{\rho\sigma}D_{\tau\rho}\tilde{q}_\rho\bar{q}_\sigma D_{\sigma\tau}^\tran,
\end{equation}
where $\tilde q$, $\bar q$ refer to the occupation number basis. For a general basis the diagonal elements need not be positive. (For a general matrix positive eigenvalues do not imply positive diagonal elements.) Only if all diagonal elements of $\rho_D^\prime$ are positive semidefinite, they can be interpreted as probabilities, $p_\tau^{(D)} \geq 0$, with
\begin{equation}\label{eqn:OP16}
	p_\tau^{(D)}
	= (\rho_D^\prime)_{\tau\tau},
	\qquad
	\sum_\tau p_\tau^{(D)}
	= \tr\rho_D^\prime
	= \tr\rho^\prime
	= 1.
\end{equation}

There are interesting cases for which all diagonal elements of $\rho_D^\prime$ are positive. The first is quantum mechanics where $\bar{q}$ and $\tilde{q}$ can be identified, such that the positivity of $(\rho_D^\prime)_{\tau\tau}$ follows automatically from \cref{eqn:OP15}. A more general case occurs for a symmetric density matrix,
\begin{equation}\label{eqn:symmetric density}
	\rho_{\tau\rho}^\prime
	= \rho_{\tau\rho}^\prime
	= \tilde{q}_\tau \bar{q}_\rho
	= \tilde{q}_\rho \bar{q}_\tau.
\end{equation}
Multiplication with $\tilde{q}_\rho$ and summation over $\rho$ yields
\begin{equation}
	\tilde{q}_\tau
	= a \bar{q}_\tau,
	\qquad
	a
	= \sum_\rho (\tilde{q}_\rho)^2
	> 0,
\end{equation}
or
\begin{equation}
	\rho_{\tau\rho}^\prime
	= \frac{1}{a} \tilde{q}_\tau \tilde{q}_\rho
	= a \bar{q}_\tau \bar{q}_\rho.
\end{equation}
One concludes $\rho_{\tau\tau}^\prime > 0$ for an arbitrary basis in which \cref{eqn:symmetric density} is realized. In turn, $\rho_D^\prime$ is also symmetric and
\begin{equation}\label{eqn:trace greater zero}
	(\rho_D^\prime)_{\tau\tau}
	= a (D \bar{q}_\tau)^2
	\geq 0.
\end{equation}
For symmetric density matrices arbitrary symmetric operators $A^\prime(t)$ correspond to local observables. They can be diagonalized by orthogonal matrices $D$, such that the condition \labelcref{eqn:71E} is realized by virtue of \cref{eqn:trace greater zero}.

\subsection{Generalized von Neumann equation}

The time evolution of the density matrix can be inferred from \cref{eqn:AF13},
\begin{equation}\label{eqn:D9a}
	\rho_{\tau\rho}^\prime(t + \epsilon)
	= S_{\tau\alpha}(t) \, \rho_{\alpha\beta}^\prime(t) \, S_{\beta\rho}^{-1}(t).
\end{equation}
The eigenvalues of $\rho^\prime$ and $\det\rho^\prime$ remain invariant under the evolution. The relation \labelcref{eqn:104A} for pure state density matrices is preserved by the evolution as well,
\begin{alignedeqn}
	\rho^{\prime 2}(t + \epsilon)
	&= \bigl(S \, \rho^\prime \, S^{-1} \, S \, \rho^\prime \, S^{-1}\bigr)(t)\\
	&= \bigl(S \, \rho^{\prime 2} \, S^{-1}\bigr)(t)\\
	&= \bigl(S \, \rho^\prime \, S^{-1}\bigr)(t)
	= \rho^\prime(t + \epsilon).
\end{alignedeqn}

In the continuum limit, $\epsilon \to 0$, one finds with \crefrange{eqn:AF14}{eqn:AF17} for sufficiently smooth $\tilde{q}(t) \approx \bigl(\tilde{q}(t + \epsilon) + \tilde{q}(t - \epsilon)\bigr)/2$ and similar for $\bar{q}(t)$,
\begin{alignedeqn}\label{eqn:D11a}
	\partial_t\rho_{\tau\rho}^\prime
	= W_{\tau\alpha}(t) \rho_{\alpha\rho}^\prime(t) - \rho_{\tau\alpha}^\prime(t) \tilde{W}_{\alpha\rho}(t).
\end{alignedeqn}
For $S$ independent of $t$ one has $W = \tilde{W}$. More generally, we concentrate in the following on the case $\tilde{W}(t) = W(t)$, for which
\begin{alignedeqn}\label{eqn:231}
	\partial_t \rho^\prime
	= [W,\rho^\prime].
\end{alignedeqn}
Density matrices that commute with $W$ are $t$-independent. As it should be, the evolution \labelcref{eqn:231} leaves a pure state pure, e.g. $\rho^{\prime 2} = \rho^\prime$ implies $\partial_t \rho^{\prime 2} = \partial_t \rho^\prime$.

\Cref{eqn:231} constitutes the central evolution equation for the density matrix of classical statistical systems. It generalizes the von Neumann equation since $W$ is not necessarily antisymmetric. We observe the conservation of the trace
\begin{equation}\label{eqn:trace conservation}
	\partial_t \tr \rho^\prime
	= 0.
\end{equation}
For classical statistical systems in the occupation number basis both $\tilde{q}_\tau$ and $\bar{q}_\tau$ are positive, such that $p_\tau(t) = \rho_{\tau\tau}^\prime(t) \geq 0$ for all $t$ and \cref{eqn:trace conservation} amounts to the conservation of the normalization of the local probabilities.

We may split $W$ and $\rho^\prime$ into symmetric and antisymmetric parts. The evolution of the antisymmetric part $\rho_A^\prime$ obeys
\begin{equation}
	\partial_t \rho_A^\prime
	= [W_S,\rho_S^\prime] + [W_A,\rho_A^\prime].
\end{equation}
A symmetric density matrix $\rho^\prime = \rho_S^\prime$ remains symmetric if it commutes with the symmetric part of $W$,
\begin{equation}\label{eqn:remain symmetric condition}
	[W_S,\rho^\prime]
	= 0.
\end{equation}
This condition is much weaker than the condition $W_S = 0$ for a pure quantum evolution. We will see in \cref{sec:evolution for subsystems} that symmetric density matrices obeying \cref{eqn:remain symmetric condition} describe subsystems that follow a quantum evolution.

If the model admits a complex structure (cf. \cref{sec:complex structure}), the complex pure state density matrix is defined as
\begin{equation}\label{eqn:239A}
	\rho_{\alpha\beta}(t)
	= \psi_\alpha(t) \, \bar{\psi}_\beta(t),
\end{equation}
with $\psi$ and $\bar\psi$ the complex wave function and conjugate wave function. (For classical statistics $\bar\psi$ differs from $\psi^\ast$, with $\bar\psi = \psi^\ast$ realized for the special case of quantum mechanics.) The evolution of the complex density matrix is again given by a generalized von Neumann equation
\begin{equation}\label{eqn:239B}
	i\partial_t\rho
	= [G,\rho]
	= [\hat{H},\rho] + i[\hat{J},\rho].
\end{equation}
Local observables that are compatible with the complex structure are represented by complex operators $\hat{A}$ such that
\begin{equation}\label{eqn:239C}
	\langle A(t)\rangle
	= \tr\bigl\{\hat{A}(t) \, \rho(t)\bigr\}.
\end{equation}

\subsection{Pure state density matrix from functional integral}

One may formally obtain the density matrix by partial functional integration or performing the configuration sum partially. Since the density matrix has two indices we will have to distinguish formally between two sets of occupation numbers at $t$, depending if they are connected to the range $t^\prime > t$ or $t^\prime < t$. For this purpose we formally distinguish for the wave functions the arguments $n_\gamma(t)$ of $f(t)$ and $\bar{n}_\gamma(t)$ for $\bar{f}(t)$, and correspondingly for $\bar{h}_\rho(t)$ and $h_\tau(t)$,
\begin{equation}\label{eqn:D3a}
	f(t)
	= \tilde{q}_\tau(t) \, h_\tau(t),
	\qquad
	\bar{f}(t)
	= \bar{q}_\rho(t) \, \bar{h}_\rho(t).
\end{equation}
The density matrix can then equivalently be expressed as a function of occupation numbers
\begin{alignedeqn}\label{eqn:D4a}
	\hat{\rho}^\prime(t)
	&= f(t) \, \bar{f}(t)
	= \tilde{q}_\tau(t) \, \bar{q}_\rho(t) \, h_\tau(t) \, \bar{h}_\rho(t)\\
	&= \rho_{\tau\rho}^\prime(t) \, h_\tau(t) \, \bar{h}_\rho(t).
\end{alignedeqn}
It depends on $n(t)$ and $\bar{n}(t)$. Local operators take the form
\begin{equation}\label{eqn:D5a}
	\hat{A}^\prime(t)
	= A_{\tau\rho}^\prime(t) \, \bar{h}_\tau(t) \, h_\rho(t).
\end{equation}
They are again functions of $n(t)$ and $\bar{n}(t)$. (We employ various expressions for local observables that should not be confounded. While $A(t)$ in \cref{eqn:AF10B} depends on $\{n(t)\}$, $\hat{A}^\prime(t)$ depends on $\{n(t)\}$ and $\{\bar{n}(t)\}$. For classical local observables $A(t)$ the values they take for the local spin configuration $\{n(t)\} = \tau$ are given by $A_\tau$. The expansion coefficients of $A(t)$ in basis functions $h_\tau$ are denoted by $\alpha_\tau(t)$. The matrix $A^\prime(t)$ in \cref{eqn:AF20} has elements $A_{\tau\rho}^\prime(t)$. For diagonal $A^\prime(t)$ in the occupation number basis the diagonal elements equal $\alpha_\tau = A_\tau$.)

From the quantum relation \labelcref{eqn:quantum rule} for the expectation value we infer
\begin{alignedeqn}\label{eqn:D6a}
	\langle A(t)\rangle
	&= \int \dif \bar{n}(t) \, \dif n(t) \, \hat{A}^\prime(t) \, \hat{\rho}^\prime(t)\\
	&= A_{\tau\rho}^\prime \, \rho_{\rho\tau}^\prime
	= \tr(A^\prime \, \rho^\prime).
\end{alignedeqn}
The classical local observables at $t$ discussed previously are expressed in the occupation number basis by diagonal operators
\begin{alignedeqn}\label{eqn:D7a}
	\hat{A}^\prime(t)
	&= A(t)\delta \bigl(\bar{n}(t),n(t)\bigr)
	= A_\tau(t) \, f_\tau(t) \, \delta \bigl(\bar{n}(t),n(t)\bigr),\\
	A_{\tau\rho}^\prime(t)
	&= A_\rho(t) \, \delta_{\tau\rho}.
\end{alignedeqn}
Their expectation values \labelcref{eqn:D6a} realize the standard definition
\begin{equation}\label{eqn:D8a}
	\langle A(t)\rangle
	= \sum_\tau A_\tau(t) \, p_\tau(t),
\end{equation}
with $A_\tau$ the value of the classical local observable $A(t)$ for the spin configuration $\tau$. Classical local observables at $t^\prime \neq t$ and more general local observables typically correspond to non-diagonal $A^\prime$ and therefore to a non-trivial dependence of $\hat{A}^\prime(t)$ on $n(t)$ and $\bar{n}(t)$.

We can formally express $\hat\rho^\prime(t)$ by the functional integral
\begin{alignedeqn}\label{eqn:D12a}
	\hat{\rho}^\prime(t)
	&= \int \mathcal{D} \bar{n}(t^\prime > t) \, \mathcal{D} n(t^\prime < t) \, \bar{f}_\text{f}\bigl(\bar{n}(t_\text{f})\bigr)\\
	&\times \bar{K}_>(t) \, K_<(t) \, f_\text{in} \bigl(n(t_\text{in})\bigr),
\end{alignedeqn}
where the bar on $\bar{K}_>$ indicates that the arguments are taken as $\bigl\{\bar{n}(t^\prime)\bigr\}$. For $t^\prime > t$ the distinction between $\bar{n}(t^\prime)$ and $n(t^\prime)$ does not matter since these are integration variables. As compared to the definition of $Z$ in \cref{eqn:partition function def} the integration over $n(t)$ is missing in \cref{eqn:D12a}. We may view this as a ``cut'' at $t$. Furthermore, at the cut we distinguish between $\bar{n}(t)$ and $n(t)$, such that $\hat{\rho}^\prime(t)$ depends on these two types of arguments, $\hat{\rho}^\prime(t) = \hat{\rho}^\prime \bigl(\bar{n}(t),n(t)\bigr)$.

The expression \labelcref{eqn:D12a} is independent of the choice of basis and may be considered as a ``functional integral definition'' of the density matrix. For a pure state it is composed of two factors, one depending only on $n(t)$ the other on $\bar{n}(t)$. The expansion \labelcref{eqn:D4a} and the factorized form \labelcref{eqn:D1a} hold in an arbitrary basis. This extends to the quantum rule \labelcref{eqn:D6a} for the expectation values of local observables. Symmetric density matrices correspond to $\hat{\rho}[n,\bar{n}] = \hat{\rho}[\bar{n},n]$.

\subsection{Boundary conditions and mixed states}

We can generalize the boundary conditions by replacing $f_\text{in}(n_\text{in}) \, \bar{f}_\text{f}(n_\text{f})$ by $b(n_\text{in},n_\text{f})$,
\begin{alignedeqn}\label{eqn:B5-D6}
	b(n_\text{in},n_\text{f})
	= b\bigl(n(t_\text{in}), n(t_\text{f})\bigr)
	= b_{\tau\rho}^\prime \, h_\tau(t_\text{in}) \, h_\rho(t_\text{f}).
\end{alignedeqn}
The functional integral for $Z$ becomes
\begin{equation}\label{eqn:B5-D7}
	Z
	= \int \mathcal{D} n \, b(n_\text{in},n_\text{f}) \, K[n],
\end{equation}
This construction corresponds to the notion of mixed states in quantum mechanics. In particular, we can implement periodic boundary conditions for $t$-points on a circle by the choice
\begin{alignedeqn}\label{eqn:B5-D9}
	b(n_\text{in},n_\text{f})
	&= \mathcal{K} (n_\text{in},n_\text{f}),\\
	b_{\tau\rho}^\prime
	&= S_{\tau\rho}.
\end{alignedeqn}
In case of translation symmetry \cref{eqn:partition function def} becomes
\begin{equation}\label{eqn:77Aa}
	Z
	= b_{\rho\tau}^\prime (S^G)_{\tau\rho}
	= \tr\{b^\prime S^G\},
\end{equation}
such that the periodic boundary condition \labelcref{eqn:B5-D9} yields the well known formula for the transfer matrix
\begin{equation}\label{eqn:77B}
	Z
	= \tr\{S^{G+1}\}.
\end{equation}

An arbitrary boundary condition $b$ can be written as a linear superposition of ``pure state boundary conditions'',
\begin{equation}\label{eqn:B5-D10}
	b
	= \sum_\alpha w_\alpha^{(b)} \,
	f^{(\alpha)}(n_\text{in}) \,
	\bar{f}^{(\alpha)}(n_\text{f})
	= \sum_\alpha w_\alpha^{(b)} \, b^{(\alpha)},
\end{equation}
with suitable coefficients $w_\alpha$. If the boundary terms $f^{(\alpha)}$ and $\bar{f}^{(\alpha)}$ are such that they define positive probability distributions $w^{(\alpha)}[n]$, we can take $w_\alpha^{(b)}$ as probabilities for boundary conditions, $w_\alpha^{(b)} \geq 0$, $\sum_\alpha w_\alpha^{(b)} = 1$, with $w[n] = \sum_\alpha w_\alpha^{(b)} w^{(\alpha)}[n]$ again a probability distribution.

Since the definition of $\hat{\rho}^\prime$ in \cref{eqn:D12a} is linear in $b^{(\alpha)}$,
\begin{alignedeqn}\label{eqn:D13a}
	\hat{\rho}^{\prime(\alpha)}(t)
	= \int &\mathcal{D} \bar{n}(t^\prime > t) \, \mathcal{D} n(t^\prime < t)\\
	&\times b^{(\alpha)}(t_\text{f},t_\text{in}) \, \bar{K}_>(t) \, K_<(t),
\end{alignedeqn}
it is straightforward to extend the definition of $\hat{\rho}^\prime$ to ``mixed state boundary conditions'' by replacing in \cref{eqn:D13a} $b^{(\alpha)}(t_\text{f},t_\text{in})$ by $b(t_\text{f},t_\text{in})$. We can formally write \cref{eqn:B5-D7} as
\begin{alignedeqn}\label{eqn:D14a}
	Z
	&= \int \dif \bar{n}(t) \, \dif n(t) \, \delta \bigl(\bar{n}(t),n(t)\bigr)\, \hat{\rho}^\prime(t)\\
	&= \sum_\tau p_\tau(t)
	= 1.
\end{alignedeqn}
This amounts to the normalization condition
\begin{equation}\label{eqn:80A}
	\tr\rho^\prime
	= 1.
\end{equation}

For a mixed state boundary condition \cref{eqn:104A} no longer holds. In the following we assume
\begin{equation}\label{eqn:136A}
	\rho^\prime
	= \sum_\alpha w_\alpha^{(b)} \, \rho^{\prime(\alpha)},
	\quad
	w_\alpha^{(b)}
	\geq 0,
	\quad
	\sum_\alpha w_\alpha^{(b)}
	= 1,
\end{equation}
with $\rho^{\prime(\alpha)}$ pure state classical density matrices that can be realized by appropriate boundary conditions. The simple argument of quantum mechanics that all eigenvalues of the density matrix are positive cannot be carried over. For an arbitrary vector $v$ one has
\begin{equation}\label{eqn:136B}
	v_\tau \, \rho_{\tau\rho}^\prime \, v_\rho
	= \sum_\alpha w_\alpha^{(b)} \, v_\tau \, \tilde{q}_\tau^{(\alpha)} \, \bar{q}_\rho^{(\alpha)} \, v_\rho.
\end{equation}
While for $\tilde q = \bar q$ this expression is always positive, negative values seem possible for $\tilde q \neq \bar q$. In the occupation number basis all elements of a pure state density matrix are positive, $\rho_{\tau\rho}^{\prime(\alpha)} \geq 0$. This extends to the mixed states density matrix
\begin{equation}\label{eqn:136C}
	\rho_{\tau\rho}^\prime
	\geq 0.
\end{equation}
In a different bases, or for submatrices obtained by a partial trace in a given basis, the relation \labelcref{eqn:136C} does not need to hold. For symmetric $\rho^{\prime(\alpha)}$ also $\rho^\prime = \sum_\alpha w_\alpha^{(b)} \, \rho^{\prime(\alpha)}$ is symmetric. In this case all eigenvalues of $\rho^\prime$ are positive.

The linear evolution equations \labelcref{eqn:D9a,eqn:231}, as well as the quantum rule for expectation values \labelcref{eqn:D6a} carries over to mixed-state density matrices. The condition for static density matrices, $[W,\rho^\prime] = 0$, admits many solutions. (We use here a ``time-language'' where static stands for $t$-independent.) For $\rho^\prime = f(W)$, with $f$ an arbitrary function, the density matrix is static. If there are some operators $B_i$ commuting with $W$, the general static density matrix can be extended to
\begin{equation}\label{eqn:80B}
	\rho^\prime
	= f(W,B_i).
\end{equation}
We will discuss static solutions for a particular model in the next section.

\subsection{Allowed density matrices and local probabilities}

The evolution law \labelcref{eqn:D9a,eqn:231} for the density matrix is linear. Solutions to this evolution equation therefore obey the superposition principle. A linear combination of two solutions of this evolution equation is again a solution of this evolution equation. The simplicity of the linear evolution law constitutes a central advantage of the use of density matrices for the problem of information transport in classical statistical systems. No such simple law can be formulated for the evolution of the local probabilities. We will see below, however, that not all solutions of the evolution equation for the density matrix are compatible with the boundary conditions. In particular, arbitrary initial $\rho^\prime(t_\text{in})$ may not be permitted.

The general evolution of $\rho^\prime$ can be understood by diagonalizing $W$ with a regular matrix $D$
\begin{alignedeqn}\label{eqn:RS1}
	W_d
	&= D W D^{-1}
	= \diag(\tilde{\lambda}_\tau),\\
	\rho_d
	&= D\rho^\prime D^{-1}.
\end{alignedeqn}
We assume here that $W$ does not depend on $t$ such that the general solution for the matrix elements of $\rho_d$ reads
\begin{equation}\label{eqn:RS2}
	(\rho_d)_{\tau\rho}
	= (\rho_d^0)_{\tau\rho} e^{(\tilde{\lambda}_\tau - \tilde{\lambda}_\rho)t},
\end{equation}
implying
\begin{equation}\label{eqn:RS3}
	\rho_{\sigma\lambda}^\prime
	= \sum_{\tau,\rho}D_{\sigma\tau}^{-1}(\rho_d^0)_{\tau\rho}D_{\rho\lambda}e^{(\tilde{\lambda}_\tau-\tilde{\lambda}_\rho)t}.
\end{equation}
The differences $\tilde{\lambda}_\tau-\tilde{\lambda}_\rho$ can have positive or negative real parts, corresponding to increasing matrix elements in the positive or negative time direction. This restricts the allowed values of $(\rho_d^0)_{\tau\rho}$.

In the occupation number basis the allowed solutions for $\rho^\prime$ must obey the condition that the diagonal elements are local probabilities for all $t$ (no sum over $\tau$)
\begin{equation}\label{eqn:RS4}
	0
	\leq \rho_{\tau\tau}^\prime(t)
	\leq 1,
	\qquad
	\tr\rho^\prime
	= 1.
\end{equation}
At $t_\text{in}$ the diagonal elements are given by the initial probabilities $\rho_{\tau\tau}^\prime(t_\text{in}) = p_\tau(t_\text{in})$. The off diagonal elements $\rho_{\tau\rho}^\prime(t_\text{in}),\tau\neq \rho$, have to be chosen such that the condition \labelcref{eqn:RS4} holds for all $t$. This demonstrates in a simple way that arbitrary $\rho^\prime(t_\text{in})$ may be in conflict with the boundedness of $\rho^\prime(t_\text{f})$. Similar arguments extend to other choices of basis.

Further restrictions on the allowed solutions of the evolution equations arise from the condition \labelcref{eqn:136C} for classical density matrices in the occupation number basis. An example will be discussed in the next section.

\section{Solution of boundary value problem by use of generalized von Neumann equation}
\label{sec:solution of bvp}

The practical use of the classical density matrix will become more apparent by the investigation of a few simple examples. In particular, this will shed light on the restrictions for allowed solutions of the evolution equation from the requirement of positive local probabilities for all $t$.

\subsection{One-dimensional Ising model}

Characteristic features of the evolution of the density matrix can be illustrated by the one-dimensional Ising model where $\rho^\prime$ is a $2 \times 2$ matrix. In the absence of a magnetic field one has \cite{CWIT}
\begin{equation}
	\mathcal{K}(t)
	= \exp\bigl\{\beta s(t + \epsilon) s(t)\bigr\},
\end{equation}
corresponding in the occupation number basis to a step evolution operator
\begin{equation}
	S
	= e^{-\varphi} \begin{pmatrix}
		e^\beta & e^{-\beta}\\
		e^{-\beta} & e^\beta
	\end{pmatrix},
\end{equation}
with
\begin{equation}
	\varphi
	= \ln[2 \cosh \beta].
\end{equation}
Here $\beta$ defines the ratio of interaction strength $J$ and temperature $T$, $\beta = J/(k_\text{B} T)$. For large $\beta$ one infers
\begin{equation}
	S
	= \begin{pmatrix}
		1 - e^{-2 \beta} & e^{-2 \beta}\\
		e^{-2 \beta} & 1 - e^{-2 \beta}
	\end{pmatrix}.
\end{equation}
We can take the continuum limit, with
\begin{equation}\label{eqn:80C}
	W
	= \omega (\tau_1 - \mathds{1}_2),
	\qquad
	\omega
	= \frac{e^{-2 \beta}}{\epsilon}.
\end{equation}

The general static density matrix reads
\begin{equation}\label{eqn:80D}
	\bar{\rho}^\prime
	= \frac{1}{2} \begin{pmatrix}
		1 & a\\
		a & 1
	\end{pmatrix},
\end{equation}
with free parameter $a$. This can be understood in terms of the wave functions. With eigenvalues of $W$ in \cref{eqn:80C} given by $\tilde{\lambda}_1 = 0$, $\tilde{\lambda}_2 = -2 \omega$ we associate equilibrium solution for $\tilde{q}$ with the eigenvector to $\tilde{\lambda}_1$, and similarly for $\bar{q}$,
\begin{equation}\label{eqn:80E}
	\tilde{q}_\ast
	= \frac{c_1}{\sqrt{2}} \begin{pmatrix}
		1\\ 1
	\end{pmatrix},
	\qquad
	\bar{q}_\ast
	= \frac{\bar{c}_1}{\sqrt{2}} \begin{pmatrix}
		1\\ 1
	\end{pmatrix}.
\end{equation}
The pure state equilibrium density matrix $\rho_\ast^\prime$ is realized for $\bar{c}_1 c_1 = 1$ and therefore corresponds to $a = 1$, with $\rho_\ast^{\prime 2} = \rho_\ast^\prime$. The density matrix formed with the eigenvectors to $\lambda_2$
\begin{equation}\label{eqn:A20}
	\delta q(t)
	= \frac{c_2(t)}{\sqrt{2}} \begin{pmatrix}
		1\\
		-1
	\end{pmatrix},
	\qquad
	\delta\bar{q}(t)
	= \frac{\bar{c}_2(t)}{\sqrt{2}} \begin{pmatrix}
		1\\
		-1
	\end{pmatrix},
\end{equation}
namely
\begin{equation}\label{eqn:80F}
	\delta \rho^\prime
	= \frac{\bar{c}_2 c_2}{2}
	\begin{pmatrix}
		1 & -1\\
		-1 & 1
	\end{pmatrix},
\end{equation}
is also static. Indeed, with $c_2 \sim \exp(-2 \omega t)$, $\bar{c}_2 \sim \exp(2 \omega t)$ the product $c_2 \bar{c}_2$ is independent of $t$. The general static density matrix is a linear combination
\begin{equation}\label{eqn:80H}
	\bar{\rho}^\prime
	= w_1\rho_\ast^\prime + \delta \rho^\prime,
\end{equation}
with probabilities
\begin{alignedeqn}\label{eqn:80I}
	w_1
	&= \bar{c}_1 c_1,
	\qquad
	w_2
	= \bar{c}_2 c_2,\\
	w_1 + w_2
	&= 1,
	\qquad
	w_1 - w_2
	= a.
\end{alignedeqn}
This limits $-1 \leq a \leq 1$.

The reason for the static density matrix $\delta\rho^\prime$ is easy to understand. While $\delta q(t)$ decreases exponentially, the conjugate wave function $\delta\bar{q}(t)$ increases exponentially with the same rate. The exponential loss of boundary information discussed in \cref{sec:complex structure} corresponds to $\bar{c}_2 = 0$. Only if one can realize non-zero $\bar{c}_2 c_2$ a static density matrix with $a\neq 1$ can be realized. We observe that the diagonal part of $\rho^\prime$, e.g. the local probabilities $p_\tau$, are independent of $a$. This extends to expectation values of classical local observables at $t$.

The $W$ matrix \labelcref{eqn:80C} has eigenvalues $\tilde{\lambda}_\tau = 0$, $-2 \omega$. Correspondingly, the four possible combinations $\tilde{\lambda}_\tau - \tilde{\lambda}_\rho$ are $\{-2 \omega,0,0,2 \omega\}$. The general static density matrix has contributions from $\tilde{\lambda}_\tau = \tilde{\lambda}_\rho$.

The general solution of the evolution equation reads
\begin{alignedeqn}\label{eqn:RU1}
	\rho^\prime(t)
	= \frac{1}{2} \Bigl[&1 + a\tau_1 + b_\text{in} \exp\bigl\{-2 \omega(t - t_\text{in})\bigr\}(i\tau_2 + \tau_3)\\
	&+c_\text{f} \exp\bigl\{-2 \omega(t_\text{f} - t)\bigr\}(-i\tau_2 + \tau_3)\Bigr].
\end{alignedeqn}
The local probability $\rho_{11}^\prime$ is given by
\begin{alignedeqn}\label{eqn:RU2}
	\rho_{11}^\prime(t)
	= \frac{1}{2}\Bigl[1 + b_\text{in} \, \exp\bigl\{-2 \omega(t - t_\text{in})\bigr\}\\
	+ c_\text{f} \, \exp\bigl\{-2 \omega(t_\text{f} - t)\bigr\}\mathrlap{\Bigr].}
\end{alignedeqn}
This is a local probability provided
\begin{equation}\label{eqn:RU3}
	\bigl|b_\text{in} \exp\bigl\{-2 \omega(t - t_\text{in})\bigr\} + c_\text{f} \exp\bigl\{-2 \omega(t_\text{f} - t)\bigr\}\bigr|
	\leq 1.
\end{equation}
In particular, the condition \labelcref{eqn:RU3} has to hold for $t = t_\text{in}$ and $t = t_\text{f}$,
\begin{alignedeqn}\label{eqn:RU4}
	&\bigl| b_\text{in} + c_\text{f} \, \exp\bigl\{-2 \omega(t_\text{f} - t_\text{in})\bigr\}\bigr|
	\leq 1,\\
	&\bigl|c_\text{f} + b_\text{in} \, \exp\bigl\{-2 \omega(t_\text{f} - t_\text{in})\bigr\}\bigr|
	\leq 1.
\end{alignedeqn}
For $t_\text{f} - t_\text{in} \gg 1/2 \omega$ this restricts
\begin{equation}\label{eqn:156A}
	|b_\text{in}| \leq 1,
	\qquad
	|c_\text{f}| \leq 1.
\end{equation}

This is an example of how the requirement of local probabilities restricts the possible solutions of the evolution equation of the density matrix. These restrictions occur as restrictions for the allowed boundary conditions. With the restriction \labelcref{eqn:156A} \cref{eqn:RU1} provides upper bounds for the deviation from the equilibrium occupation number $\bar n = 1/2$,
\begin{alignedeqn}\label{eqn:156B}
	\Delta n(t)
	= \langle n(t)\rangle - \frac{1}{2}
	&= \frac{b_\text{in}}{2} \exp\bigl\{-2 \omega (t - t_\text{in})\bigr\}\\
	&\hphantom{{}=}+ \frac{c_\text{f}}{2} \exp\bigl\{-2 \omega (t_\text{f} - t)\bigr\}.
\end{alignedeqn}
Far inside the bulk, e.g. for $\omega (t - t_\text{in}) \ll 1$, $\omega (t_\text{f} - t) \gg 1$ one finds an exponential suppression of $\Delta n(t)$. This demonstrates the loss of memory of boundary conditions for the Ising model.

From \cref{eqn:RU2} we can compute the expectation values for the occupation number at $t_\text{f}$ and $t_\text{in}$,
\begin{alignedeqn}\label{eqn:RU5}
	\langle n(t_\text{in})\rangle
	&= \rho_{11}^\prime(t_\text{in})\\
	&= \frac{1}{2} \Bigl[1 + b_\text{in} + c_\text{f} \exp\bigl\{-2 \omega(t_\text{f} - t_\text{in})\bigr\}\Bigr],\\[1ex]
	\langle n(t_\text{f})\rangle
	&= \rho_{11}^\prime(t_\text{f})\\
	&= \frac{1}{2}\Bigl[1 + c_\text{f} + b_\text{in} \exp\bigl\{-2 \omega(t_\text{f} - t_\text{in})\bigr\}.
\end{alignedeqn}
For large $\omega (t_\text{f} - t_\text{in})$ one has
\begin{equation}\label{eqn:RU6}
	\bigl\langle n(t_\text{in})\bigr\rangle
	\approx \frac{1}{2} \bigl(1 + b_\text{in}\bigr),
\end{equation}
while $\bigl\langle n(t_\text{f})\bigr\rangle$ becomes almost independent of $b_\text{in}$. This again reflects the exponential loss of memory of the boundary information. It also tells us that the occupation number at one boundary becomes independent of the boundary conditions at the other boundary. This changes if for finite $t_\text{f} - t_\text{in}$ we take the limit $\omega \to 0$. In this limit effects from both boundaries are mixed.

From \cref{eqn:RU1} we can also infer the evolution of the off-diagonal elements
\begin{alignedeqn}\label{eqn:158A}
	&\rho_{12}^\prime(t)
	= \frac{1}{2} \bigl(a + b(t) - c(t)\bigr),\\
	&\rho_{21}^\prime(t)
	= \frac{1}{2} \bigl(a - b(t) + c(t)\bigr),
\end{alignedeqn}
where we employ the shorthands
\begin{alignedeqn}\label{eqn:158CB}
	b(t)
	= b_\text{in} \, \exp\bigl\{-2 \omega (t - t_\text{in})\bigr\},\\
	c(t)
	= c_\text{f} \, \exp\bigl\{-2 \omega (t_\text{f} - t)\bigr\}.
\end{alignedeqn}
Both off-diagonal elements approach the static density matrix \labelcref{eqn:80D} for $\omega (t - t_\text{in}) \ll 1$, $\omega (t_\text{f} - t) \gg 1$.

For the determinant one obtains
\begin{align}\label{eqn:158C}
	\det \rho^\prime(t)
	&= \frac{1}{4} \bigl[1 - a^2 - 4 \, b(t) \, c(t)\bigr]\\
	&= \frac{1}{4} \Bigl[1 - a^2 - 4 b_\text{in} \, c_\text{f} \, \exp\bigl\{-2 \omega (t_\text{f} - t_\text{in})\bigr\}\Bigr].\notag
\end{align}
It is independent of $t$. Positive eigenvalues of $\rho^\prime$ are obtained for $\det \rho^\prime \geq 0$. A requirement $\det \rho^\prime \geq 0$ would limit the allowed values of $a$ for $b_\text{in} \, c_\text{f} \neq 0$, but the modification as compared to limits for static solutions is tiny for large $\omega (t_\text{f} - t_\text{in}) \gg 1$. A pure state density matrix requires $\det \rho^\prime = 0$ or
\begin{equation}\label{eqn:158D}
	a^2
	= 1 - 4 b_\text{in} \, c_\text{f} \, \exp\bigl\{-2 \omega (t_\text{f} - t_\text{in})\bigr\}.
\end{equation}

We may finally express $b_\text{in}$ and $c_\text{f}$ in terms of the expectation values at the boundaries
\begin{equation}\label{eqn:158E}
	\Delta n_\text{in}
	= \langle n(t_\text{in})\rangle - \frac{1}{2},
	\quad
	\Delta n_\text{f}
	= \langle n(t_\text{f})\rangle - \frac{1}{2},
\end{equation}
resulting in a complete solution of the boundary problem for arbitrary mixed states
\begin{alignedeqn}\label{eqn:158F}
	\Delta n(t)
	&= \frac{\Delta n_\text{in} - r \, \Delta n_\text{f}}{1 - r^2} \, \exp\bigl\{-2 \omega (t - t_\text{in})\bigr\}\\
	&\hphantom{{}=}+ \frac{\Delta n_\text{f} - r \, \Delta n_\text{in}}{1 - r^2} \, \exp\bigl\{-2 \omega (t_\text{f} - t)\bigr\},
\end{alignedeqn}
where
\begin{equation}\label{eqn:158G}
	r
	= \exp\bigl\{-2 \omega (t_\text{f} - t_\text{in})\bigr\}.
\end{equation}
For small $r$ this yields the almost independent exponential decay of information from both boundaries. On the other hand, for small $\omega$ and $r$ close to one,
\begin{equation}\label{eqn:158H}
	r
	\approx 1 - 2 \omega (t_\text{f} - t_\text{in}),
\end{equation}
the solution interpolates linearly between $\Delta n_\text{in}$ and $\Delta n_\text{f}$,
\begin{alignedeqn}\label{eqn:158I}
	\Delta n(t)
	&= \frac{1}{2} \, (\Delta n_\text{in} + \Delta n_\text{f})\\
	&\hphantom{{}=}+ \frac{1}{2} \, (\Delta n_\text{in} - \Delta n_\text{f}) \, \frac{t_\text{f} + t_\text{in} - 2t}{t_\text{f} - t_\text{in}}.
\end{alignedeqn}

\subsection{Four-state oscillator chain}

We next present a complete solution of the boundary problem for the four state oscillator chain. This is interesting since the general solution of the evolution equation for the density matrix admits undamped oscillatory behavior of local probabilities. We show, however, that the initial conditions necessary for this type of solution cannot be realized. They are in conflict with the positive matrix elements of the density matrix in a classical statistical ensemble. The deeper reason that arbitrary initial values of the density matrix cannot be realized arises from the fact that for pure states $\bar q(t_\text{in})$ cannot be freely chosen. This extends to $\rho^\prime(t_\text{in})$.

From the general solution \labelcref{eqn:RS3} one infers that a $N \times N$ matrix $\rho^\prime$ admits $N$ linearly independent static solutions, corresponding to $\tilde{\lambda}_\tau = \tilde{\lambda}_\rho$. They are the equivalent of the static density matrices in quantum mechanics that can be constructed from the $N$ energy eigenstates. In case of imaginary eigenvalues there will be, in addition, oscillatory solutions. If $\tilde{\lambda}_\tau$ and $\tilde{\lambda}_\tau^\ast$ are both eigenvalues, with $\tilde{\lambda}^\ast_\tau \neq \tilde{\lambda}_\tau$, the difference $\tilde{\lambda}_\tau - \tilde{\lambda}_\tau^\ast$ is purely imaginary. This entails the existence of solutions to the evolution equation \labelcref{eqn:231} that describe undamped oscillations.

If the static solutions or the undamped oscillations could be realized by suitable boundary conditions, this would permit to store in the bulk memory of the boundary conditions - either in a static form by specifying a particular static density matrix, or in the form of oscillations. We expect that such memory materials can indeed be realized if the step evolution operator has more than one eigenvalue with $|\lambda| = 1$. The associated eigenvalues of $W$ are then zero or purely imaginary. However, one finds static or undamped oscillatory solutions for $\rho^\prime$ also in cases were the real part of $\tilde{\lambda}_\tau$ does not vanish. They correspond to situations where $\tilde{q}$ decreases exponentially while $\bar{q}$ increases exponentially, such that the bilinear $\sim \tilde{q} \bar{q}$ is static or oscillates. We will see that for an infinite number of $t$-steps, $G \to \infty$, it is not possible to realize this type of static or oscillatory density matrices by suitable boundary conditions. The case for finite $G$ is more subtle.

We will illustrate this general properties by an explicit solution of the boundary problem for the four-state oscillator chain given by the step evolution operator \labelcref{eqn:GD} or continuous evolution operator \labelcref{eqn:HE}. We will find that the most general solution of the evolution equation for the density matrix indeed admits undamped oscillations. The family of density matrices corresponding to static solutions or undamped oscillations obeys the quantum evolution according to the von Neumann equation. At first sight this seems to realize a memory material even in case of a unique largest eigenvalue of the step evolution operator. We will see, however, that the positivity of the overall probability distribution $w[n]$ restricts the allowed values of the density matrix at $t_\text{in}$. As a consequence, the amplitude of the undamped oscillations vanishes in the limit $\omega(t_\text{f} - t_\text{in}) \to \infty$.

Since the issue of a possible realization of a memory material is subtle we discuss the four-state oscillator chain in detail. For this model the $W$ matrix reads
\begin{equation}\label{eqn:80J}
	W
	= -\omega(\mathds{1}_4 - V),
\end{equation}
with
\begin{equation}\label{eqn:80K}
	V
	= \begin{pmatrix}
		\tau_- & \tau_\uparrow\\
		\tau_\downarrow & \tau_+
	\end{pmatrix}
\end{equation}
and
\begin{alignedeqn}\label{eqn:80L}
	&\tau_+
	= \begin{pmatrix}
		0 & 1\\
		0 & 0
	\end{pmatrix},
	\qquad
	&&\tau_-
	= \begin{pmatrix}
		0 & 0\\
		1 & 0
	\end{pmatrix},\\
	&\tau_\uparrow
	= \begin{pmatrix}
		1 & 0\\
		0 & 0
	\end{pmatrix},
	\quad
	&&\tau_\downarrow
	= \begin{pmatrix}
		0 & 0\\
		0 & 1
	\end{pmatrix}.
\end{alignedeqn}
We will next establish the general solution of the evolution equation \labelcref{eqn:231}.

We begin with the general statics solution. There are three non-trivial operators commuting with $W$,
\begin{equation}\label{eqn:80M}
	B_1
	= \begin{pmatrix}
		0 & \tau_1\\
		\tau_1 & 0
	\end{pmatrix},
	\quad
	B_2
	= V^\tran,
	\quad
	B_3
	= V.
\end{equation}
They obey
\begin{alignedeqn}\label{eqn:80N}
	&B_1^\tran
	= B_1,
	\qquad
	B_2^\tran
	= B_3,
	\quad
	&&B_2 B_3
	= B_3 B_2
	= 1,\\
	&B_1 B_2
	= B_2 B_1
	= B_3,
	\quad
	&&B_1 B_3
	= B_3 B_1
	= B_2.
\end{alignedeqn}
The general static density matrix takes the form
\begin{equation}\label{eqn:80O}
	\bar{\rho}^\prime
	= \frac{1}{4}(\mathds{1}_2 + b_k B_k).
\end{equation}
We observe that the diagonal elements of $B_k$ vanish such that the static local equilibrium probabilities show equal distribution
\begin{equation}\label{eqn:80P}
	\bar{p}_\tau
	= \frac{1}{4},
\end{equation}
independently of $b_k$.

For the time evolution \labelcref{eqn:231} we have decomposed $W$ into a ``symmetric part'' $J$ and antisymmetric part corresponding to a Hamiltonian $H$ according to \cref{eqn:FF3}
\begin{equation}\label{eqn:80Q}
	W
	= J - i H.
\end{equation}
The subclass of density matrices $\rho_H^\prime$ that commutes with $J$ does not ``feel'' the symmetric part. Density matrices in this subclass obey the von Neumann equation
\begin{equation}\label{eqn:80R}
	\partial_t\rho_H^\prime
	= -i[H,\rho_H^\prime],
	\qquad
	[J,\rho_H^\prime]
	= 0.
\end{equation}
This subclass therefore follows the time evolution of quantum mechanics.

In order to determine $\rho_H^\prime$ we write
\begin{alignedeqn}\label{eqn:80S}
	J
	&= \frac{1}{2} \bigl(W + W^\tran\bigr)
	= -\omega + \frac{\omega}{2} \bigl(V + V^\tran\bigr)\\
	&= -\omega + \frac{\omega}{2} (B_2 + B_3),
\end{alignedeqn}
and
\begin{alignedeqn}\label{eqn:80T}
	H
	&= \frac{i}{2} \bigl(W - W^\tran\bigr)
	= \frac{i\omega}{2} \bigl(V - V^\tran\bigr)\\
	&= -\frac{i\omega}{2} \bigl(B_2 - B_3\bigr),
\end{alignedeqn}
where we use $V = B_3$, $V^\tran = B_2$. We observe that $H$ and $J$ commute
\begin{equation}\label{eqn:NX2}
	[H,J]
	= 0.
\end{equation}
The condition $[J,\rho_H^\prime] = 0$ is obeyed for
\begin{equation}\label{eqn:NX3}
	\rho_H^\prime
	= \frac{1}{4} + b_1 B_1 + b_2 B_2 + b_3 B_3 + c_1 C_1 + c_2 C_2,
\end{equation}
with
\begin{equation}\label{eqn:NX4}
	C_1
	= \begin{pmatrix}
		\tau_3 & -i \tau_2\\
		i \tau_2 & -\tau_3
	\end{pmatrix},
	\qquad
	C_2
	= \begin{pmatrix}
		-\tau_1 & 1\\
		1 & -\tau_1
	\end{pmatrix}.
\end{equation}
This establishes the existence of a five-dimensional family of density matrices, parametrized by $b_i$, $c_i$, that follows the unitary evolution of quantum mechanics. In contrast to $b_i$ the coefficients $c_i(t)$ depend on $t$.

We next discuss the explicit solution for density matrices in this family. With
\begin{equation}\label{eqn:NX5}
	[H,C_1]
	= -2 i \omega C_2,
	\qquad
	[H,C_2]
	= 2 i \omega C_1,
\end{equation}
we obtain the evolution equation
\begin{equation}\label{eqn:NX6}
	\partial_t c_1
	= 2 \omega c_2,
	\qquad
	\partial_t c_2
	= -2 \omega c_1.
\end{equation}
It describes an oscillation with period $\pi/\omega$. The part $\sim C_1$ contributes to the diagonal elements of $\rho^\prime$. If one can realize a density matrix with $c_1 \neq 0$ or $c_2 \neq 0$ the local probabilities $p_\tau(t)$ will oscillate and transport information. This differs from the Ising model where all local probabilities approach in the bulk the equilibrium distribution.

We next turn to the general solution for $\rho^\prime$ that we may write in the form
\begin{alignedeqn}\label{eqn:NX7}
	\rho^\prime
	= \frac{1}{4} + \sum_{k=1}^3 b_k B_k + \bar{c}\bigl[&\sin(2 \omega t + \alpha)C_1\\
	&+ \cos (2 \omega t + \alpha) C_2\bigr] + \delta\rho^\prime,
\end{alignedeqn}
with $\rho_H^\prime$ realized for $\delta\rho^\prime = 0$. Our task is the computation of $\delta \rho$, as well as the integration constants $b_k$, $\bar{c}$ and $\alpha$, for arbitrary boundary conditions.

The eigenvalues $\tilde{\lambda}$ of $W$ are $0$, $-2 \omega$, $-(1 + i) \omega$, and $-(1 - i) \omega$. Correspondingly, one finds for the combinations $\tilde{\lambda}_\tau-\tilde{\lambda}_\rho$:
\begin{alignedeqn}\label{eqn:NX8}
	\frac{\tilde{\lambda}_\tau - \tilde{\lambda}_\rho}{\omega}
	\in \Bigl\{\overset{\overset{\mathclap{\times 4}}{\downarrow}}{0},2 i,-2 i,-\overbrace{1 + i}^{\times 2},-\overbrace{1 - i}^{\times 2},\\
	\underbrace{1 - i}_{\times 2},\underbrace{1 + i}_{\times 2},2,-2\mathrlap{\Bigr\}.}
\end{alignedeqn}
The four zeros correspond to the general static solution, and $\pm 2 i \omega$ accounts for the undamped oscillations. The other ten differences of eigenvalues correspond to the decreasing or increasing solutions contained in $\delta\rho^\prime$. We make the ansatz
\begin{alignedeqn}\label{eqn:NX9}
	\delta \rho^\prime
	&= D_{++} e^{(1 + i)\omega t} + D_{+-} e^{(1-i)\omega t} + D_{-+} e^{(-1 + i)\omega t}\\
	&\hphantom{={}}+ D_{--} e^{-(1 + i)\omega t} + E_+ e^{2 \omega t} + E_- e^{-2 \omega t},
\end{alignedeqn}
where
\begin{alignedeqn}\label{eqn:NX10}
	[\tilde{J},D_{\pm\gamma}]
	&= \pm 2 D_{\pm\gamma},
	\qquad
	[\tilde{J},E_{\pm}]
	= \pm 4 E_\pm,\\
	\lbrack\tilde{H},D_{\beta\pm}\rbrack
	&= \mp 2 i D_{\beta\pm},
	\qquad
	[\tilde{H},E_\pm]
	= 0,
\end{alignedeqn}
with
\begin{equation}\label{eqn:NX11}
	\tilde{J}
	= B_2 + B_3,
	\qquad
	\tilde{H}
	= B_2-B_3.
\end{equation}
The matrices $D$ and $E$ obey
\begin{equation}\label{eqn:NX11A}
	D_{\beta-}
	= (D_{\beta+})^\ast,
	\qquad
	E_\beta^\ast
	= E_\beta.
\end{equation}
If $D_{+\gamma}$ and $E_+$ solve \cref{eqn:NX10}, one obtains possible solutions for $D_{-\gamma}$ and $E_-$ as
\begin{equation}\label{eqn:NX12}
	D_{-\gamma}
	= (D_{+\gamma})^\tran,
	\qquad
	E_-
	= E_+^\tran.
\end{equation}

The relations \labelcref{eqn:NX10} require
\begin{equation}\label{eqn:NX13}
	[B_3,D_{++}]
	= (1 + i) D_{++},
	\qquad
	[B_3,E_+]
	= 2 E_+.
\end{equation}
The general solution of \cref{eqn:NX10} reads
\begin{alignedeqn}\label{eqn:NX14}
	D_{++}
	&= d_1 D_1 + d_2 D_2,\\
	D_{-+}
	&= d_3 D_1^\tran + d_4 D_2^\tran,\\
	E_+
	&= e_+ E,
	\qquad
	E_-
	= e_- E^\tran,
\end{alignedeqn}
with
\begin{alignedeqn}\label{eqn:NX15}
	D_1
	&= \begin{pmatrix}
		1 & -i & i & -1\\
		1 & -i & i & -1\\
		1 & -i & i & -1\\
		1 & -i & i & -1
	\end{pmatrix}\!,\\
	\enskip
	D_2
	&= \begin{pmatrix}
		1 & -1 & -1 & 1\\
		-i & i & i & -i\\
		i & -i & -i & i\\
		-1 & 1 & 1 & -1
	\end{pmatrix}\!,\\
	E
	&= \begin{pmatrix}
		1 & -1 & -1 & 1\\
		1 & -1 & -1 & 1\\
		1 & -1 & -1 & 1\\
		1 & -1 & -1 & 1
		\end{pmatrix}\!,
\end{alignedeqn}
and $d_i$ complex and $e_\beta$ real.

We conclude that the most general solution of the evolution equation for $\rho^\prime$ is given by \cref{eqn:NX7}, with
\begin{alignedeqn}\label{eqn:NX15A}
	\delta\rho
	&= \bigl[2 d_1 F_1 \cos(\omega t + \beta_1) + 2 d_2 F_2 \cos(\omega t + \beta_2)\\
	&+2 d_1 F_3 \sin(\omega t + \beta_1) + 2 d_2 F_4 \sin(\omega t + \beta_2)\bigr]e^{\omega t}\\
	&+\bigl[2 d_3 F_1^\tran \cos(\omega t + \beta_3) + 2 d_4 F_2^\tran \cos(\omega t + \beta_4)\\
	&+2 d_3 F_3^\tran \sin(\omega t + \beta_3) + 2 d_4 F_4^\tran \sin(\omega t + \beta_4)\bigr] e^{-\omega t}\\
	&+e_ + E e^{2 \omega t} + e_- E^\tran e^{-2 \omega t},
\end{alignedeqn}
where
\begin{alignedeqn}\label{eqn:NX16}
	&F_1
	= \begin{pmatrix}
		1 & 0 & 0 & -1\\
		1 & 0 & 0 & -1\\
		1 & 0 & 0 & -1\\
		1 & 0 & 0 & -1
	\end{pmatrix}\!,
	&&F_2
	= \begin{pmatrix}
		1 & -1 & -1 & 1\\
		0 & 0 & 0 & 0\\
		0 & 0 & 0 & 0\\
		-1 & 1 & 1 & -1
	\end{pmatrix}\!,\\
	&F_3
	= \begin{pmatrix}
		0 & 1 & -1 & 0\\
		0 & 1 & -1 & 0\\
		0 & 1 & -1 & 0\\
		0 & 1 & -1 & 0
	\end{pmatrix}\!,
	&&F_4
	= \begin{pmatrix}
		0 & 0 & 0 & 0\\
		1 & -1 & -1 & 1\\
		-1 & 1 & 1 & -1\\
		0 & 0 & 0 & 0
	\end{pmatrix}\!.
\end{alignedeqn}

From the general solution \labelcref{eqn:NX7,eqn:NX15A} one can compute the local probabilities $p_\tau = \rho_{\tau\tau}^\prime$ for any $t$. For suitable differences one finds
\begin{alignedeqn}\label{eqn:NX17}
	&p_1-p_2-p_3 + p_4\\
	&= 4\bigl\{\bar{c} \sin(\omega t + \alpha) + \tilde{e}_+ e^{-2 \omega(t_\text{f} - t)} + \tilde{e}_- e^{-2 \omega(t - t_\text{in})}\bigr\},\\
	&p_1-p_4\\
	&= 4\Bigl\{\bigl[\tilde{d}_1 \cos(\omega t + \beta_1) + \tilde{d}_2 \cos(\omega t + \beta_2)\bigr]
	e^{-\omega(t_\text{f} - t)}\\
	&\hphantom{=4\Bigl\{}+ \bigl[\tilde{d}_3 \cos(\omega t + \beta_3) + \tilde{d}_4 \cos (\omega t + \beta_4)\bigr]
	e^{-\omega(t - t_\text{in})}\Bigr\},\\
	&p_2-p_3\\
	&= 4\Bigl\{\bigl[\tilde{d}_1 \sin(\omega t + \beta_1)-\tilde{d}_2 \sin(\omega t + \beta_2)\bigr]
	e^{-\omega(t_\text{f} - t)}\\
	&\hphantom{=4\Bigl\{}+ \bigl[\tilde{d}_3 \sin(\omega t + \beta_3)-\tilde{d}_4 \sin(\omega t + \beta_4)\bigr]
	e^{-\omega(t - t_\text{in})}.
\end{alignedeqn}
Here the constants $\tilde{d}_i$ are obtained from $d_i$ by multiplication with suitable factors $e^{-\omega t_\text{in}}$ or $e^{\omega t_\text{f}}$, and similar for $\tilde{e}_\pm$. All three differences must lie in the range between $-1$ and $1$ for all $t$, restricting the allowed values of the twelve integration constants $\bar{c}$, $\alpha$, $\tilde{d}_i$, $\beta_i$, $\tilde{e}_\pm$. More precisely, the restrictions are given by $0 \leq p_\tau(t) \leq 1$. Together with the three integration constants $b_k$ the general solution has 15 integration constants, as appropriate for a first order differential equation for a $4 \times 4$ matrix $\rho^\prime$ with $\tr \rho^\prime = 1$.

For large $t_\text{f} - t_\text{in} \gg 1/\omega$ we observe that the constants $\tilde{d}_1$, $\tilde{d}_2$, $\beta_1$, $\beta_2$, $\tilde{e}_+$ have a sizeable influence only for $t$ in the vicinity of $t_\text{f}$. Similarly, the constants $\tilde{d}_3,\tilde{d}_4,\beta_3,\beta_4,\tilde{e}_-$ only influence the probabilities near $t_\text{in}$. For $t$ inside the bulk, e.g. for $\omega(t - t_\text{in}) \gg 1$, $\omega(t_\text{f} - t) \gg 1$, none of these ten integration constants matters and one has $p_4(t) = p_1(t)$, $p_3(t) = p_2(t)$. Similar to the Ising model $\delta \rho^\prime$ vanishes exponentially as one moves from the boundaries to the bulk. We observe, however, that even far inside the bulk the oscillations persist for the difference
\begin{equation}\label{eqn:NX18}
	p_1 - p_2 - p_3 + p_4
	\approx 4 \bar{c} \sin(\omega t + \alpha).
\end{equation}
This constitutes an important difference as compared to the one-dimensional Ising model. It could suggest that information can be transported into the bulk or from one boundary to the other if suitable boundary conditions can realize $\bar{c}\neq 0$. We will show next that this is not the case.

\subsection{Boundary conditions and initial value for density matrix}

One would like to characterize the classical statistical system by $\mathcal{K}(t)$ and the density matrix at the initial boundary, $\rho^\prime(t_\text{in})$. Then the boundary values of expectation values, as $\langle n_\gamma(t_\text{in})\rangle$ or $\langle h_\tau(t_\text{in})\rangle$, are fixed. Subsequently, one can follow the evolution in order to compute $\rho^\prime(t)$, and therefore the local probabilities $p_\tau(t)$, for arbitrary $t$.

The question arises if suitable boundary conditions can realize arbitrary initial values $\rho^\prime(t_\text{in})$. This is not a priori obvious since a pure state density matrix involves both $\tilde{q}(t_\text{in})$ and $\bar{q}(t_\text{in})$. Since $\bar{q}(t_\text{in})$ depends on $\bar{q}(t_\text{f})$, properties of both the initial and final boundary influence the initial density matrix $\rho^\prime(t_\text{in})$. In other words, one may ask if there exist initial and final boundary conditions $\tilde{q}(t_\text{in})$ and $\bar{q}(t_\text{f})$ such that an arbitrary pure state density matrix $\rho^\prime(t_\text{in})$ is realized. The answer will be negative. We will first investigate this question for the four-state oscillator chain and subsequently generalize the discussion.

The important features for our discussion can be seen already for pure state density matrices, with a rather obvious generalization to mixed states. We therefore investigate first the pure state density matrices by discussing the corresponding classical wave functions. The key point will be that $\bar q(t_\text{in})$ has to be realized by an appropriate $\bar q(t_\text{f})$. If there is only one largest eigenvalue, only the equilibrium conjugate wave function survives at $t_\text{in}$, $\bar q(t_\text{in}) = \bar q_\ast$. The restriction on $\bar q(t_\text{in})$ in turn restricts $\rho^\prime(t_\text{in})$. We will discuss this argument in detail.

For the four-state oscillator chain an arbitrary classical wave function can be written as
\begin{equation}\label{eqn:ID1}
	\tilde{q}(t)
	= g_j(t)v_j,
\end{equation}
with $v_j$ the eigenstates of $W$, where $v_0,v_{-1},v_2$ are given by \cref{eqn:GF}. Solutions of the evolution equation \labelcref{eqn:AF14} obey
\begin{equation}\label{eqn:ID2}
	g_j(t)
	= g_j e^{\tilde{\lambda}_j t},
\end{equation}
with eigenvalues of $W$ reading
\begin{equation}\label{eqn:ID3}
	\tilde{\lambda}_0
	= 0,
	\quad
	\tilde{\lambda}_{1\pm}
	= (-1 \pm i)\omega,
	\quad
	\tilde{\lambda}_2
	= -2 \omega.
\end{equation}
Similarly, one has for the conjugate wave function
\begin{equation}\label{eqn:ID4}
	\bar{q}(t)
	= \bar{g}_j(t)v_j^\ast,
	\qquad
	\bar{g}_j(t)
	= \bar{g}_j e^{-\tilde{\lambda}_j t}.
\end{equation}
The most general pure state density matrix therefore takes the form
\begin{equation}\label{eqn:ID5}
	\rho_{\tau\rho}^\prime(t)
	= \sum_{j,k}g_j\bar{g}_k
	e^{(\tilde{\lambda}_j - \tilde{\lambda}_k)t}v_{j,\tau}v_{k,\rho}^\ast.
\end{equation}

As an example for an oscillating pure state density matrix we may consider
\begin{alignedeqn}\label{eqn:ID6}
	g_{1+}
	&= g_{1-}
	= \frac{e^{\omega\bar{t}}}{2 \sqrt{2}},
	\qquad
	\bar{g}_{1+}
	= \bar{g}_{1-}
	= \frac{e^{-\omega\bar{t}}}{2 \sqrt{2}},\\
	g_0
	&= g_2
	= \bar{g}_0
	= \bar{g}_2
	= 0,
\end{alignedeqn}
where
\begin{alignedeqn}\label{eqn:ID7}
	\tilde{q}(t)
	&= e^{-\omega(t - \bar{t})}Q(t),\\
	\bar{q}(t)
	&= e^{\omega(t - \bar{t})}Q(t).
\end{alignedeqn}
Here $\bar{t}$ is arbitrary and $Q(t)$ reads
\begin{equation}\label{eqn:ID8}
	Q(t)
	= \frac{1}{\sqrt{2}}
	\begin{pmatrix}
		\tilde{c}(t)\\
		\tilde{s}(t)\\
		-\tilde{s}(t)\\
		-\tilde{c}(t)
	\end{pmatrix},
\end{equation}
using the abbreviations
\begin{equation}\label{eqn:ID9}
	\tilde{c}(t)
	= \cos \omega t,
	\qquad
	\tilde{s}(t)
	= \sin\omega t.
\end{equation}
The corresponding pure state density matrix is given by
\begin{equation}\label{eqn:ID10}
	\rho^\prime(t)
	= \frac{1}{2}
	\begin{pmatrix}
		\tilde{c}^2 & \tilde{c}\tilde{s} & -\tilde{c}\tilde{s} & -\tilde{c}^2\\
		\tilde{c}\tilde{s} & \tilde{s}^2 & -\tilde{s}^2 & -\tilde{c}\tilde{s}\\
		-\tilde{c}\tilde{s} & -\tilde{s}^2 & \tilde{s}^2 & \tilde{c}\tilde{s}\\
		-\tilde{c}^2 & -\tilde{c}\tilde{s} & \tilde{c}\tilde{s} & \tilde{c}^2
	\end{pmatrix}\!.
\end{equation}
Its diagonal elements describe oscillating local probabilities. The wave functions \labelcref{eqn:ID7} are the most general ones that yield the pure state density matrix \labelcref{eqn:ID10}. They have the initial and final values
\begin{alignedeqn}\label{eqn:ID11}
	\tilde{q}(t_\text{in})
	&= e^{\omega(\bar{t} - t_\text{in})}Q(t_\text{in}),\\
	\bar{q}(t_\text{f})
	&= e^{\omega(t_\text{f}-\bar{t})}Q(t_\text{f}).
\end{alignedeqn}

We will show that the conjugate wave function $\bar q(t)$ in \cref{eqn:ID7} cannot be realized since $g_0 = 0$ is in conflict with a positive overall probability distribution. For this purpose we present in detail the argument that the density matrix \labelcref{eqn:ID10} cannot be realized by a classical probability distribution $w[n]$. The reality of $\tilde{q}(t)$ and $\bar{q}(t)$ imposes real $g_0,g_2,\bar{g}_0$ and $\bar{g}_2$, while $g_{1+} = g_{1-}^\ast$, $\bar{g}_{1+} = \bar{g}_{1-}^\ast$, as obeyed by \cref{eqn:ID6}. The crucial point is that $w[n] \geq 0$ requires positive wave functions $\tilde{q}_\tau(t) \geq 0$, $\bar{q}_\tau(t) \geq 0$, as we will establish next. The positivity condition $w[n] \geq 0$ amounts to $w_{\rho_1 \rho_2 \dots} \geq 0$ in \cref{eqn:28B,eqn:28C}. Since $S_{\rho\tau}(t)$ is positive a positive $w$ is realized if for all $\rho$ and $\tau$ one has
\begin{equation}\label{eqn:ID12}
	b_{\rho\tau}
	= \tilde{q}_\rho(t_\text{in})\bar{q}_\tau(t_\text{f})
	\geq 0.
\end{equation}
Indeed, if in \cref{eqn:28C} some element $b_{\rho_1 \rho_{G+1}}$ is negative, the product of $S$-factors with the same $(\rho_1,\rho_{G+1})$-pair would have to vanish for arbitrary $\rho_2,\dots \rho_G$. This is not the case and we require \cref{eqn:ID12} as a necessary condition for the positivity of $w[n]$. In turn, this imposes $\tilde{q}_\rho(t_\text{in}) \geq 0$ and $\bar{q}_\tau(t_\text{f}) \geq 0$. From these inequalities we conclude $\tilde{q}_\rho(t) \geq 0$, $\bar{q}_\tau(t) \geq 0$ for all $t$.

We next employ the positivity of $\tilde{q}_\tau$ and $\bar{q}_\tau$ for all $t$ in order to show that the pure state density matrix \labelcref{eqn:ID6,eqn:ID10} cannot be realized for positive $w[n]$. This gives the first example that the possible values of the initial density matrix $\rho^\prime[t_\text{in}]$ are restricted. The explicit form of the wave function \labelcref{eqn:ID1} reads
\begin{equation}\label{eqn:ID13}
	\tilde{q}(t)
	= \begin{pmatrix}
		g_0 + g_2(t) + \tilde{c}(t) \, g_R(t) - \tilde{s}(t) \, g_I(t)\\
		g_0 - g_2(t) + \tilde{s}(t) \, g_R(t) + \tilde{c}(t) \, g_I(t)\\
		g_0 - g_2(t) - \tilde{s}(t) \, g_R(t) - \tilde{c}(t) \, g_I(t)\\
		g_0 + g_2(t) - \tilde{c}(t) \, g_R(t) + \tilde{s}(t) \, g_I(t)
	\end{pmatrix}\!,
\end{equation}
where
\begin{alignedeqn}\label{eqn:ID14}
	&g_2(t)
	= g_2 e^{-2 \omega t},
	\qquad
	g_{1+}
	= g_R^\prime + i g_I^\prime,\\
	&g_{R,I}(t)
	= 2 g_{R,I}^\prime \, e^{-\omega t}.
\end{alignedeqn}
The conjugate wave function $\bar{q}(t)$ takes a form similar to \cref{eqn:ID13}, with $g_0,g_2(t), g_{R,I}(t)$ replaced by $\bar{g}_0,\bar{g}_2(t),\bar{g}_{R,I}(t)$. For the time dependence of $\bar{g}_2(t),\bar{g}_{R,I}(t)$ one replaces by $e^{2 \omega t}$ or $e^{\omega t}$ the corresponding factors $e^{-2 \omega t},e^{-\omega t}$ in $g_2(t),g_{R,I}(t)$. Due to the alternating signs of the various contributions $\sim g_2,g_{R,I}$ the coefficient $g_0$ must be positive and cannot vanish. Indeed, for $g_0 = 0$ the only non-negative $\tilde{q}$ would be $\tilde{q} = 0$. By the same argument $\bar{g}_0$ must be positive and differ from zero. In contrast, the density matrix \labelcref{eqn:ID10} requires $g_0 = \bar{g}_0 = 0$. The oscillating density matrix given by \cref{eqn:ID10} can therefore not be realized for positive $w[n]$.

More generally, the positivity of $\bar{g}_0$ implies that for $\omega (t_\text{f} - t_\text{in}) \to \infty$ the conjugate wave function $\bar{q}(t_\text{in})$ will be dominated by the equilibrium wave function, thus restricting the possibilities for $\rho^\prime(t_\text{in})$. Indeed, for $\bar{g}_0 > 0$ the conjugate wave function approaches for $\omega(t_\text{f} - t) \to \infty$ the simple form $\bar{q}_\tau(t) = \bar{g}_0$. This follows since the relative differences of components of $\bar{q}$ vanish in this limit
\begin{equation}\label{eqn:ID15}
	\lim_{\omega (t_\text{f} - t) \to \infty}
	\frac{\bar{q}_\tau(t)-\bar{q}_\rho(t)}{\bar{q}_\tau(t) + \bar{q}_\rho(t)}
	\to 0.
\end{equation}
Indeed we observe that the numerator always contains factors $\exp\bigl(-\omega(t_\text{f} - t)\bigr)$, while the denominator has a contribution $2 \bar{c}_0$ for all $\rho$ and $\tau$. In the limit $\omega(t_\text{f} - t_\text{in}) \to \infty$ a pure state initial density matrix has therefore the form
\begin{equation}\label{eqn:ID16}
	\rho_{\tau\rho}^\prime(t_\text{in})
	= \bar{g}_0\tilde{q}_\tau(t_\text{in})
	= \langle h_\tau(t_\text{in})\rangle,
\end{equation}
independent of $\rho$. In this case the initial expectation values of occupation numbers and their products fix $\rho^\prime(t_\text{in})$ uniquely.

The specific family of initial density matrices \labelcref{eqn:ID16} cannot realize memory materials. We finally establish that far inside the bulk, $\omega (t - t_\text{in}) \gg 1$, expectation values of all local observables are given by equilibrium values. For given expectation values at $t_\text{in}$ the expectation values for all $t > t_\text{in}$ can be computed by following the evolution of $\rho^\prime(t)$. For $\rho^\prime(t_\text{in})$ given by \cref{eqn:ID16} and $g_0 > 0$ one obtains for $\omega (t - t_\text{in}) \gg 1$ the unique form
\begin{equation}\label{eqn:ID17}
	\rho_{\tau\rho}^\prime(t)
	= \bar{g}_0 g_0
	= \frac{1}{4},
\end{equation}
independent of $\tau$ and $\rho$. Similar to \cref{eqn:ID15} one has for all $\tau$ and $\rho$
\begin{equation}\label{eqn:ID18}
	\frac{\tilde{q}_\tau(t)-\tilde{q}_\rho(t)}{\tilde{q}_\tau(t) + \tilde{q}_\rho(t)}
	\sim \exp\bigl(-\omega(t - t_\text{in})\bigr),
\end{equation}
or
\begin{equation}\label{eqn:ID19}
	\lim_{\omega (t - t_\text{in}) \to \infty}
	\tilde{q}_\tau(t)
	= g_0,
\end{equation}
establishing \cref{eqn:ID17}.

We conclude that for $\omega(t_\text{f} - t_\text{in}) \gg 1$ the system approaches in the bulk the unique equilibrium state with local probabilities $p_\tau(t) = 1/4$ and $\langle f_\tau(t)\rangle = 1/4$, according to the density matrix \labelcref{eqn:ID17}. The boundary information is lost in the bulk. We can extend the discussion to arbitrary mixed states. For mixed states the initial density is a weighed sum of density matrices of the form \labelcref{eqn:ID16}. For $\omega(t - t_\text{in}) \gg 1$ the evolution of these mixed state density matrices leads to the pure state density matrix \labelcref{eqn:ID17}. This is an example of ``syncoherence'' \cite{CWQSS} - a mixed initial state $\rho^\prime(t_\text{in})$ develops into a pure state $\rho^\prime(t)$.

With the general solution \labelcref{eqn:ID13} for the wave function $\tilde{q}(t)$, and a similar general solution for the conjugate wave function $\bar{q}(t)$, we have the most general solution for a pure state density matrix. Boundary values for $\tilde{q}$ can be set at $t_\text{in}$, while boundary values for $\bar{q}$ may be established at $t_\text{f}$. For $\omega(t_\text{f} - t_\text{in}) \gg 1$ the boundary values $\tilde{q}(t_\text{in})$ can be related directly to expectation values of local observables $A(t_\text{in})$, and $\bar{q}(t_\text{f})$ is connected similarly to local observables at $t_\text{f}$. The boundary conditions determine all coefficients in the general solutions for $\tilde{q}$ and $\bar{q}$. This solves the boundary problem completely, since $\rho^\prime(t)$ is determined for arbitrary $t$.

The local probabilities $p_\tau(t)$ for the four-state oscillator chain follow damped oscillations as one moves from the boundary into the bulk. Such a behavior is expected in general if the bulk can be characterized by a unique equilibrium state, all eigenvalues of the step evolution operator $S_{\tau\rho}$ except one obey $|\lambda_i| < 1$, and some eigenvalues have imaginary parts. Quite generally, such damped oscillations can be characterized by a frequency $\omega$ and a damping rate $\gamma$, with typical time evolution
\begin{equation}\label{eqn:ID20}
	p_\tau(t)
	= p_\ast + c_\tau \cos(\omega t + \alpha_\tau)e^{-\gamma(t - t_\text{in})}.
\end{equation}
(There may be more than one frequency and damping rate - we consider here the smallest $\gamma$.) For the four-state oscillator chain one has $\gamma = \omega$.

For more general models $\gamma$ differs from $\omega$. An interesting situation arises if $\gamma \ll \omega$. In this case the local probabilities perform many oscillations before they are damped towards the equilibrium distribution. For small enough $\gamma$, such that $\gamma(t_\text{f} - t_\text{in}) \ll 1$, this realizes effective memory materials since the damping can be neglected. One has both damped oscillations with amplitudes $e^{-\gamma (t - t_\text{in})}$ or $e^{-\gamma (t_\text{f} - t)}$, and undamped oscillations with amplitudes $e^{-\gamma (t_\text{f} - t_\text{in})}$. Exact memory materials are realized in the limit $\gamma \to 0$. If simultaneously $\omega(t_\text{f} - t_\text{in}) \gg 1$ the effective memory material is characterized by oscillating local probabilities. The limit $\gamma \ll \omega$ may by realized in the asymmetric diagonal Ising model for large but finite $\beta$. In the absence of a unique equilibrium state the damping can be absent for certain states and one may find oscillations with $\gamma = 0$.

\section{Evolution for subsystems}
\label{sec:evolution for subsystems}

Static memory materials are realized whenever the step evolution operator has more than one largest eigenvalue with $|\lambda| = 1$. One expects in this case the existence of a suitable subsystem that obeys a unitary evolution. In this section we formulate the subsystem in terms of the density matrix. This will also provide examples for a unitary evolution of systems that are not unique jump chains. In \cref{sec:Three-spin chains} we have encountered quantum subsystems that indeed obey a unitary evolution. In the present section we will discuss a large class of systems for which the subsystem obeys a unitary evolution while the environment approaches a unique equilibrium state. This setting is sufficient for a quantum evolution, but not necessary. Indeed, the environment could follow its own non-trivial evolution without affecting the quantum subsystem. We will discuss this setting in the continuum limit.

\subsection{Von Neumann equation for subsystems}

The four-state oscillator chain provides a first glimpse how quantum evolution could arise for a suitable subsystem. Indeed, far inside the bulk the contribution $\delta \rho^\prime$ in \cref{eqn:NX7} can be neglected. The remaining part $\rho_H^\prime$ obeys the von Neumann equation
\begin{equation}\label{eqn:NY1}
	\partial_t\rho_H^\prime
	= -i[H,\rho_H^\prime],
\end{equation}
and therefore the evolution of quantum mechanics.

For the four-state oscillator chain the boundary conditions provide restrictions on $\rho_H^\prime$. For $\omega(t_\text{f} - t_\text{in}) \to \infty$ these restrictions force $\rho_H^\prime$ to be the equilibrium density matrix. Such a strong restriction of $\rho_H^\prime$ to a single matrix does not occur if more than one eigenvalue of $S$ obeys $|\lambda| = 1$. We ask here what are the conditions on $\rho'_H$ for realizing \cref{eqn:NY1} for a general evolution given by $W = J - i H$.

The family of density matrices $\rho_H^\prime$ that obey the condition
\begin{equation}\label{eqn:NY2}
	[J,\rho_H^\prime]
	= 0
\end{equation}
is subject to the quantum evolution \labelcref{eqn:NY1}. The property \labelcref{eqn:NY2} is preserved by the evolution if
\begin{alignedeqn}\label{eqn:NY3}
	[J,\partial_t\rho_H^\prime]
	= -i\bigl[J,[H,\rho_H^\prime]\bigr]
	= -i\bigl[[J,H],\rho_H^\prime\bigr]
	= 0.
\end{alignedeqn}
One infers that density matrices obeying conditions \labelcref{eqn:NY2,eqn:NY3} constitute a ``quantum subsystem'' \cite{CWQSS}. For the one-dimensional Ising model the diagonal part of $\rho_H^\prime$ is unique. For the four-state oscillator chain $\rho_H^\prime$ is not unique and constitutes a dynamical subsystem. \Cref{eqn:NY3} is obeyed by virtue of \cref{eqn:NX2}. For the memory materials discussed in \cref{sec:simple models} the full system obeys a unitary evolution and one has $\rho_H^\prime = \rho^\prime$.

For $J \neq 0$ the typical evolution of $\delta \rho^\prime = \rho^\prime - \rho_H^\prime$ is an attraction towards zero as one moves from the boundary into the bulk. This part of the information on the boundary conditions is lost far away from the boundaries. In contrast, the remaining information contained in $\rho_H^\prime$ is preserved by the quantum evolution. A classical statistical system with non-unique diagonal part of $\rho_H^\prime$ in the bulk constitutes a memory material. For memory materials a subsystem follows the quantum evolution.

Conditions \labelcref{eqn:NY2,eqn:NY3} are not sufficient for the realization of a memory material. To establish a non-unique $\rho_H^\prime$ in the bulk it is not sufficient that the general solution of the evolution equation admits non-unique $\rho_H^\prime$. As we have seen, the positivity of the overall probability distribution $w[n]$ may severely restrict the allowed space of solutions. For the four-state oscillator chain this restriction will enforce an essentially unique $\rho_H^\prime$ in the bulk if $t_\text{f} - t_\text{in}$ is large enough. More precisely, one finds $\bar{c} \sim \exp\bigl\{-\omega(t_\text{f} - t_\text{in})\bigr\}$ in \cref{eqn:NX18}. The restrictions from the positivity of $w[n]$ are actually obtained most easily by studying the solutions of the evolution equation for the wave function, while they are less directly apparent on the level of the density matrix. Still, the density matrix remains a very useful object if we want to focus on the unitary subsystem. The coarse graining to a subsystem may result in a mixed state, even if the density matrix for the whole system describes a pure state.

\subsection{Quantum subsystem and environment}

A simple example for a quantum subsystem that is not a unique jump chain is a block-diagonal evolution
\begin{equation}\label{eqn:QS1}
	W
	= \begin{pmatrix}
		W_1 & 0\\
		0 & -i H_2
	\end{pmatrix},
	\qquad
	H_2^\tran
	= -H_2.
\end{equation}
A block diagonal density matrix remains block diagonal under the evolution
\begin{equation}\label{eqn:QS2}
	\rho^\prime
	= \begin{pmatrix}
		\rho_1^\prime & 0\\
		0 & \rho_2^\prime
	\end{pmatrix}.
\end{equation}
The quantum subsystem is described by the block $\rho_2^\prime$ and obeys a unitary evolution. The other block $\rho_1^\prime$ may be associated with an environment. The environment could, for example, approach a unique equilibrium state.

The particular setting \labelcref{eqn:QS1} may look rather trivial. We argue however, that it reflects a rather general case if the eigenvalues of $S_{\tau\rho}$ with $|\lambda| = 1$ are not unique. Indeed, rather arbitrary situations describing a quantum subsystem and an environment can be brought to the form \labelcref{eqn:QS1} by a suitable similarity transformation.

Assume for simplicity that $S_{\tau\rho}$ does not depend on $t$. We can diagonalize $S$ by a regular complex matrix $D$, cf. \cref{eqn:Q10},
\begin{equation}\label{eqn:325A}
	S^\prime
	= D S D^{-1}
	= \diag(\lambda_i).
\end{equation}
The evolution equation \labelcref{eqn:AF14} remains preserved,
\begin{equation}\label{eqn:325B}
	\tilde{q}^\prime(t + \epsilon)
	= S^\prime\tilde{q}^\prime(t),
\end{equation}
if we employ
\begin{equation}\label{eqn:325C}
	\tilde{q}^\prime(t)
	= D \tilde{q}(t).
\end{equation}
In the new basis $(S^\prime,\tilde{q}^\prime)$ the evolution of sectors with different $\lambda_i$ decomposes into separate blocks. With block diagonal $S^\prime$ also $W^\prime$ is block diagonal.

Let us concentrate on the sector with $|\lambda_i| = 1$. For the present discussion this will be associated with the quantum subsystem, while the sectors with eigenvalues $|\lambda_i| < 1$ constitute the environment. (This is not the most general case. Some of the eigenvalues for the environment may also obey $|\lambda_i|=1$.) For
\begin{equation}\label{eqn:325D}
	\lambda
	= e^{i\alpha}
\end{equation}
the corresponding part of the matrix $W^\prime$ is purely imaginary, cf. \cref{eqn:FF1},
\begin{equation}\label{eqn:325E}
	W^\prime
	= \frac{1}{2 \epsilon}(e^{i\alpha}-e^{-i\alpha})
	= \frac{i \sin\alpha}{\epsilon}.
\end{equation}
For real $S$ and complex $\lambda_i$ the complex conjugate $\lambda_i^\ast$ is also an eigenvalue. The subsector of $W$ with the eigenvalues $e^{i\alpha}$ and $e^{-i\alpha}$ reads
\begin{equation}\label{eqn:325F}
	W^\prime
	= \frac{i}{\epsilon}
	\begin{pmatrix}
		\sin\alpha & 0\\
		0 & -\sin\alpha
	\end{pmatrix}.
\end{equation}
It can be brought to real form by defining
\begin{equation}\label{eqn:325G}
	W^{\prime\prime}
	= G W^\prime G^{-1}
	= \frac{\sin\alpha}{\epsilon}
	\begin{pmatrix}
		0 & 1\\
		-1 & 0
	\end{pmatrix},
\end{equation}
where
\begin{equation}\label{eqn:325H}
	G
	= \frac{1}{\sqrt{2}}
	\begin{pmatrix}
		1 & 1\\
		i & -i
	\end{pmatrix}.
\end{equation}
For the corresponding decomposition $W^{\prime\prime} = J^{\prime\prime} - i H^{\prime\prime}$ one finds for this sector
\begin{equation}\label{eqn:325I}
	J^{\prime\prime}
	= 0,
	\qquad
	H^{\prime\prime}
	= -\frac{\sin\alpha}{\epsilon} \, \tau_2.
\end{equation}
In this basis the setting \labelcref{eqn:QS1} is realized.

The same argument applies to all sectors with eigenvalues $|\lambda| = 1$. (For $\lambda_i = 1$, $\alpha = 0$ one has $W^\prime = W^{\prime\prime} = 0$, corresponding to $H = 0$ and compatible with $W = -i H$.) We may combine all blocks with $|\lambda| = 1$ to the quantum subsector with $W = -i H_2$. Within the quantum subsector we can further perform rotations without affecting the real and antisymmetric form of $H_2$. We may also perform arbitrary regular transformations in the environment, e.g. the subsector with eigenvalues $|\lambda| < 1$. This demonstrates that one can find a suitable similarity transformation $\tilde{D}$ such that the block diagonal form \labelcref{eqn:QS1} is realized. If there are complex eigenvalues \labelcref{eqn:325D} with $\alpha\neq 0,\pi$ the eigenvalues of $H_2$ differ from zero and occur in pairs with opposite signs.

We conclude that undamped oscillations occur rather generically in suitable subsectors if more than one eigenvalue of $S$ obeys $|\lambda| = 1$. It is sufficient that at least one of those eigenvalues has an imaginary part, which can happen in many circumstances. In this sense static memory materials can be realized in very general settings, extending far beyond the unique jump chains discussed in \cref{sec:simple models}. The identification of the quantum subsector may not always be easy, however. Formally, it is given by wave functions obeying
\begin{equation}\label{eqn:325J}
	\tilde{q}_Q
	= \tilde{D}^{-1} \tilde{q}_Q^\prime,
\end{equation}
where $\tilde{q}_Q^\prime$ are the wave functions corresponding to $|\lambda| = 1$ in the basis for which $\tilde{D} S\tilde{D}^{-1}$ leads to $W$ of the form \labelcref{eqn:QS1}. The expression of $\tilde{q}_Q$ in terms of the original components $\tilde{q}_\tau$ may be rather involved. The situation could be sometimes analogous to a pure quantum state that has effectively thermalized: while the initial information is still present, it is not visible in simple correlation functions of low order. In this sense the memory of initial conditions may sometimes be hidden. The most interesting realizations of static memory materials are those where the boundary information directly affects measurable observables.

\subsection{Incomplete statistics for subsystems}

For a given $\mathcal{K}$ or step evolution operator $S_{\tau\rho}$ the evolution of the density matrix is fixed uniquely. Knowledge of $\rho^\prime(t)$ contains all the information needed in order to compute expectation values of local observables at $t^\prime\neq t$. One has to solve the evolution with initial value $\rho^\prime(t)$ in order to compute $\rho^\prime(t^\prime)$, and from there to extract the expectation values by \cref{eqn:D6a}. The density matrix $\rho^\prime(t)$ contains only a small subset of the information contained in the full probability distribution $w[n]$. It constitutes an example for a subsystem characterized by ``incomplete statistics'' \cite{CWOQM}. If $\mathcal{K}(t^\prime)$ is known for a given interval in $t^\prime$ including $t$, all local observables in this interval are, in principle, determined by $\rho'(t)$. The details of $w[n]$ related to $t^\prime$ outside this interval do not matter. All probability distributions $w[n]$ that lead to the same density matrix $\rho^\prime(t)$ constitute an equivalence class which yields the same expectation values for local observables inside the interval.

The statistical information contained in the density matrix $\rho_{\tau\rho}^\prime(t)$ exceeds the local probabilities $p_\tau(t) = \rho_{\tau\tau}^\prime(t)$. The local information related to the off-diagonal elements of $\rho^\prime(t)$ is necessary for the computation of the time evolution of the local probabilities. In our formalism it arises naturally from the overall probability distribution $w[n]$. This ``off-diagonal information'' allows the computation of expectation values of additional observables, as represented by off-diagonal operators $A^\prime$. On the other hand the local information in $\rho^\prime(t)$ is incomplete in the sense that joint probabilities for observables represented by diagonal and off-diagonal operators can, in general, not be computed from $\rho^\prime(t)$. This incompleteness is the basic property underlying the non-commutative properties of observables in quantum mechanics \cite{CWQM}.

Furthermore, if only a subclass of density matrices $\rho_H^\prime$ survives in the bulk this defines again a subsystem. For the ``subsystem observables'' whose expectation values can be computed from $\rho_H^\prime$ alone, without the need to know the ``environment'', we encounter again a situation of incomplete statistics. Expectation values can be computed from the information in $\rho_H^\prime$, while this information is insufficient for the computation of joint probabilities.

\subsection{Decoherence and syncoherence}

Decoherence \cite{ZE,JZ,ZU} describes the transition from a pure state to a mixed state for some subsystem that is described by a coarse-grained density matrix. It cannot be accounted for by the evolution equation \labelcref{eqn:D9a} or \labelcref{eqn:231} since for these equations a pure state remains a pure state. There is no difference in this respect between classical statistics and quantum mechanics. If the subsystem is defined by $\rho_H^\prime$, the reduction to the subsystem may lead to a mixed state. The evolution \labelcref{eqn:NY1} within the subsystem does not describe decoherence, however. The situation is different if we project to other subsystems. In this case the evolution equation projected onto the subsystem typically contains additional terms beyond \cref{eqn:231}. The modifications will be of similar nature as for the Lindblad equation \cite{KO,LI,ZO} in quantum mechanics. The modified evolution equations can describe decoherence.

In the opposite, syncoherence \cite{CWQSS} describes the transition from a mixed state to a pure states. A simple physical example for a subsystem is a mixed state of excited atoms that evolves into a pure ground state by emission of radiation. Syncoherence can already occur within the evolution equation \labelcref{eqn:231}. It is sufficient that the reduced density matrix $\rho_H^\prime$ is a pure state density matrix. The term $\sim J$ in the evolution equations damps out the part $\delta\rho^\prime = \rho^\prime - \rho_H^\prime$, such that a mixed state density matrix $\rho^\prime$ evolves into a pure state density matrix $\rho_H^\prime$. This happens whenever the equilibrium state is unique. Only a single component of the wave functions $\tilde q_\ast$, $\bar q_\ast$ survives far inside the bulk such that the density matrix becomes the pure state density matrix formed from $\tilde q_\ast$, $\bar q_\ast$. The evolution of the models discussed in \cref{sec:solution of bvp} describes syncoherence of this type. Other forms of syncoherence are possible as well.

\section{Complex structure}
\label{sec:complex structure}

This section discusses complex classical wave functions and a complex classical density matrix. We deal here with complex structures in classical statistics that apply to the whole system. As we have seen in the explicit example of \cref{sec:Three-spin chains}, it is also possible to realize a complex structure only for a quantum subsystem, without extending it necessarily to the environment. These additional possibilities for complex structures are not addressed in the present section. 

So far, we only dealt with real quantities, even though the generalized Schrödinger equation \labelcref{eqn:46K,eqn:FF4} has been written formally in the usual complex form. In the occupation number basis the purely imaginary $H$ always occurs as $i H = -W_A$. This extends to every basis for which $\tilde q$ and $\bar q$ are real, but does not hold for an arbitrary basis. For example, in a basis where $S$ or $W$ are diagonal the wave functions are typically complex. Imaginary and real parts of the wave functions are not independent, however, since they originate from common real wave functions in the occupation number basis.

Quite often one encounters a genuine complex structure where imaginary and real parts of the wave functions are independent. The $N$-component real wave function $\tilde{q}$ is then associated to a $N/2$-component complex wave function $\psi$, while the $N$ real components of $\bar{q}$ correspond to $N/2$ complex components of $\bar{\psi}$. For general classical statistical systems $\psi$ and $\bar{\psi}$ are independent wave functions. In the presence of a genuine complex structure we deal with complex quantities also in the occupation number basis.

A genuine complex structure can be introduced whenever there exists a basis for which the real $N \times N$ matrix $W$ can be written in terms of $N/2 \times N/2$ matrices $W_1$ and $W_2$ in the form
\begin{equation}\label{eqn:CS1}
	W
	= \begin{pmatrix}
		W_1 & W_2\\
		-W_2 & W_1
	\end{pmatrix}.
\end{equation}
The antisymmetric and symmetric parts of $W$ read
\begin{alignedeqn}\label{eqn:CS6}
	W_A
	= -i H
	&= \begin{pmatrix}
		W_{1A} & W_{2S}\\
		-W_{2S} & W_{1A}
	\end{pmatrix},\\
	W_S
	= J
	&= \begin{pmatrix}
		W_{1S} & W_{2A}\\
		-W_{2A} & W_{1S}
	\end{pmatrix},
\end{alignedeqn}
with $W_{iS},W_{iA}$ the symmetric and antisymmetric parts of $W_i$. According to the block structure \labelcref{eqn:CS1} we also group the components of $\tilde{q}$ and $\bar{q}$ as
\begin{equation}\label{eqn:CS4}
	\tilde{q}
	= \begin{pmatrix}
		\tilde{q}_R\\ \tilde{q}_I
	\end{pmatrix},
	\qquad
	\bar{q}
	= \begin{pmatrix}
		\bar{q}_R\\ \bar{q}_I
	\end{pmatrix}.
\end{equation}

We next define the complex wave function as
\begin{equation}\label{eqn:CS5}
	\psi
	= \tilde{q}_R + i\tilde{q}_I,
\end{equation}
while for the conjugate wave function we employ
\begin{equation}\label{eqn:CS7}
	\bar{\psi}
	= \bar{q}_R - i \bar{q}_I.
\end{equation}
In terms of the $N/2$-component complex wave function $\psi$ the evolution equation \labelcref{eqn:AF14} takes the form of a complex generalized Schrödinger equation,
\begin{equation}\label{eqn:CS2}
	i \partial_t \psi
	= G\psi,
	\qquad
	G
	= \hat{H} + i \hat{J}.
\end{equation}
The Hermitian and antihermitian parts of $G$ are associated to $\hat{H} = \hat{H}^\dagger$ and $\hat{J} = \hat{J}^\dagger$, respectively, with
\begin{equation}\label{eqn:CS3}
	\hat{H}
	= W_{2S} + i W_{1A},
	\qquad
	\hat{J}
	= W_{1S} - i W_{2A}.
\end{equation}
The equivalence of \cref{eqn:CS2} with \cref{eqn:AF14} is established by insertion of \cref{eqn:CS5}. For $\tilde{W} = W$, \cref{eqn:AF16} transforms to
\begin{equation}\label{eqn:CS8}
	-i\partial_t\bar{\psi}
	= G^\tran \bar{\psi}
	= (\hat{H}^\ast + i\hat{J}^\ast) \bar{\psi}.
\end{equation}
For quantum systems with $\bar{q} = \tilde{q}$ one has $\bar{\psi} = \psi^\ast$. For $\hat{J} = 0$, \cref{eqn:CS8} is indeed the complex conjugate of \cref{eqn:CS2}.

For general classical statistical systems the conservation of the norm for the local probabilities is now reflected by the identity
\begin{equation}\label{eqn:CS9}
	\partial_t (\bar{\psi}^\tran \psi)
	= 0.
\end{equation}
On the other hand, the length of the complex vector $\psi$ changes according to
\begin{equation}\label{eqn:CS10}
	\partial_t (\psi^\dagger \psi)
	= 2 \psi^\dagger \hat{J} \psi.
\end{equation}
The form of \cref{eqn:CS10} shows that the antihermitian part of $G$ acts as a generalized damping term that can change the norm $|\psi|$. A unitary evolution is realized if the r.h.s. of \cref{eqn:CS10} vanishes.

If $W$ admits a complex structure according to \cref{eqn:CS1} the pure state complex density matrix $\rho$ obtains from the complex wave function $\psi$ and conjugate wave function $\bar{\psi}$ as
\begin{equation}\label{eqn:163AA}
	\rho_{\lambda\sigma}(t)
	= \psi_\lambda(t)\bar{\psi}_\sigma(t).
\end{equation}
This generalizes to mixed states according to the generalized boundary condition \labelcref{eqn:B5-D10},
\begin{equation}\label{eqn:164AB}
	\rho_{\lambda\sigma}(t)
	= \sum_\alpha w_\alpha\psi_\lambda^{(\alpha)}(t) \bar{\psi}_\sigma^{(\alpha)}(t).
\end{equation}
The complex evolution equation is the generalization of the von Neumann equation to non-hermitian $G$,
\begin{equation}\label{eqn:164AC}
	i \partial_t \rho
	= [G,\rho]
	= [\hat{H},\rho] + i [\hat{J},\rho].
\end{equation}
In the complex language subsystems with a unitary evolution can be realized if $[\hat{J},\rho] = 0$.

Consider next operators that take in the basis \labelcref{eqn:CS1}, \labelcref{eqn:CS4} the form
\begin{equation}\label{eqn:CSA}
	A^\prime
	= \begin{pmatrix}
		\hat A_R & -\hat A_I\\
		\hat A_I & \hat A_R
	\end{pmatrix},
\end{equation}
with $\hat A_R$ and $\hat A_I$ real $N/2 \times N/2$ matrices. Such operators are compatible with the complex structure. We define in the basis \labelcref{eqn:CS1}, \labelcref{eqn:CS4} an $N \times N$ matrix
\begin{equation}\label{eqn:CSB}
	I
	= \begin{pmatrix}
		0 & -1\\
		1 & 0
	\end{pmatrix},
	\qquad
	I^2
	= -1.
\end{equation}
Multiplication by $I$ corresponds to multiplication by $i$ in the complex basis. Both $W$ in \cref{eqn:CS1} and $A^\prime$ in \cref{eqn:CSA} commute with $I$, and we can write
\begin{alignedeqn}\label{eqn:CSC}
	&W
	= W_1 - W_2 \, I,
	\qquad
	&&A^\prime
	= \hat A_R + \hat A_I \, I,\\
	&[W,I]
	= 0,
	&&[A^\prime,I]
	= 0.
\end{alignedeqn}
The complex operator $\hat A$ is an $N/2 \times N/2$ matrix given by
\begin{equation}\label{eqn:CSD}
	\hat A
	= \hat A_R + i \, \hat A_I.
\end{equation}

We observe the isomorphism between the real and complex formulation
\begin{alignedeqn}\label{eqn:CSE}
	&\tilde q
	\to \psi
	\quad\Rightarrow\quad
	&&I \, \tilde q
	\to i \psi,
	\quad
	&&A^\prime \, \tilde q
	\to \hat A \, \psi,\\
	&\bar q
	\to \bar\psi
	\quad\Rightarrow\quad
	&&\bar q^\tran \, I
	\to i \bar\psi^\tran,
	\quad
	&&\bar q^\tran \, A^\prime
	\to \bar\psi^\tran \, \hat A.
\end{alignedeqn}
The expectation value of $A$ becomes in the complex formulation
\begin{equation}\label{eqn:CSF}
	\langle A\rangle
	= \bar q^\tran \, A^\prime \, \tilde q
	= \Re(\bar\psi^\tran \hat A \, \psi).
\end{equation}
For the particular case of quantum mechanics where $\bar\psi^\tran = \psi^\dagger$ only the hermitian part of $\hat A$ contributes. For the general classical statistical setting also the antihermitian part of $\hat A$ can contribute to the expectation value. The expectation values of observables that are not compatible with the complex structure, e.g. for which the associated operator is not of the form \labelcref{eqn:CSA}, can still be computed in the complex formulation. It mixes, however, real and imaginary parts in a form that cannot be expressed by the simple relation \labelcref{eqn:CSF}.

\section{Change of basis, similarity transformations and symmetries}
\label{sec:change of basis}

Crucial ingredients of the quantum formalism, as the change of basis, are actually not restricted to the particular unitary quantum evolution. They can be formulated for arbitrary classical statistical systems with quasi-local weight distribution
\begin{equation}\label{eqn:207A}
	w[n]
	= \prod_t\mathcal{K}(t) \, b(t_\text{in},t_\text{f}).
\end{equation}
(Here $b$ is a generalized boundary term depending on $n_\gamma(t_\text{in})$ and $n_\gamma(t_\text{f})$.) Exploiting these concepts for classical statistics offers a new view on otherwise perhaps hidden properties.

In the next three sections we explore how some of the formal structures of quantum mechanics extend to our classical statistical setting. This issue is rather trivial for the classical statistical systems that directly realize quantum systems, e.g. those with orthogonal $S$ or antisymmetric $W$. We will be concerned here with the general case where $\tilde{q}$ and $\bar{q}$ are different and the length of the way factors is not conserved by the evolution.

In this section we first discuss the change of basis. The possibility of a basis change is obviously realized for classical statistical systems in the transfer matrix formalism. All matrix expressions, e.g. the rules for the evaluation of expectation values of local observables, can be evaluated for an arbitrary choice of basis for the matrices. We extend here the change of basis to the concepts of the classical wave functions and density matrix.

A change of basis transforms \textit{both} the basis functions $h_\tau$ on one side, and the wave functions, density matrix and operators on the other side, such that $w[n]$, $f(t)$ or $A(t)$ remain invariant. In contrast, similarity transformations act on the wave functions, density matrix and operators, while keeping the basis functions $h_\tau$ fixed. Expectation values of operators and correlations, evolution equations and the partition function are invariant under such similarity transformations. On the other hand, the weight distribution $w[n]$, the wave functions $f(t)$ and local observables $A(t)$ transform non-trivially. This allows us to represent the same set of expectation values and correlations by a different weight distribution $w^\prime$, different step evolution operators $S^\prime$ and different observables. The weight distribution $w^\prime$ no longer needs to be positive. Even in the occupation number basis neither the elements of the step evolution operator $S^\prime$ nor the components of the wave vector $\tilde{q}^\prime$ need be positive. We discuss both global similarity transformations that leave the eigenvalues of $S$ unchanged, and local similarity transformations that may change the eigenvalues of $S$.

Symmetry transformations leave the weight distribution invariant, while keeping the basis functions fixed. Beyond the symmetry transformations acting on the variables $s_\gamma(t)$ or $n_\gamma(t)$ which leave $K[s]$ invariant, we discuss here general transformations acting on the step evolution operator, accompanied by corresponding transformations of wave functions and operators for observables. In this respect we concentrate on the subclass of similarity transformations that leave the weight distribution $w[n]$ invariant.

The results of this section are instructive for the general question which classical statistical systems are quantum systems, and the opposite question which quantum systems can be represented as classical statistical systems. We do not address here the possibility of quantum subsystems discussed in \cref{sec:evolution for subsystems}, but rather ask these questions for the overall system.

Classical statistical systems have a non-negative overall probability distribution $w[n] \geq 0$. For the subclass of quantum systems with $w[n] \geq 0$ the step evolution operator $S$ is a rotation, $S^\tran S = \mathds{1}$. Let us first work in the occupation number basis and consider step evolution operators for which all elements are non-negative, $S_{\tau\rho} \geq 0$. For finite $N$ it follows \cite{CWIT} that $S$ must describe a unique jump chain, with only one element equal to one in each column and row of the matrix, and all other matrix elements zero. This seems to constitute a severe restriction for the realization of quantum systems by classical statistical systems.

The condition $S_{\tau\rho} \geq 0$ is not necessary for a positive $w[n]$. By symmetry transformations we can map a step evolution operator with negative elements to a form with $S_{\tau\rho} \geq 0$, without changing $w[n]$. For finite $N$ we find that the form of $S$ in the occupation number basis remains still rather restricted. The combination of the conditions $w[n] \geq 0$, $S^\tran S = \mathds{1}$ admits as solutions only rescaled unique jump chains, where the unique non-zero elements in each row and column no longer equal one. For an infinite number of local states, $N \to \infty$, additional possibilities are expected to open up.

The answer to the question which quantum systems can be represented by a classical system is more complex. A given quantum system may be specified by a step evolution operator $S$. What are the conditions for $S$ such that expectation values and correlations of observables of this quantum system can be computed from a classical statistical system? First of all, a given form of $S$ does not specify the basis functions. If a change of basis, $S^\prime = V \, S \, V^\tran$, $V^\tran V = \mathds{1}$, results in the (rescaled) unique jump operator $S^\prime$, we can interpret $S$ as the step evolution operator in a basis related to the occupation number basis by a basis transformation $V$. There is no need anymore for all elements to be positive or for there to be only one non-zero element in each column and row of $S$. Physically such $S$ still represents a unique jump chain - only the observables may not be the obvious observables of the unique jump chain in the occupation number basis.

The apparent conclusion that for finite $N$ the unique jump chains are the only quantum systems that can be represented by a classical statistical system is not correct, however. In the occupation number basis, a given orthogonal $S$ that is not a (rescaled) unique jump operator typically results for some configurations $[n]$ in negative values of the weight distribution $w[n]$, as computed from \cref{eqn:28B,eqn:28C}. Consider now a similarity transformation that transforms the indefinite weight distribution $w[n]$ to a non-negative probability distribution $w^\prime[n] \geq 0$. Since expectation values and correlations remain invariant under similarity transformations we can compute them from the classical statistical system specified by $w^\prime[n]$. In this way we can map the quantum system to an equivalent classical statistical system. All weight distributions that originate from each other by similarity transformations belong to an equivalence class. Expectation values and correlations of observables are only properties of the equivalence class. (This restricts the choice of the physical correlations, cf. refs. \cite{CWQM,CWPT}.) If the indefinite weight distribution $w[n]$ of the quantum system belongs to the same equivalence class as the probability distribution $w^\prime[n]$ of a classical statistical system, we can compute all expectation values and correlations of the quantum system by the values of corresponding observables in the classical statistical system. In the opposite direction, expectation values and correlations in the classical statistical system can be computed from an equivalent quantum system.

\subsection{Change of basis}

The choice of the occupation number basis is convenient because the basis functions $h_\tau$ obey the simple rules \labelcref{eqn:AF22}. This choice, however, is not unique and we may consider any other choice $h_\tau^\prime$,
\begin{equation}\label{eqn:BB1}
	h_\tau
	= h_\alpha^\prime V_{\alpha\tau},
	\qquad
	V^\tran V
	= 1,
\end{equation}
with orthogonal $V$. Observable results cannot depend on the choice of basis functions. A change of the basis functions results in a corresponding change of the step evolution operator and operators for observables,
\begin{equation}\label{eqn:BB2}
	S^\prime
	= VSV^\tran,
	\qquad
	A_V^\prime
	= VA^\prime V^\tran.
\end{equation}
The weight distribution $w[n]$ remains invariant.

The evolution factors $\mathcal{K}(t)$,
\begin{equation}\label{eqn:BB3}
	\mathcal{K}(t)
	= h_\tau(t + \epsilon) \, S_{\tau\rho}(t) \, h_\rho(t)
	= h_\alpha^\prime(t + \epsilon) \, S_{\alpha\beta}^\prime(t) \, h_\beta^\prime(t)
\end{equation}
remain the same in the new basis \labelcref{eqn:BB1,eqn:BB2}. If we also transform the wave functions in the boundary terms
\begin{equation}\label{eqn:BB4}
	\tilde{q}^\prime(t_\text{in})
	= V\tilde{q}(t_\text{in}),
	\qquad
	 \bar{q}^\prime(t_\text{f})
	= \bar{q}(t_\text{f})V^\tran,
\end{equation}
the boundary factor $f_\text{in}$ in \cref{eqn:AF2A} remains unchanged,
\begin{equation}\label{eqn:BB5}
	f(t_\text{in})
	= h_\tau(t_\text{in}) \, \tilde{q}_\tau(t_\text{in})
	= h_\alpha^\prime(t_\text{in}) \, \tilde{q}_\alpha^\prime(t_\text{in}),
\end{equation}
and similar for $\bar{f}(t_\text{f})$. This confirms that the weight factor $w[n]$ as given by \cref{eqn:AF1,eqn:28A} is invariant under a change of basis. The relations \labelcref{eqn:28B,eqn:28C} no longer hold for arbitrary $h^\prime$, which obey only the weaker condition
\begin{equation}\label{eqn:BB6}
	\delta_{\alpha\beta} \, h_\alpha^\prime \, h_\beta^\prime
	= 1.
\end{equation}
As a result, the expression \labelcref{eqn:28B} for $w[n]$ is no longer valid in the new basis.

The wave functions $f(t)$ and $\bar{f}(t)$ in \cref{eqn:AF5,eqn:AF6} are independent of the basis, such that \cref{eqn:AF11} implies for all $t$
\begin{equation}\label{eqn:BB7}
	\tilde{q}^\prime(t)
	= V\tilde{q}(t),
	\qquad
	\bar{q}^\prime(t)
	= \bar{q}(t)V^\tran.
\end{equation}
Together with the transformation \labelcref{eqn:BB2} of operators the expression \labelcref{eqn:AF20} for the expectation value is basis independent for orthogonal $V$,
\begin{equation}\label{eqn:BB8}
	\langle A(t)\rangle
	= \langle \bar{q}_\tau^\prime(t) A_{V,\tau\rho}^\prime(t) \tilde{q}_\rho^\prime(t)\rangle.
\end{equation}
Also the evolution equations \labelcref{eqn:AF13} remain the same in every basis. Since orthogonal transformations do not change the eigenvalues of a matrix, a step evolution operator $S$ is transformed into a step evolution operator $S^\prime$ with the same eigenvalues. For the particular case of quantum mechanics, $S^\tran S = 1$, one also has $S^{\prime T} S^\prime = 1$. With $\bar{q} = \tilde{q}$ the change of basis \labelcref{eqn:BB1,eqn:BB2} corresponds to the familiar change of basis in quantum mechanics by orthogonal transformations. In the presence of a complex structure a suitable subgroup of the orthogonal transformations reproduces the change of basis by unitary transformations.

Orthogonal transformations do not preserve the positivity of a matrix (in the sense that not all matrix elements remain positive). If $S$ has only positive elements, the transformed matrix $S^\prime$ will, in general, also have negative elements. Furthermore, diagonal operators $A^\prime$ typically result in non-diagonal operators $A_V^\prime$ in another basis, and vice versa. This has interesting consequences for the representation of quantum systems by classical statistical probability distributions. Assume a discrete version of a quantum evolution, $q^\prime(t + \epsilon) = S^\prime q^\prime(t)$, with orthogonal $S^\prime$ not necessarily positive. Observables in this quantum system are represented by operators $A_{QM}^\prime$. Assume now the existence of an orthogonal matrix $V$ such that $S = V^\tran S^\prime V$ is a positive matrix. We can then use the basis with positive $S$, introduce basis functions $h_\tau$ for the occupation number basis and represent the system by the classical probability distribution $w[n]$ by use of \cref{eqn:28B,eqn:28C}. The expectation values of the quantum observables can be measured in this classical statistical system by determining for this classical probability distribution the expression \labelcref{eqn:AF20} for $A^\prime = V^\tran A_{QM}^\prime V$. In summary, a quantum system with indefinite step evolution operator $S^\prime$ can correspond to a positive weight distribution if a basis transformation to positive $S$ is possible.

\subsection{Similarity transformations}

A different type of transformations transforms $S$ and $A^\prime$, while keeping fixed basis functions $h_\tau$. In the occupation number basis \cref{eqn:28B,eqn:28C} remain valid and the weight distribution $w[n]$ is transformed to a new weight distribution $w^\prime[n]$ by use of $S^\prime$ and transformed boundary conditions. In general, similarity transformations map different weight distributions or ``different models'' into each other. For the fixed basis we focus here on the occupation number basis.

The partition function \labelcref{eqn:partition function def} or \labelcref{eqn:partition function} remains invariant under similarity transformations with arbitrary complex regular matrices ${D}_{\tau\rho}$ if we transform
\begin{alignedeqn}\label{eqn:Q10}
	S^\prime(t)
	&= D S(t) D^{-1},\\
	\tilde{q}^\prime(t_\text{in})
	&= D \tilde{q}(t_\text{in}),
	\qquad
	\bar{q}^\prime(t_\text{f})
	= (D^\tran)^{-1} \bar{q}(t_\text{f}).
\end{alignedeqn}
The evolution equations for $\tilde{q}^\prime(t)$ and $\bar{q}^\prime(t)$ take the same form as \cref{eqn:AF13} if we replace $S$ by $S^\prime$ and define
\begin{equation}\label{eqn:208A}
	\tilde{q}^\prime(t)
	= D\tilde{q}(t),
	\qquad
	\bar{q}^\prime(t)
	= (D^\tran)^{-1}\bar{q}(t).
\end{equation}
This extends to the expectation values of local observables if $A_{\tau\rho}^\prime(t)$ is transformed as
\begin{equation}\label{eqn:Q11}
	A_D^\prime(t)
	= D A^\prime(t) D^{-1}.
\end{equation}
(Diagonal observables $A^\prime(t)$ may be transformed by \cref{eqn:Q11} to non-diagonal $A_D^\prime(t)$.) We conclude that all local structures discussed in this paper and all expectation values of local observables are invariant under the similarity transformation \labelcref{eqn:Q10,eqn:208A,eqn:Q11}. This extends to correlation functions which are represented as operator products.

We emphasize that the equivalence induced by similarity transformations is restricted here to the local subsystem. The overall weight distributions $w[n]$ and $w^\prime[n]$ differ. We have made no statement about the transformation of the most general classical observables, such that the expectation values of general observables in the system specified by $w^\prime[n]$ are not related to observables in the system given by $w[n]$. In contrast, for local observables for which expectation values and correlations can be evaluated from the generalized quantum expression \labelcref{eqn:quantum rule} for the associated operators, the transformation rule \labelcref{eqn:Q11} is sufficient for relating expectation values and correlations in the systems specified by $w^\prime[n]$ and $w[n]$. For every set of local observables expressed by a set of operators $A^\prime(t)$ in the system $w[n]$ there exists for the system $w^\prime[n]$ an associated set of local observables expressed by the set of operators $A_D^\prime(t)$. Expectation values and correlations for the two sets are identical. The wave functions, density matrix and evolution equation needed for the computation of expectation values are transformed according to \cref{eqn:Q10,eqn:208A}.

For the particular case where $D = V$ is a rotation the transformation of the wave functions and operators is the same as for the change of basis discussed previously. On the level of the local objects $\tilde{q},\bar{q}, A^\prime$ we can associate the general transformation \labelcref{eqn:Q10,eqn:208A,eqn:Q11} with a generalized change of basis. Since ${D}$ needs not to have only positive elements the new step evolution operator $S^\prime$ can have negative elements even if $S$ has only positive elements. For complex $D$ the new $S^\prime$ may be complex. For real $D$ the weight function $w^\prime[n]$ obtained by replacing in \cref{eqn:28C} $S \to S^\prime$, $q \to q^\prime$, $\bar{q} \to \bar{q}^\prime$ is real. However, it needs no longer to be an overall probability distribution. Indeed, the negative elements in $S^\prime$ can lead to negative $w_{\rho_1 \rho_2 \dots}$ in \cref{eqn:28C}, such that $w^\prime[n]$ can be negative for some configurations $[n]$. This typically happens if ${D}$ is a rotation. Nevertheless, the expectation values and correlations of all local observables computed with $w^\prime$ are the same as the ones obtained from $w$. More precisely, the expectation value of the observable represented by the operator $A_D^\prime$, computed from $w^\prime$, is the same as the expectation value of the observable represented by the operator $A^\prime$, computed from $w$.

Under certain circumstances the similarity transformations can be used to transform $S$ to a rotation matrix $S^\prime$, even though $S$ is not a rotation. This holds if one can find a matrix $B$ such that
\begin{equation}\label{eqn:209A}
	S^\tran B
	= B \, S^{-1},
\end{equation}
and if $B$ can be represented as
\begin{equation}\label{eqn:209B}
	B
	= D^\tran D.
\end{equation}
In this case the wave vector $\tilde{q}^\prime(t)$, as defined in \cref{eqn:208A}, obeys a unitary quantum evolution. Accordingly, the wave vector $\tilde{q}(t)$ follows a type of ``rescaled unitary evolution'' as inferred from $\tilde{q} = D^{-1} \tilde{q}^\prime$.

One may specialize to the particular case where $S$ is already a rotation. For real $D$ and $D^\tran D = 1$ also $S^\prime$ is a rotation. If ${D}$ is a rotation, $D^{-1} = D^\tran$, the norm of $q$ is preserved. This corresponds to the standard change of basis in quantum mechanics by unitary (in our case orthogonal) transformations. If, furthermore, $D = D_s$ commutes with $S$ and therefore with $H$,
\begin{equation}\label{eqn:Q12}
	[D_s,S]
	= 0,
	\qquad
	[D_s,H]
	= 0,
\end{equation}
the transformation describes a symmetry of the corresponding quantum system. This symmetry may be ``spontaneously broken'' by a particular solution or initial condition if ${D}_s q(t_\text{in})\neq q(t_\text{in})$.

\subsection{Local similarity transformations}

We can extend the concept of similarity transformations \labelcref{eqn:Q10} to local similarity transformations
\begin{alignedeqn}\label{eqn:Q17}
	&S^\prime(t)
	= D(t + \epsilon) \, S(t) \, D^{-1}(t),\\
	&q^\prime(t_\text{in})
	= D(t_\text{in}) \, q(t_\text{in}),\\
	&\bar{q}^\prime(t_\text{f})
	= \bigl({D}^\tran(t_\text{f})\bigr)^{-1} \, \bar{q}(t_\text{f}),
\end{alignedeqn}
with different transformation matrices $D(t)$ for different $t$. Local similarity transformations leave again the partition function invariant and constitute transformations within the local equivalence class. The wave functions at $t$ transform as
\begin{alignedeqn}\label{eqn:Q18}
	&q^\prime(t)
	= D(t) q(t),\\
	&\bar{q}^\prime(t)
	= \bigl(D^\tran(t)\bigr)^{-1} \bar{q}(t),\\
	&\rho_D^\prime(t)
	= D(t) \, \rho^\prime(t) \, D^{-1}(t),
\end{alignedeqn}
and local observables as
\begin{equation}\label{eqn:Q19}
	A_D^\prime(t)
	= D(t)A^\prime(t) D^{-1}(t).
\end{equation}
It is straightforward to verify that evolution equations, expectation values and correlations remain invariant under local similarity transformations.

Local similarity transformations are a powerful tool for establishing structural aspects of our formalism, as we will discuss in \cref{sec:classical and quantum statistics}. In particular, we will establish that arbitrary quantum systems can be represented as classical statistical systems with an appropriate choice of operators. In contrast to global similarity transformations the local similarity transformations can change the eigenvalues of the step evolution operator, while the eigenvalues of operators for observables remain invariant.

\subsection{Sign of the wave function}

We concentrate here on the occupation number basis. With positive step evolution operator $S_{\tau\rho} \geq 0$ all components of the wave function $\tilde{q}_\tau(t_\text{in})$ remain positive if initially one has $\tilde{q}_\tau(t_\text{in}) \geq 0$. A similar property holds for the conjugate wave function if all components $\bar{q}_\tau(t_\text{f})$ are positive. It is not always convenient to work with positive real wave functions. As an example we may consider the memory materials which correspond to propagating fermions that we have discussed in \cref{sec:simple models}. We may impose periodic boundary conditions in $x$. Simple periodic functions such as $\tilde{q}(t_\text{in},x) \sim \sin(\omega x)$ have alternating signs. Typically, $\tilde{q}(t,x)$ will then have alternating signs for all $t$.

Negative signs of some components $\tilde{q}_\tau(t)$ are, in principle, no problem. They only need to be accompanied by negative signs of the corresponding components $\bar{q}_\tau(t)$ of the conjugate wave function, such that the local probabilities $p_\tau(t) = \tilde{q}_\tau(t) \, \bar{q}_\tau(t)$ are positive.

We will establish that the choice of signs for the components $\tilde{q}_\tau(t)$ of the wave function is arbitrary, provided that a change of signs is accompanied by a corresponding change in $\bar{q}_\tau(t)$. For this purpose we use the local similarity transformations \crefrange{eqn:Q17}{eqn:Q19}. Consider the case where ${D}(t)$ in \cref{eqn:Q17} is diagonal,
\begin{equation}\label{eqn:Q20}
	{D}(t)_{\tau\rho}
	= \hat{s}_\tau(t)\delta_{\tau\rho}
\end{equation}
with eigenvalues $\hat{s}_\tau(t) = \pm 1$. This realizes the simultaneous change of sign of $\tilde{q}_\tau(t)$ and $\bar{q}_\tau(t)$. Diagonal operators $\hat{A}$ are invariant under this transformation. We can view the transformation \labelcref{eqn:Q18,eqn:Q20} as a discrete local gauge transformation. With (no index sums)
\begin{alignedeqn}\label{eqn:Q21}
	S_{\tau\rho}^\prime(t)
	&= \hat{s}_\tau(t + \epsilon) \, S_{\tau\rho}(t)\hat{s}_\rho(t),\\
	q_\tau^\prime(t)
	&= \hat{s}_\tau(t)q_\tau(t),
	\qquad
	\bar{q}_\tau^\prime(t)
	= \hat{s}_\tau(t)q_\tau(t),
\end{alignedeqn}
the weight function $w[n]$ is invariant under this gauge transformation. In other words, the local sign changes \labelcref{eqn:Q21} do not transform among different members of a local equivalence class but keep $w[n]$ fixed.

The step evolution operators $S^\prime$ obtained from the sign transformation in \cref{eqn:Q21} constitute examples how a positive and normalized overall probability distribution, and therefore a classical statistical model, can be realized by step evolution operators that involve negative elements. It is sufficient that local sign changes \labelcref{eqn:Q21} exist that make $S$ positive. The step evolution operators $S^\prime(t)$ may not be the same for all $t$. In this case they do not realize standard translation symmetry in $t$. Still, if the associated positive $S(t)$ is independent of $t$, there is a hidden translation symmetry in the formulation with $S^\prime$ as well. It involves appropriate multiplications with the matrix ${D}$ in \cref{eqn:Q20} at every time step.

While the choice of signs of the components of the wave function $\tilde{q}_\tau$ (with associated signs of $\bar{q}_\tau$) is, a priori, arbitrary, not every choice will be consistent with the continuum limit. Indeed, a differentiable time dependence of $\tilde{q}_\tau(t)$ will be destroyed by arbitrary sign changes. Requirements of a continuous and differentiable description single out particular choices of signs. They actually fix the gauge freedom of local choices of signs almost completely. The analogue holds if we require a continuous and differentiable dependence of the wave function on some other variable, as for $\tilde{q}_\tau(t,x)$ in the fermion model of \cref{sec:simple models}.

For quantum systems, with $\tilde{q} = \bar{q} = q$, this has an interesting consequence. From $p_\tau = q_\tau^2$ one infers that the local probabilities $p_\tau$ determine the wave function up to a sign $\hat{s}_\tau$. If this sign ambiguity can be fixed by continuity requirements, the information contained in the local probability distribution $[p_\tau(t)]$ is sufficient for a determination of the density matrix $\rho^\prime(t)$, and therefore for the formulation of a linear evolution law in terms of $\rho^\prime(t)$. In this case the evolution is uniquely determined by the local probabilities. We recall in this context that in the usual complex formulation of quantum mechanics not only $\psi_\tau^\ast \psi_\tau$ have the properties of probabilities, but also the separate expressions $\psi_{R\tau} \psi_{R\tau}$ and $\psi_{I\tau} \psi_{I\tau}$ for the real and imaginary components. In this extended probabilistic setting the ambiguity in the wave function does not concern phases, but only signs.

\subsection{Symmetry transformations}

Let us consider transformations in the space of weight distributions, represented by invertible maps $w[n] \to w^\prime[n]$. We are interested in transformations induced by maps between step evolution operators $S(t) \to S^\prime(t^\prime)$, accompanied by appropriate transformations of boundary terms. Transformations in the space of step evolution operators may be induced, in turn, by replacements of occupation numbers $n_\gamma(t) \to \tilde n_\gamma(t^\prime)$. These are, however, not the only possibilities. Examples for other transformations are the local similarity transformations \labelcref{eqn:Q17}.

Symmetry transformations of the weight distribution are those transformations that leave $w[n]$ invariant. An example are the local sign transformations \labelcref{eqn:Q21}. Symmetry transformations of the partition function leave $Z$ invariant. They constitute a much larger class of transformations than the symmetries of the weight distribution. Examples are arbitrary local similarity transformations \labelcref{eqn:Q17}, in particular the global similarity transformations \labelcref{eqn:Q10}.

One may investigate the conditions for a general local similarity transformation \labelcref{eqn:Q17} to be a symmetry of the weight distribution. This is realized if (no sum over $\rho$)
\begin{equation}\label{eqn:ST1}
	D_{\alpha\rho}^{-1}(t) D_{\rho\beta}(t)
	= \delta_{\alpha\rho}\delta_{\beta\rho},
\end{equation}
which means that $D(t)$ is a diagonal matrix
\begin{equation}\label{eqn:ST2}
	D_{\alpha\rho}(t)
	= d_\rho(t)\delta_{\alpha\rho}.
\end{equation}
Local similarity transformations with arbitrary diagonal matrices leave the weight distribution invariant.

We next ask which orthogonal step evolution operators $S$ can be transformed by the symmetry transformations \labelcref{eqn:Q17,eqn:ST2} to a positive matrix $S^\prime$. We consider finite $N$. Since only the signs of $S^\prime$ matter we can take $d_\rho(t) = \hat{s}_\rho(t) = \pm 1$. Thus $D(t)$ are orthogonal matrices and $S^\prime$ is also orthogonal. Since $S^\prime$ is a positive orthogonal matrix it must be a unique jump operator. We conclude that $S$ must be a (rescaled) unique jump operator with arbitrary signs for the unique non-zero elements in each row and column. They can be brought by the symmetry transformation \labelcref{eqn:Q17,eqn:ST2} to the standard unique jump operator for which all elements equal one or zero.

We can use this result in order to establish in the occupation number basis that for finite $N$ orthogonal step evolution operators (not necessarily positive) are compatible with a positive weight distribution $w[n]$ only if they are (rescaled) unique jump operators. The necessary condition for positive $w[n]$ discussed in \cite{CWIT}, namely that $S_{\alpha\rho}(t)S_{\rho\beta}(t - \epsilon)$ has the same sign for all $\rho$, for arbitrary $\alpha,\beta$ and $t$, implies that we can assign a sign $\hat{s}_\rho$ to every column of $S(t)$, and the same sign to every row $S(t - \epsilon)$. This means that all elements in the corresponding column or row have the same sign or vanish. In other words, for a positive weight function it is necessary that a sign matrix \labelcref{eqn:Q20} exists such that both $\hat{S}(t)$ and $\hat{S}(t - \epsilon)$,
\begin{equation}\label{eqn:ST3}
	\hat{S}(t)
	= S(t)D(t),
	\qquad
	\hat{S}(t - \epsilon)
	= D(t)S(t - \epsilon),
\end{equation}
are positive matrices. This argument holds for all $t$ and we conclude that for positive $w[n]$ all $S(t)$ can be made positive by appropriate sign transformations. For orthogonal $S(t)$ this requires $S(t)$ to be unique jump matrices.

The symmetry transformation \labelcref{eqn:ST2} can also be used for arbitrary complex $d_\rho(t)$. In consequence, rescaled unique jump operators with arbitrary complex elements can be brought to the positive standard form where all non-zero elements equal one. As we have mentioned already, diagonal operators $A^\prime(t)$ are left invariant by the transformation \labelcref{eqn:Q19}. On the other hand, off-diagonal operators experience changes according to \cref{eqn:Q19,eqn:ST2}.

\section{Equivalent local probabilistic systems}
\label{sec:equivalent local probabilistic systems}

Suppose someone presents a quantum Hamiltonian and asks: is it possible to compute expectation values and correlations of observables in the quantum system from some classical statistical system? The answer is yes if there exists a positive classical statistical probability distribution $w^\prime[n]$ that belongs to the same equivalence class as the (in general indefinite) weight distribution $w[n]$ of the quantum system, and no if not. We will give a general answer to this question in the next section. In the present section we explore in more detail which systems belong to the same equivalence class as the ones defined by the Schrödinger or von Neumann equations for the evolution according to a given quantum Hamiltonian.

Notions of equivalence can be formulated on different levels. We always will be concerned with the local system, as expressed by the density matrix $\rho^\prime(t)$ (or wave functions in case of a pure state), its evolution law, and the set of operators $A^\prime(t)$ for local observables. ``Strong equivalence'' is realized if two systems lead to the same $\rho^\prime(t)$ for all $t$. The expectation values and correlations of local observables expressed by the same operator $A^\prime(t)$ are then identical. On the level of observables, two observables have for all $t$ the same expectation values if they are represented by the same operator $A^\prime(t)$. In order to qualify for equivalent local observables one needs, in addition, the ``probabilistic requirement'' that the probabilities to find possible measurement values corresponding to the spectrum of $A^\prime(t)$ can be computed from $\rho^\prime(t)$ and are positive.

``Weak equivalence'' is given if two systems can be mapped onto each other by a local similarity transformation. In case of weak equivalence also the operators have to be mapped consistently between two equivalent systems. For a weak equivalence between a quantum system and a classical statistical system the expectation values and correlations of observables in the quantum system can, in principle, be computed in the classical statistical system. In practice, this may often be forbiddingly complicated if the map between operators in the quantum and classical systems is not sufficiently simple. In particular, simple operators in the quantum system may be mapped to rather complicated off-diagonal operators in the classical statistical system for which no measurement prescription can realistically be conceived. An ``intermediate equivalence'' of two systems occurs if their density matrices are related by a global ($t$-independent) similarity transformation. This corresponds to a generalized global change of basis.

Furthermore, for many questions we may relax the requirement for equivalence by demanding only that the continuum limits of $\rho^\prime(t)$ and $A^\prime(t)$ coincide, or can be mapped into each other by an equivalence transformation.

\subsection{Complex evolution equation}

The first issue concerns the complex formulation of quantum mechanics. A Hermitian Hamiltonian is obtained for a unitary step evolution operator. We therefore ask for the conditions on similarity transformations to map a real step evolution operator $S$ onto a unitary step evolution operator $S^\prime$. We restrict the discussion here to global similarity transformations and investigate weak equivalence between a complex quantum system and a real classical statistical system. With complex $D$ we want to realize
\begin{equation}\label{eqn:209C}
	S^\prime
	= D \, S \, D^{-1},
	\qquad
	S^{\prime\dagger} S^\prime
	= 1.
\end{equation}
This can be achieved if one can find a matrix $B$ obeying \cref{eqn:209A}, with \cref{eqn:209B} generalized to
\begin{equation}\label{eqn:209D}
	B
	= D^\dagger D.
\end{equation}
For unitary $S^\prime$ the complex wave function $\tilde{q}^\prime = D\tilde{q}$ follows a unitary evolution. We realize quantum mechanics if the boundary condition \labelcref{eqn:Q7} is replaced by
\begin{alignedeqn}\label{eqn:209E}
	\bar{q}^\prime(t_\text{f})
	&= \tilde{q}^{\prime\ast}(t_\text{f}),\\
	\bar{q}(t_\text{f})
	&= D^\tran D^\ast\tilde{q}(t_\text{f})
	= B^\ast\tilde{q}(t_\text{f}).
\end{alignedeqn}
For real $\bar{q},\tilde{q}$ this requires $B$ to be real and symmetric.

For unitary $S^\prime$ both $\bar{q}^\prime$ and $\tilde{q}^{\prime\ast}$ obey the same evolution, such that the boundary condition \labelcref{eqn:209E} implies for all $t$
\begin{equation}\label{eqn:209F}
	\bar{q}^\prime(t)
	= \tilde{q}^{\prime\ast}(t)
	= q^{\prime\ast}(t),
\end{equation}
where we identify $q^\prime$ with $\tilde{q}^\prime$. Expectation values of local observables can be computed by the quantum rule
\begin{alignedeqn}\label{eqn:209G}
	\langle A(t)\rangle
	&= q^\prime(t)^\dagger A_D^\prime(t) q^\prime(t)\\
	&= \langle q^\prime|A_D^\prime|q^\prime\rangle
	= \bar q^\tran(t) \, A^\prime(t) \, \tilde q(t),
\end{alignedeqn}
with $\langle q^\prime| = (q^\prime)^\dagger$. With the translation \labelcref{eqn:Q19} of the operator $A_D^\prime(t)$ in the quantum system to the operator $A^\prime(t)$ in the classical statistical system the expectation value can be computed in both systems. We can identify $q^\prime(t)$ with the usual complex wave function. A further change of basis by a unitary matrix $D^\prime$ amounts to the standard change of basis in quantum mechanics.

The transition from a real to a complex formulation by a change of basis is a well known phenomenon, familiar from Fourier transforms. If we start with a real description the complex wave functions obey a constraint since only half of the components are independent. Often it is possible to introduce a complex structure already at a deeper level, for example by grouping $N$ real components of a wave function to $N/2$ complex components. The wave function and step evolution operator become then complex quantities without a constraint. Our formalism covers this case as well, cf. \cref{sec:complex structure}.

We conclude that a map from a real evolution law for real wave functions in classical statistics to the complex formulation of quantum mechanics does not encounter an obstacle in principle. We will see in the next section that the use of local complex similarity transformations greatly enhances the possibilities.

\subsection{Evolution operator and continuum limit}

We next ask if it is necessary that the step evolution operator $S$ of a classical statistical system can be mapped to a unitary evolution operator of an associated quantum system. The answer is no. It is only needed that $S$ can be mapped to some $S^\prime$ which generates in the continuum limit a Hermitian Hamiltonian. This can be achieved even if the step evolution operator $S^\prime$ is not unitary. The continuum limit ``forgets'' parts of the information contained in the step evolution operator. More precisely, $S^\prime$ must coincide with the unitary evolution operator of the quantum system only up to corrections that do not contribute $\sim \epsilon$ in the limit $\epsilon \to 0$, e.g. terms $\sim \epsilon^2$. This additional freedom further enlarges the equivalence class of systems that lead to the same expectation values and correlations of observables.

In quantum mechanics the time evolution of the wave function is described by the evolution operator
\begin{equation}\label{eqn:Q13}
	q(t)
	= U(t,t^\prime) \, q(t^\prime).
\end{equation}
The evolution operator can be obtained by integration of the Schrödinger equation. In particular, for small $\epsilon$ one has
\begin{alignedeqn}\label{eqn:Q14}
	U_\epsilon(t)
	&= U(t + \epsilon,t)
	= 1-i\epsilon H(t),\\
	q(t + \epsilon)
	&= U_\epsilon(t) \, q(t)
	= q(t)-i\epsilon H(t) \, q(t).
\end{alignedeqn}
In general, $U_\epsilon(t)$ can differ from the step evolution operator $S(t)$. An example is a memory material describing the motion of a single particle in two dimensions \cite{CWIT}. It is characterized by a step evolution operator
\begin{equation}\label{eqn:Q15}
	S(x,y)
	= \delta(x-\epsilon,y),\\
\end{equation}
with corresponding Hamiltonian
\begin{equation}
	H(x,y)
	= \frac{i}{2 \epsilon} \bigl(\delta(x-\epsilon,y) - \delta(x,y-\epsilon)\bigr).
\end{equation}
To order $\epsilon$ the evolution operator of quantum mechanics is given by
\begin{alignedeqn}
	U_\epsilon(x,y)
	&= \delta(x,y) - i \epsilon H(x,y)\\
	&= \delta(x,y) + \frac{1}{2} \bigl(\delta(x-\epsilon,y) - \delta(x,y-\epsilon)\bigr).
\end{alignedeqn}
Both $S$ and $S^\prime = U_\epsilon$ lead indeed to the same $W$ or $H$, and therefore to the same evolution of the wave function. We observe, however, that in contrast to $S$ the matrix $U_\epsilon$ is neither positive nor has it the unit jump character.

This highlights an important property of the continuum limit. In the continuum limit only the operator $W$ defined by \cref{eqn:FF1} matters. Different $S$ can lead to identical $W$ and therefore an identical continuum time evolution. This reflects the fact that many different discrete formulations can lead to the same continuum limit. Once one restricts the system to sufficiently smooth wave functions the detailed properties of the discrete formulation no longer matter for the expectation values of local observables. We can again construct a weight distribution $w^\prime[n]$ in analogy to \cref{eqn:28C}, replacing $S \to S^\prime = U_\epsilon$, without changing $q(t_\text{in}) = \bar{q}(t_\text{in})$. The result is not necessarily a classical overall probability distribution, but it entails the same ``local physics'' as $w[n]$, e.g. the same expectation values of local observables.

These considerations extend to a continuum limit in other variables, such as $x$ in the Ising-type model for fermions in \cref{sec:simple models}. The continuum limit in the $x$-direction is associated to an infinite number of states $N \to \infty$. The ``loss of memory'' of the continuum limit is one of the reasons why the restriction to (rescaled) unit jump chains for a realization of quantum systems by classical statistical systems no longer holds for $N \to \infty$.

\subsection{Strong local equivalence classes}

In the occupation number basis we may consider general weight functions $w[n]$, as defined by \cref{eqn:28B,eqn:28C}. For arbitrary $S_{\rho\tau}(t)$ they may no longer correspond to overall classical probability distributions. We can define ``strong local equivalence classes'' of all $w[n]$ that lead to the same density matrix $\rho^\prime(t)$ for all $t$. (A different and weaker notion of equivalence requires only the same local probability distribution $p_\tau(t)$ \cite{CWQM,CWPT}.) Since expectation values and correlations of local observables are completely determined by $\rho^\prime(t)$, all members of a given equivalence class will lead to the same local physics. For example, one may envisage many different ``histories'' of $t$-dependent $S(t)$ that all lead for $\bar t$ in a given interval to the same $\rho^\prime(\bar{t})$. The different weight distributions $w[n]$ corresponding to these different sequences of $S(t)$ all belong to the same equivalence class. Furthermore, for sufficiently smooth wave functions a change $S \to S^\prime$ that leads to the same continuum evolution of $\tilde{q}(t)$ and $\bar{q}(t)$ relates again different members of the same local equivalence class. With a given evolution of $\tilde{q}(t)$ and $\bar{q}(t)$ it is sufficient to identify one particular member of the local equivalence class to be a positive probability distribution, $\bar{w}[n] = p[n]$, in order to realize this evolution by a classical statistical system.

Local equivalence classes are a central concept for the understanding why classical statistical systems can describe quantum systems despite common belief that this is impossible due to no-go theorems. First, we know that many different classical observables are mapped to the same local operator $A^\prime(t)$. We may again group all observables that are represented by the same $A^\prime(t)$ into a local equivalence class of observables. The expectation values of all observables in the given equivalence class are the same. If the result of $n$ immediately consecutive measurements of the observable is represented by the operator $[A^\prime(t)]^n$, also the possible outcomes of measurements of all observables in a given equivalence class are the same. The Kochen–Specker theorem \cite{KS} assumes that to each local quantum operator a unique classical observable can be associated. This is not realized in our setting, since the map from classical observables to quantum operators $A^\prime(t)$ is not invertible. The existence of local equivalence classes circumvents \cite{CWQM} the contradictions of the Kochen–Specker theorem.

Bell's inequalities \cite{BE} follow if correlations of observables are given by the classical correlation between classical observables, e.g.
\begin{equation}\label{eqn:cl correlation func}
	\langle A B\rangle_\text{cl}
	= \int \mathcal{D} n \, w[n] \, A[n] \, B[n].
\end{equation}
We first note that not all local observables $A(t)$ have to correspond to classical observables. For example, the derivative operator $\hat{P} = -i \partial_x$ in the Ising-type model for fermions of \cref{sec:simple models} has no obvious expression as a classical observable $P[n]$. In this case a central assumption for Bell's inequalities is not realized. Furthermore, even if $A[n]$ and $B[n]$ can be realized as classical observables, the correlation function relevant for measurements may not be given \cite{CWQM} by the classical correlation function \labelcref{eqn:cl correlation func}. Indeed, if the conditional probabilities relevant for sequences of measurements are computable from the local probabilistic information, the classical correlation function is typically excluded \cite{CWQM}. The classical correlation functions $\langle A_1 B\rangle_\text{cl}$ and $\langle A_2 B\rangle_\text{cl}$ of two observables $A_1$ and $A_2$ in the same equivalence class typically differ from each other. In contrast, conditional correlations based on the operator product
\begin{equation}
	\langle A B\rangle_m
	= \langle A^\prime B^\prime\rangle
	= \frac{1}{2} \tr\bigl(\rho^\prime \{A^\prime,B^\prime\}\bigr)
\end{equation}
are compatible with the local equivalence class. They are more suitable for describing local measurements \cite{CWQM} and need not obey Bell's inequalities.

\section{Connections between classical and quantum statistics}
\label{sec:classical and quantum statistics}

In this section we employ the local similarity transformation \labelcref{eqn:Q17,eqn:Q18,eqn:Q19} in order to clarify some general structural features of the quantum formalism for classical statistical systems. This sheds light on the connection between quantum statistics and classical statistics.

In this section we keep fixed basis functions $h_\tau$ corresponding to the occupation number basis. For the unitarity evolution of arbitrary quantum systems we construct a ``classical basis'' for which the weight distribution is positive. We also discuss for arbitrary classical statistical systems a ``unitary basis'' for which the evolution is unitary. Classical and quantum systems related by these similarity transformations are weakly equivalent. Furthermore, classical statistical systems admit a Heisenberg picture for which the wave function is constant while the evolution is expressed in terms of $t$-dependent operators.

\subsection{Classical basis for quantum systems}

For the unitary evolution of an arbitrary quantum system there exists a choice of $t$-dependent local basis for which the transfer matrices or step evolution operators have only positive or zero elements. In this ``classical basis'' the weight distribution is a classical statistical probability distribution. More in detail, we start with the complex weight distribution $w[n]$, constructed according to \cref{eqn:28B,eqn:28C} for some sequence of unitary matrices $\{S(t)\}$. For orthogonal $S(t)$ the weight distribution is real, but not positive. We construct here a local similarity transformation that transforms $w[n]$ to a ``classical basis'' for which $w^\prime[n] = p[n]$ is real and positive. In this classical basis the density matrix $\hat{\rho}^\prime(t)$ can be computed from the classical statistical probability distribution according to \cref{eqn:D12a}. \Cref{eqn:D4a} permits the extraction of the density matrix $\rho^\prime(t)$, from which expectation values and correlations of local observables can be computed.

We start with the case where the step evolution operator $S$ of the quantum system is a time independent unitary matrix. As a first step we transform $S$ to a rescaled unique jump operator $S^{\prime\prime}$ for which all non-zero elements obey $|S_{ij}^{\prime\prime}| = 1$. This can be done by a unitary matrix. Any unitary (in particular also orthogonal) matrix $S$ can be diagonalized by a unitary transformation
\begin{alignedeqn}\label{eqn:CB1}
	\tilde{S}
	&= W S \, W^\dagger,
	\qquad
	W^\dagger \, W
	= 1,\\
	\tilde{S}
	&= \diag (\lambda_i),
	\qquad
	|\lambda_i|
	= 1.
\end{alignedeqn}
The rescaled unit jump operators $S^{\prime\prime}$ are also unitary, such that
\begin{alignedeqn}\label{eqn:CB2}
	\tilde{S}^{\prime\prime}
	&= \tilde{W}^\dagger \, S^{\prime\prime} \, \tilde{W},
	\qquad
	\tilde{W}^\dagger \, \tilde{W}
	= 1,\\
	\tilde{S}^{\prime\prime}
	&= \diag(\lambda_i^{\prime\prime}),
	\qquad
	|\lambda_i^{\prime\prime}|
	= 1.
\end{alignedeqn}
If we choose $S^{\prime\prime}$ such that the sequence of eigenvalues $\lambda^{\prime\prime}$ coincides with the sequence $\lambda$, one has $\tilde{S}^{\prime\prime} = \tilde{S}$ and therefore
\begin{equation}\label{eqn:CB3}
	S^{\prime\prime}
	= V S \, V^\dagger,
	\qquad
	V
	= \tilde{W} \, W,
	\qquad
	V^\dagger \, V
	= 1.
\end{equation}
If furthermore both $S$ and $S^{\prime\prime}$ are real, $V^\tran V$ commutes with $S$
\begin{equation}\label{eqn:CB4}
	[V^\tran \, V,S]
	= 0.
\end{equation}
In this case one may be particularly interested in choices of $S^{\prime\prime}$ for which $V$ is orthogonal.

The second step brings the rescaled step evolution operator $S^{\prime\prime}$ to a standard form where all elements equal one or zero. This is achieved by a local change of basis \labelcref{eqn:ST2}, where all diagonal elements are phase factors,
\begin{equation}\label{eqn:CB5}
	\tilde{D}(t)
	= \diag \bigl(d_i(t)\bigr),
	\qquad
	|d_i(t)|
	= 1.
\end{equation}
Thus the matrix
\begin{equation}\label{eqn:CB6}
	S^\prime(t)
	= \tilde{D}(t + \epsilon) \, V \, S \, V^\dagger \, \tilde{D}^\ast(t)
\end{equation}
has only positive elements. For suitable boundary conditions the weight distribution $w^{\prime\prime}[n]$ is a normalized positive probability distribution. It defines the ``classical basis''. (This construction can be generalized for $t$-dependent $S(t)$.)

The local similarity transformation
\begin{equation}\label{eqn:CB7}
	D(t)
	= \tilde{D}(t) \, V
\end{equation}
transforms the operator for local observables according to \cref{eqn:Q19}. Typically, diagonal operators in one basis become non-diagonal in the other basis. Thus simple diagonal operators in the ``quantum basis'' with unitary $S$ may turn to rather complex operators in the ``classical basis'' with positive $S^\prime$, and vice versa. Nevertheless, our explicit construction of the classical basis demonstrates that arbitrary quantum systems with $t$-independent $S$ can be represented as classical probability distributions with the use of appropriate operators for observables.

The practical use of the map between the quantum and classical basis is greatest if the map \labelcref{eqn:CB7} is simple enough. If $S^{\prime\prime}$ is already a positive unique jump operator we can take $\tilde{D}(t) = 1$. For orthogonal $V$ the transformation \labelcref{eqn:Q19} is simply a global rotation. The elements $(A_D^\prime)_{\tau\rho}$ are then real linear combinations of the elements $A_{\tau\rho}^\prime$ or, for diagonal $A^\prime$, of the elements $\alpha_\tau = A_\tau$. If $A^\prime$ or $A_\tau$ are independent of $t$ these linear combinations are the same for all $t$. For practical use it is further needed that non-diagonal $(A_D^\prime)_{\tau\rho}$ can be determined directly from the classical statistical probability distribution, cf. ref. \cite{CWFGI}. One can then ``measure'' the expectation values of quantum observables by use of a classical probability distribution.

The condition for $S^{\prime\prime}$ to be a positive unique jump matrix requires the existence of such a matrix with the same eigenvalues as the step evolution operator $S$ of the quantum system. The eigenvalues of a unitary step evolution operator are phases
\begin{equation}\label{eqn:spp eigenvalues}
	\lambda_i
	= \exp(-i \epsilon \, E_i),
\end{equation}
where $E_i$ correspond to the eigenvalues of the Hamiltonian in the continuum limit. We conclude that the construction of the classical basis for quantum system is particularly simple if a positive unique jump operator $S^{\prime\prime}$ exists that has the spectrum of eigenvalues \labelcref{eqn:spp eigenvalues}. In the continuum limit this requirement is weakened - only the realization of an appropriate antisymmetric $W^{\prime\prime}$ by a positive step evolution operator $S^{\prime\prime}$ is needed. It is possible to construct positive step evolution operators with a rather rich spectrum of eigenvalues of the type \labelcref{eqn:spp eigenvalues} \cite{CWIT}.

The construction of a classical basis for arbitrary quantum systems is not limited to a $t$-independent $S$. Quite generally, we can construct iteratively a local similarity transformation by
\begin{equation}\label{eqn:explicit construction}
	D(t + \epsilon)
	= S^\prime(t) \, D(t) \, S^\dagger(t)
\end{equation}
with $D(t_\text{in}) = 1$. It transforms a chain of arbitrary unitary matrices $S(t)$ to an arbitrary chain of positive step evolution operators $S^\prime$. Each quantum system characterized by $S$ is weakly equivalent to a classical statistical system characterized by $S^\prime$.

The practical use of the existence of the transformation \labelcref{eqn:explicit construction} may be limited since simple operators for observables in the quantum system may transform to rather complicated operators in the equivalent classical statistical system. Nevertheless, the explicit construction \labelcref{eqn:explicit construction} shows that there can be no no-go theorems that obstruct the construction of a classical statistical system equivalent to a quantum system. One only has to admit sufficiently complex local observables $A(t)$ in the classical statistical system that typically do not take the form of classical local observables at $t$.

\subsection{Unitary basis for classical statistics}

The local similarity transformations \labelcref{eqn:Q17} leave the partition function invariant. They may be interpreted as local basis changes. If we admit a different basis at every $t$ the evolution of any quasi-local classical statistical system can be made unitary by an appropriate choice of a $t$-dependent ``unitary basis''.

In order to establish this result we show that for an arbitrary sequence of regular real positive matrices $\bar{S}(t)$ one can find sequences of matrices $D(t)$ such that $S^\prime(t)$ is unitary for all $t$. With
\begin{alignedeqn}\label{eqn:GB1}
	S^\prime(t)
	&= D(t + \epsilon) \, \bar{S}(t) \, D^{-1}(t),\\
	S^{\prime\dagger}(t)
	&= (D^{-1})^\dagger(t) \, \bar{S}^\tran(t) \, D^\dagger(t + \epsilon),
\end{alignedeqn}
the condition for unitarity of $S^\prime$ reads $S^\prime S^{\prime\dagger} = 1$,
\begin{equation}\label{eqn:GB2}
	B(t + \epsilon)
	= \bar{S}(t) \, B(t) \, \bar{S}^\tran(t),
\end{equation}
with
\begin{equation}\label{eqn:GB3}
	B(t)
	= D^{-1}(t) \, (D^{-1})^\dagger(t)
	= B^\dagger(t).
\end{equation}
\Cref{eqn:GB1} can be solved iteratively, starting at $t_\text{in}$ with $B(t_\text{in}) = 1$. What remains to be shown is that for arbitrary regular Hermitian $B$ there exists a suitable matrix $D^{-1}$ such that \cref{eqn:GB3} is obeyed. We can diagonalize $B$ by a unitary transformation,
\begin{equation}\label{eqn:GB4}
	B_d
	= U \, B \, U^\dagger
	= \diag(\lambda_b),
\end{equation}
with real positive eigenvalues $\lambda_b > 0$. The matrix $D^{-1}$ can then be chosen as
\begin{equation}\label{eqn:GB5}
	D^{-1}
	= U^\dagger \diag(\sqrt{\lambda_b}) \, U.
\end{equation}

While the ``unitary basis'' brings the evolution of the wave function $\tilde{q}^\prime$ to the unitary evolution of quantum mechanics, the differences between arbitrary classical statistical systems and quantum systems is now encoded in the form of the transformed operators \labelcref{eqn:Q19}. The matrices $D(t)$ are not unitary, cf. \cref{eqn:GB5}. In particular, for a unique equilibrium state the loss of information of boundary conditions appears now in the form of the transformed operators.

\subsection{Heisenberg picture for classical statistics}

For an arbitrary sequence of step evolution operators $\{S(t)\}$ one can always find a basis where the evolution is trivial. Indeed, for the choice
\begin{equation}\label{eqn:HP1}
	D(t + \epsilon)
	= D(t) \, S^{-1}(t)
\end{equation}
the transformed step evolution operator is trivial,
\begin{equation}\label{eqn:HP2}
	S^\prime(t)
	= D(t + \epsilon) \, S(t) \, D^{-1}(t) =1.
\end{equation}
For this choice of local basis the classical wave function is constant
\begin{equation}\label{eqn:HP3}
	\tilde{q}^\prime(t)
	= \tilde{q}^\prime(t_\text{in}),
	\qquad
	\bar{q}^\prime(t)
	= \bar{q}^\prime(t_\text{f}).
\end{equation}
\Cref{eqn:HP1} can be solved iteratively with $D(t_\text{in}) = 1$.

As for the Heisenberg picture in quantum mechanics, the evolution concerns now only the operators,
\begin{alignedeqn}\label{eqn:HP4}
	A_H^\prime(t)
	&= D(t) \, A^\prime(t) \, D^{-1}(t)\\
	&= \bar{U}^{-1}(t,t_\text{in}) \, A^\prime(t) \, \bar{U}(t,t_\text{in}),
\end{alignedeqn}
where we employ the iterative solution of \cref{eqn:HP1},
\begin{alignedeqn}\label{eqn:HP5}
	D^{-1}(t)
	&= \bar{U}(t,t_\text{in})\\
	&= S(t - \epsilon) \, S(t - 2 \epsilon) \dots S(t_\text{in} + \epsilon) \, S(t_\text{in}).
\end{alignedeqn}
If $A_H^\prime(t)$ can be computed, this offers a particularly simple way for extracting the dependence of the expectation value $\langle A(t)\rangle$ on the boundary conditions,
\begin{equation}\label{eqn:HP6}
	\langle A(t)\rangle
	= \bar{q}^{\prime\tran}(t_\text{f}) \, A_H^\prime(t) \, \tilde{q}^\prime(t_\text{in}).
\end{equation}
\Cref{eqn:HP6} corresponds to a standard formula in the transfer matrix formalism. We conclude that the transition between the Schrödinger and Heisenberg pictures, familiar from quantum mechanics, carries over to classical statistical systems.

\section{Conclusions}
\label{sec:conclusions}

The transfer matrix formalism in classical statistics can be regarded as a type of Heisenberg picture. If we employ operators for observables this introduces into the classical statistical description the notion of non-commuting operators. The analogy to the formalism of quantum mechanics becomes very direct if we choose a normalization of the transfer matrix where it plays the role of a step evolution operator.

The present paper supplements the Heisenberg picture for classical statistics by the associated Schrödinger picture. For this purpose we introduce a pair of classical wave function $\tilde{q}(t)$ and conjugate wave function $\bar{q}(t)$ that embody the \textit{local} probabilistic information available on a given hypersurface $t$. This can be extended to a classical density matrix $\rho^\prime(t)$. Similar to quantum mechanics, the advantage of the Schrödinger picture is that it gives direct access to the local probabilistic information. The $t$-dependence of the classical wave function and density matrix follows a linear evolution law. The superposition principle for wave functions in quantum mechanics is shared by the classical wave functions in classical statistics.

A key element of the present paper is the discussion of local observables $A(t)$ that are not classical local observables at $t$. In other words, they are not functions of Ising spins or occupation numbers at $t$. Nevertheless, these local observables admit a local probabilistic interpretation in terms of probabilities $p^{(A)}(t)$ to find given values in measurements. The probabilities $p^{(A)}(t)$ are computable from the classical density matrix $\rho^\prime(t)$. Expectation values and correlations of this extended class of local observables can therefore be computed from the local probabilistic information.

Similar to quantum mechanics we can associate to each local observable $A(t)$ a local operator $A^\prime(t)$. Expectation values and correlations of $A(t)$ can be computed from the standard quantum formula $\langle A(t)\rangle = \tr[A^\prime(t) \rho^\prime(t)]$. While classical local observables at $t$ are represented by operators that commute with each other, this is no longer the case for more general local observables. These new non-commuting structures representing observables (and not only non-commutativity between the transfer matrix and operators for observables) make the close analogy between quantum mechanics and classical statistics even more striking.

In quantum mechanics the knowledge of the wave function at a given $t$ allows not only the computation of expectation values of observables at $t$, like the position or the spin of a particle at $t$. The information stored in the wave function is sufficient for computing as well expectation values of time derivatives of local observables, or of local observables at different times. The same holds for classical statistical systems. The local information stored in the wave functions or density matrix at $t$ is sufficient for the determination of expectation values of observables at $t^\prime \neq t$. The operators associated to these additional observables typically do not commute with the ones for the classical local observables.

On the formal level the expectation values of the extended class of local observables can be computed in the transfer matrix formalism as well. To understand the probabilistic interpretation of possible measurements for this extended set of local observables the Schrödinger picture in terms of $t$-dependent wave functions and density matrix is crucial.

In contrast to the phase factor $e^{i S_\text{Q}}$ in Feynman's path integral for quantum mechanics, the corresponding weight factor $e^{-S_\text{cl}}$ for a classical statistical system is positive. The main difference between an arbitrary quantum system and an arbitrary classical statistical system concerns the presence of two different classical wave functions for the classical statistical system. The norm of a given wave function needs no longer be conserved, and the evolution is not unitary. Correspondingly, the classical density matrix needs not be symmetric. The corresponding modifications of the Schrödinger or von Neumann equation allow classical wave functions and density matrices to approach equilibrium values in the bulk of a material, even if they differ from their equilibrium values at the boundaries. This approach to equilibrium is characteristic for all systems for which the transfer matrix has a unique largest eigenvalue.

For static memory materials the set of largest eigenvalues of the transfer matrix contains more than one element. The corresponding step evolution operator $S$ has more than one eigenvalue with $|\lambda_i| = 1$. In this case memory of the boundary conditions is present in the bulk - the expectation values of observables in the bulk depend on the boundary conditions. Information can be transported in a static system from one boundary to another. For memory materials the subsystem where the memory is stored follows the unitary evolution of quantum mechanics. Such memory materials are quantum simulators. The dependence of expectation values of observables on the location of hypersurfaces in space reproduces the time-dependence of corresponding observables in the quantum system with the same Hamiltonian.

We have discussed standard concepts of quantum mechanics such as basis transformations, similarity transformations and symmetry transformations in the context of classical statistical systems. In particular, we have shown that it is possible to find local similarity transformations that map an arbitrary quantum system to an equivalent classical statistical system with positive weight distribution. For a suitable choice of classical observables the expectation values and correlations of observables in the quantum system can be evaluated in the corresponding classical statistical system. This ``existence proof'' of a map between quantum and classical statistical systems demonstrates that no no-go theorem excludes weak equivalence. Beyond general statements of equivalence one would like to realize a practical implementation of classical statistical memory materials as ``quantum simulators''. For this purpose it will be crucial to find simple enough classical probability distributions and sets of observables that account for interference, entanglement and similar phenomena.

\paragraph{Acknowledgement} This work is part of and supported by the DFG Collaborative Research Center ``\href{http://isoquant.uni-heidelberg.de/index_de.php}{SFB 1225 (ISOQUANT)}''.

\begin{appendices}
\crefalias{section}{appendix}
\renewcommand{\theequation}{\Alph{section}.\arabic{equation}}
\numberwithin{equation}{section}

\section{Boundary conditions and initial values}
\label{app:bcs and ivs}

For particular systems or subsystems we have encountered a unitary evolution. In this case the boundary problem can be described and solved in terms of quantum mechanics if the final boundary terms are chosen such that the conjugate wave function $\bar{q}$ coincides with the wave function $\tilde{q}$. In this appendix we generalize the discussion to arbitrary boundary terms. For simplicity we concentrate on models where the whole system follows a unitary evolution. If necessary, this discussion can be adapted to quantum subsystems.

\subsection{Overlap of wave functions}

For an arbitrary rotation matrix $S = R$ the time evolution is ``unitary'' and the norm of the wave function is preserved.
So far, we have concentrated on a choice of the final time boundary term $\bar{f}_\text{f} = f(t_\text{f})$ for which $\bar{f}(t) = f(t)$. One may ask what happens for an arbitrary value of $\bar{f}_\text{f}$. In this case $\bar{f}(t) = \bar{q}_\tau(t) \, h_\tau(t)$ can differ from $f(t)$, while the partition function retains the general form
\begin{equation}\label{eqn:108}
	Z
	= \bar{q}_\tau(t) \, \tilde{q}_\tau(t).
\end{equation}
In the language of quantum mechanics, with both $\bar{f}(t)$ and $f(t)$ normalized, $\sum_\tau\bar{q}_\tau^2 = \sum_\tau q_\tau^2 = 1$, \cref{eqn:108} denotes the overlap between two wave functions
\begin{equation}\label{eqn:109}
	Z
	= \langle \bar{q}(t)|\tilde{q}(t)\rangle.
\end{equation}
Since $\bar{q}$ and $\tilde{q}$ obey the same evolution law the overlap is the same for all $t$.

If, in addition, we require the normalization of the partition function $Z = 1$ one infers $\bar q(t) = \tilde q(t) = q(t)$. We conclude that for a unitary evolution the normalization of both $\tilde q(t_\text{in})$ and $\bar q(t_\text{f})$, together with $Z = 1$, is sufficient to imply the identification of $\bar q$ and $\tilde q$ for all $t$.

For an unrenormalized $Z$ the unrenormalized expression for the expectation value of a local occupation number involves again an overlap between two wave functions,
\begin{alignedeqn}\label{eqn:110}
	&\int\mathcal{D} n\bar{f}_\text{f} \, K \, n_\gamma(t) \, f_\text{in}
	= Z \int \mathcal{D} n \, w[n] \, n_\gamma(t)\\
	&= Z \langle n_\gamma(t)\rangle
	= \int \dif n(t) \, \bar{f}(t) \, n_\gamma(t) \, f(t)\\
	&= \langle\bar{q}(t) \, N_\gamma^\prime(t) \, q(t)\rangle.
\end{alignedeqn}
The functional integral still describes quantum mechanics, for a setting where $q(t)$ is determined by the initial wave function $f_\text{in}$, while $\bar{q}(t)$ depends on $\bar{f}_\text{f}$. The expression \labelcref{eqn:110} is now a matrix element of the operator $N_\gamma^\prime(t)$ between wave functions $\bar{q}$ and $\tilde{q}$. The expectation value needs the multiplication by $Z^{-1}$. Similarly, for arbitrary (unnormalized) $Z$ the density matrix is defined by
\begin{equation}\label{eqn:360A}
	Z \rho_{\tau\rho}^\prime(t)
	= \tilde{q}_\tau(t) \, \bar{q}_\rho(t),
\end{equation}
generalizing \cref{eqn:D3a}. \Cref{eqn:D8a} remains valid, such that
\begin{equation}\label{eqn:360B}
	\langle n_\gamma(t)\rangle
	= \tr\bigl\{\rho^\prime(t) \, N_\gamma^\prime(t)\bigr\}.
\end{equation}

The boundary problem reduces to two parts. The first is the computation of the density matrix $\rho^\prime$ for some given $t$. The second amounts to the solution of the evolution equation for $\rho^\prime$. In particular, one may compute $\rho^\prime(t_\text{in})$ at the ``initial boundary''. This involves the computation of $\bar{q}(t_\text{in})$ from $\bar{q}(t_\text{f})$. For this purpose one has to solve the evolution equation \labelcref{eqn:AF13} for the conjugate wave function. If one specifies boundary conditions directly in terms of the ``initial density matrix'' $\rho^\prime(t_\text{in})$ the first step is already done. One has to make sure, however, that a given $\rho^\prime(t_\text{in})$ is compatible with the positivity of the overall probability distribution, in particular with the positivity of $\bar{q}(t_\text{f})$.

\subsection{Initial value problem and two-field formalism}

One may be interested in functional integral expressions in terms of the initial density matrix $\rho^\prime(t_\text{in})$, constructed from $\bar{q}(t_\text{in})$ and $\tilde{q}(t_\text{in})$. So far our functional integral expressions involve $\bar{f}_\text{f}(t_\text{f})$. We therefore need an explicit expression of $\bar{f}_\text{f}(t_\text{f})$ in terms of $\bar{f}(t_\text{in})$. For a general classical statistical system this is difficult to achieve. From \cref{eqn:matrix AF13} we learn that an expression of $\bar{q}(t_\text{f})$ in terms of $\bar{q}(t_\text{in})$ involves factors of $S^{-1}$. While the relation between $S$ and quantities involving occupation numbers is provided by \cref{eqn:AF12}, no such relation exists, in general, for $S^{-1}$.

The situation changes for a unitary evolution where $\bar q$ and $\tilde q$ obey the same evolution law. We can then employ for $\bar f(t)$ a relation similar to \cref{eqn:AF5}, resulting in
\begin{equation}
	\bar{f}(t_\text{f})
	= \int \mathcal{D} \bar n \, K[\bar n] \, \bar{f}(t_\text{in}),
\end{equation}
where $\bar{f}(t_\text{in})$ depends on $\bar{n}(t_\text{in})$ and $\bar{f}(t_\text{f})$ on $\bar{n}(t_\text{f})$. We concentrate on $\bar q(t) = \tilde q(t) = q(t)$ or $\bar f(t) = f(t)$. We can then write the partition function \labelcref{eqn:partition function def} as a functional integral,
\begin{alignedeqn}\label{eqn:111}
	Z
	= \int &\mathcal{D} \bar{n} \, \mathcal{D} n \, f_\text{in} \, \bigl(\bar{n}(t_\text{in})\bigr) \, K[\bar{n}] \, K[n] \, f_\text{in}\bigl(n(t_\text{in})\bigr)\\
	&\times \delta \bigl(\bar{n}(t_\text{f}),n(t_\text{f})\bigr).
\end{alignedeqn}
The variables are effectively doubled, except for $t_\text{f}$ where we have the identification $\bar{n}(t_\text{f}) = n(t_\text{f})$.
One recognizes the analogy to the Schwinger-Keldysh formalism \cite{SK,Kel1}.
For a normalized initial wave function and normalized one step evolution operator the normalization $Z = 1$ follows from \cref{eqn:111} by construction. With $Z = 1$ and $\bar{f}_\text{f} = f\bigl(t_\text{f};n(t_\text{f})\bigr)$ the weight function \labelcref{eqn:AF1} becomes
\begin{alignedeqn}\label{eqn:186X1}
	w[n]
	= \int \mathcal{D} \bar{n} \, K[\bar{n}] \, f_\text{in} \, \bigl(\bar{n}(t_\text{in})\bigr) \, K[n] \, f_\text{in}\bigl(n(t_\text{in})\bigr)\\
	\times \delta\bigl(\bar{n}(t_\text{f}),n(t_\text{f})\bigr).
\end{alignedeqn}

Classical local observables can be made dependent on $n$ or $\bar{n}$, or one may use expressions that are symmetrized in both types of variables. As an example we consider an arbitrary function $A \bigl(\bar{n}(t),n(t)\bigr)$ that depends on $\bar{n}(t)$ and $n(t)$ at a given time $t$.
We want to compute the expression
\begin{alignedeqn}\label{eqn:113}
	\langle A(t)\rangle
	= \int &\mathcal{D} \bar{n} \, \mathcal{D} n \, f_\text{in}(\bar{n}_\text{in}) \, K[\bar{n}] \, A \bigl(\bar{n}(t),n(t)\bigr)\\
	&\times K[n] \, f_\text{in}(n_\text{in}) \delta \bigl(\bar{n}(t_\text{f}),n(t_\text{f})\bigr).
\end{alignedeqn}
Using the split
\begin{equation}
	K[n]
	= K_>(t) \, K_<(t),\\
\end{equation}
with
\begin{equation}\label{eqn:split factors}
	K_<(t)
	= \prod_{t^\prime=t_\text{in}}^{t-\epsilon} \mathcal{K}(t^\prime),
	\quad
	K_>(t)
	= \prod_{t^\prime=t}^{t_\text{f}-\epsilon} \mathcal{K}(t^\prime),
\end{equation}
and the definition \labelcref{eqn:AF5} for $f(t)$ this yields
\small
\begin{align}\label{eqn:114}
	\langle A(t)\rangle
	&= \int \mathcal{D} \bar{n}(t^\prime \geq t) \, \mathcal{D} n(t^\prime \geq t) \, \delta \bigl(\bar{n}(t_\text{f}),n(t_\text{f})\bigr)\\
	&K_>(t;\bar{n}) \, K_>(t;n) \, f\bigr(t;\bar{n}(t)\bigr) \, A \bigl(\bar{n}(t),n(t)\bigr) \, f \bigl(t;n(t)\bigr).\notag
\end{align}
\normalsize
The factors $K_>(t;\bar{n})$ and $K_>(t;n)$ correspond to \cref{eqn:split factors}, with arguments $\bar{n}(t)$ or $n(t)$, respectively.

We may employ the expression
\begin{align}\label{eqn:115}
	E \bigl(\bar{n}(t),n(t)\bigr)
	= \int & \mathcal{D} \bar{n}(t^\prime > t) \, \mathcal{D} n(t^\prime > t)\\
	&\times \delta\bigl(\bar{n}(t_\text{f}),n(t_\text{f})\bigr) \, K_>(t;\bar{n}) \, K_>(t;n)\notag
\end{align}
and obtain
\begin{align}\label{eqn:ev}
	\langle A(t)\rangle
	= \int \dif \bar{n}(t) \, \dif n(t) \, &E \bigl(\bar{n}(t),n(t)\bigr) \, A \bigl(\bar{n}(t),n(t)\bigr)\notag\\
	&\times f(t;\bar{n})f(t;n).
\end{align}
For $t$-independent $S$ one finds in the occupation number basis
\begin{equation}\label{eqn:116}
	E \bigl(\bar{n}(t),n(t)\bigr)
	= \bigl((S^\tran)^F S^F\bigr)_{\tau\rho} h_\tau(t;\bar{n}) h_\rho(t;n),
\end{equation}
with $F$ the number of time steps from $t$ to $t_\text{f}$, and matrix multiplications of $S$ and $S^\tran$.
In particular, a unitary time evolution with $S^\tran S = 1$ implies
\begin{equation}\label{eqn:117}
	E\bigl(\bar{n}(t),n(t)\bigr)
	= h_\tau(t;\bar{n}) \, h_\tau(t;n)
	= \delta\bigl(\bar{n}(t),n(t)\bigr).
\end{equation}
This expression only involves the basis functions at time $t$ and is independent of all details of the evolution.
Insertion of \cref{eqn:117} into \cref{eqn:114} yields, with expansion
\begin{equation}\label{eqn:118}
	A \bigl(\bar{n}(t),n(t)\bigr)
	= A_{\tau\rho}^\prime(t) \, h_\tau(t;\bar{n}) \, h_\rho(t;n),
\end{equation}
the expectation value
\begin{alignedeqn}\label{eqn:119}
	\langle A(t)\rangle
	&= \int \dif \bar{n}(t) \, \dif n(t) \, f(t;\bar{n}) \, \delta \bigl(\bar{n}(t),n(t)\bigr)\\
	&\hphantom{{}=\int \dif \bar{n}(t}\times A \bigl(\bar{n}(t),n(t)\bigr)f(t;n)\bigr)\\
	&= \sum_\tau q_\tau(t) \, A_{\tau\tau}^\prime(t) \, q_\tau(t).
\end{alignedeqn}
Only the diagonal elements of $A_{\tau\rho}^\prime$ contribute. The role of $E$ is an effective projection on the diagonal elements of $A^\prime$,
such that indeed various implementations of an observable in terms of $\bar{n}$ or $n$ yield the same diagonal elements and therefore the same expectation value.

The important message of the formula \labelcref{eqn:117} consists in the observation that for the initial value problem the expectation values are ``independent of the future''.
One can choose $t_\text{f}$ freely provided that it is larger than the largest time argument $\bar{t}$ of an occupation number appearing in a given observable.
The expectation values are then independent of $t_\text{f}$ and it is often convenient to take $t_\text{f} = \bar{t} + \epsilon$.
This ``independence from the future'' is a particular feature of a unitary time evolution.
It does not hold for general $S_{\tau\rho}$, as can be seen in \cref{eqn:116}.

\end{appendices}

\end{multicols}

\begin{multicols}{2}[{\printbibheading[title={References}]}]
	\printbibliography[heading=none]
\end{multicols}

\end{document}